\documentclass[aps,prl,amssymb,groupedaddress,nofootinbib,floatfix,reprint]{revtex4-1}
\pdfoutput=1
\usepackage{epsfig}
\usepackage[english]{babel}

\usepackage{amsfonts}
\usepackage{amssymb}
\usepackage{amsmath}
\usepackage{multirow}
\usepackage[lofdepth,lotdepth,caption=false]{subfig}

\usepackage{rotating} 

\def\Btheo{{B_\delta^{\textrm{th}}}}

\def\th{{\textrm{th}}}
\def\Bobs{{B_\delta^{\textrm{obs}}}}

\newcommand{\const}{\mbox{const.}}

\newcommand{\mrm}[1]{\mathrm{#1}}

\newcommand{\eq}[1]{(\ref{eq:#1})} 

\renewcommand{\to}{\rightarrow}
\renewcommand{\(}{\left(}
\renewcommand{\)}{\right)}
\renewcommand{\[}{\left[}
\renewcommand{\]}{\right]}

\renewcommand{\vec}[1]{\mathbf{#1}}
\newcommand{\vx}{\vec{x}}

\newcommand{\vk}{\vec{k}}

\newcommand{\fnl}{f_{\mbox{{\scriptsize NL}}}}

\def\={\nonumber &=}
\def\nn{\nonumber}
\def\&{{}&}

\def\({\left(}
\def\){\right)}
\def\[{\left[}
\def\]{\right]}
\def\<{\left\langle}
\def\>{\right\rangle}

\def\bk{{\bf k}}

\def\bx{{\bf x}}

\def\curl{\mathcal}

\def\eq{\begin{align}}
\def\qe{\end{align}}
\def\eqa{\begin{eqnarray}}
\def\qea{\end{eqnarray}}

\def\and{\quad \mbox{and} \quad}

\newcommand{\fig}[1]{Fig.~\ref{fig:#1}} 
\newcommand{\mycaption}[1]{\caption{\footnotesize{#1}}}

\def\fnl{{f_\textrm{NL}}}

\def\hatfnl{{ \hat f_\textrm {NL}}}

\def\bfnl{\kern2pt\overline{\kern-2ptf}_\textrm{NL}}

\def\barQ{\kern2pt\overline{\kern-2pt\curl{Q}}}

\def\barR{\kern2pt\overline{\kern-2pt\curl{R}}}

\def\setsize{\csname @setfontsize\endcsname \setsize}

\begin{document}

\title{Fast Estimation of Gravitational and Primordial
Bispectra in Large Scale Structures}

\author{M.M.~Schmittfull$^{1}$}
\author{D.M.~Regan$^{2}$}
\author{E.P.S.~Shellard$^{1}$}

\affiliation
{$^{1}$Centre for Theoretical Cosmology,
DAMTP,
University of Cambridge,
 CB3 0WA, UK\\
$^{2}$Department of Physics and Astronomy,
University of Sussex,
Brighton,
BN1 9QH, UK}

\date{\today}

\begin{abstract}
We present the implementation of a fast estimator for the full dark matter bispectrum of a three-dimensional particle distribution relying on a separable modal expansion of the bispectrum. The computational cost of accurate bispectrum estimation is negligible relative to simulation evolution, so the isotropic bispectrum can be used as a standard diagnostic whenever the power spectrum is evaluated. As an application we measure the evolution of gravitational and primordial dark matter bispectra in $N$-body simulations with Gaussian and non-Gaussian initial conditions of the local, equilateral, orthogonal and flattened shape. The results are compared to theoretical models using a 3D visualisation, 3D shape correlations and the cumulative bispectrum signal-to-noise, all of which can be evaluated extremely quickly. Our measured bispectra are determined by $\mathcal{O}(50)$ coefficients, which can be used as fitting formulae in the nonlinear regime and for non-Gaussian initial conditions. In the nonlinear regime with $k<2h\,\mathrm{Mpc}^{-1}$, we find an excellent correlation between the measured dark matter bispectrum and a simple model based on a `constant' bispectrum plus a (nonlinear) tree-level gravitational bispectrum. In the same range for non-Gaussian simulations, we find an excellent correlation between the measured additional bispectrum and a constant model plus a (nonlinear) tree-level primordial bispectrum. We demonstrate that the constant contribution to the non-Gaussian bispectrum can be understood as a time-shift of the constant mode in the gravitational bispectrum, which is motivated by the one-halo model. The final amplitude of this extra non-Gaussian constant contribution is directly related to the initial amplitude of the constant mode in the primordial bispectrum. We also comment on the effects of regular grid and glass initial conditions on the bispectrum.
\end{abstract}

\maketitle

\tableofcontents

\section{I. Introduction}

Observations of the cosmic microwave background (CMB) \cite{wmap7,shellard1006,Cooray2, FRS2} and large scale structure (LSS) \cite{citeulike:2833069,citeulike:9218385} are currently consistent with Gaussian primordial cosmological perturbations. Much effort has been undertaken in recent times to constrain a possible non-Gaussian contribution. Detection of a significant primordial bispectrum or trispectrum  would have major implications for the mechanism of inflation, possibly ruling out the simplest paradigm of canonical single-field slow-roll inflation. Given this possibility, the importance of developing methods that discriminate between the plethora of inflationary models is clear. Such general methods have been developed in the case of the CMB in \cite{shellard0812,shellard0912,shellard1006,RSF10,FRS2}. This work exploited the use of a separable expansion of the underlying bispectrum or trispectrum in order to greatly reduce the computational cost involved in analysing general shapes.

The CMB bispectrum is currently the most powerful and direct probe of primordial non-Gaussianity, and higher resolution temperature and polarisation data will soon be available from Planck. Nevertheless, interest in the possibility of using observables of large scale structure to test primordial non-Gaussianity has undergone a recent resurgence, due in part to the potentially three dimensional nature of the dataset as a result of its redshift dependence. In particular, the scale-dependent bias induced by non-Gaussian local initial conditions on the halo power spectrum has been shown in \cite{dalal08} to offer an additional and powerful probe. Much work has been undertaken to improve analytic predictions for the impact of non-Gaussian initial conditions on the matter and galaxy power spectrum and bispectrum (see for example \cite{sefusatti09,sefusatti1003,scoccimarro-couchman01,verde1111,Scocc2012,seljak-stochasticity1104,Sef2011,Scocc0906,Desj1105,kendrick-stochasticity,10035020} and references therein). In \cite{Verde2010, citeulike:9218385} constraints on local-type non-Gaussianity were found at levels competitive with those obtained using CMB data. Despite these advances it is notable that until recently only local type non-Gaussianity had been studied using large-scale structure, owing to the difficulty in generating generic initial conditions. This situation greatly improved due to work by Wagner and Verde \cite{wagner-verde1006,wagner-verde1102} and work by Scoccimarro et al.~\cite{scocci1108}. However, both of these approaches become inefficient for non-separable bispectra. In \cite{shellard1008} the possibility of using the separable expansion method, exploited to great effect in the case of the CMB \cite{shellard1006}, was explored. This approach was verified in \cite{shellard1108} to allow for a far more efficient generation of non-Gaussian initial conditions than had  previously been possible in the literature. Furthermore, the separable decomposition methodology allows for the study of non-separable shapes. This has allowed for a robust and truly general approach towards the generation of primordial non-Gaussian initial conditions for use in $N$-body simulations. In this paper we exploit this approach to set up and run $N$-body simulations with non-Gaussian initial conditions. In particular, we study initial conditions given by local, equilateral, orthogonal and flattened bispectra, respectively. We choose the flattened (or trans-Planckian) model as an explicit example of an inherently non-separable bispectrum.   We exploit the modal method to efficiently and accurately reconstruct the full 3D matter bispectrum at each redshift of interest for each of these non-Gaussian models. This allows us to correlate the results of the $N$-body simulations with analytic estimates, allowing us to identify clearly the regime at which nonlinear corrections become important. This applies first to accurately determining the gravitational bispectrum well into the nonlinear regime, which is necessary in order to differentiate the more subtle impact of primordial non-Gaussianity.  The correlation measure proves useful in distinguishing the different primordial shapes.  We also thoroughly test the impact of starting $N$-body simulations using either glass or regular grid initial conditions.   We emphasise that our purpose here is to study the detailed nature of the underlying matter bispectrum derived from $N$-body simulations.   These methods can be equally applied to efficiently extract the galaxy bispectrum from huge survey data sets or to predict the halo bispectrum from simulations in a realistic observational context, but this will be the subject of future work.  

The
paper is organised as follows.  In Section II we review the generation
of non-Gaussianity due to nonlinear gravitational evolution, together
with the physically-motivated models we will use to describe our
results.  In Section III we review primordial non-Gaussianity,
introduce the specific models studied in this paper and discuss
their impact on the matter bispectrum, proposing a new time-shift
model.  In Section IV
we discuss the bispectrum estimation methodology based on a separable
expansion which allows extremely efficient estimation of the full
$N$-body bispectrum, as well as the cumulative signal-to-noise and its
correlation to theoretically-predicted bispectra.  We discuss the
simulation setup, impact of glass initial conditions, validations and
convergence tests in Section V. In Section VI we present our results
for the gravitational bispectrum and the simple fits thereof, while
primordial bispectra in non-Gaussian simulations and their fits are
discussed in section VII.  Finally, in Section VIII we present our
conclusions. Readers familiar with the modal methodology and interested primarily in the
measured bispectra and testing of various fitting formulae may wish to start with Section VI and follow the references to the earlier sections given there. 

\section{II. The Distribution of Matter}

\subsection{Power spectrum and transfer functions}

The distribution of matter in the universe can be described by its
fractional overdensity $\delta(\vx)=(\rho(\vx)-\bar\rho)/\bar\rho$,
where $\bar \rho$ is the spatial average of the density $\rho(\vx)$.
An important prediction of cosmological models is the probability of
finding a certain configuration $\delta(\vx)$ in the universe.  The
simplest possibility is a Gaussian pdf, which is determined by the
two-point correlation function, or in Fourier space by the power
spectrum $P_\delta$,
\begin{align}
  \label{eq:power-delta}
  \langle \delta(\bk_1)\delta(\bk_2) \rangle&=(2\pi)^3 \delta_D(\bk_1+\bk_2) P_\delta(k_1).
\end{align}
Here we assumed statistical homogeneity, leading to the Dirac delta
function, and statistical isotropy, which implies that the power
spectrum only depends on the magnitude of the wavevector.

Perturbations from inflation are usually described by the comoving
curvature perturbation $\mathcal{R}$, which is related to the
primordial potential $\Phi$ by $\Phi=-3\mathcal{R}/5$, in linear
perturbation theory during matter domination.  The density
perturbation $\delta$ can be obtained using the linear Poisson
equation
\begin{align}
  \label{eq:poisson}
  \delta(\bk,z) = \frac{2}{3}\,\frac{k^2T(k)D(z)}{\Omega_mH_0^2}\,\Phi(\bk)
\equiv M(k,z)\Phi(\bk),
\end{align}
where $T(k)$ is the linear transfer function at low redshift
normalised to $T(k)=1$ on large scales and calculable with CAMB
\cite{camb}, and $D(z)$ is the linear growth function for
$\Omega_{\mathrm{rad}}=0$ \cite{citeulike:9387140} normalised to
$D(z)=1/(1+z)$ during matter domination.  Later we will also find it
convenient to use the growth function normalised to unity today, $\bar D(z)=D(z)/D(0)$.

At late times and on small scales, the linear treatment breaks down
and $N$-body simulations must be used to find the nonlinear transfer
of primordial to late time perturbations.  A widely used fit of the
nonlinear power spectrum is HALOFIT \cite{halofit}, which is included
in CAMB \cite{camb}.  However, in our own bispectrum analysis we will
generally use the actual nonlinear power spectrum measured from the
$N$-body simulations.

\subsection{Matter bispectrum}

\begin{figure}[b]
\centering
\includegraphics[width=0.5\textwidth]{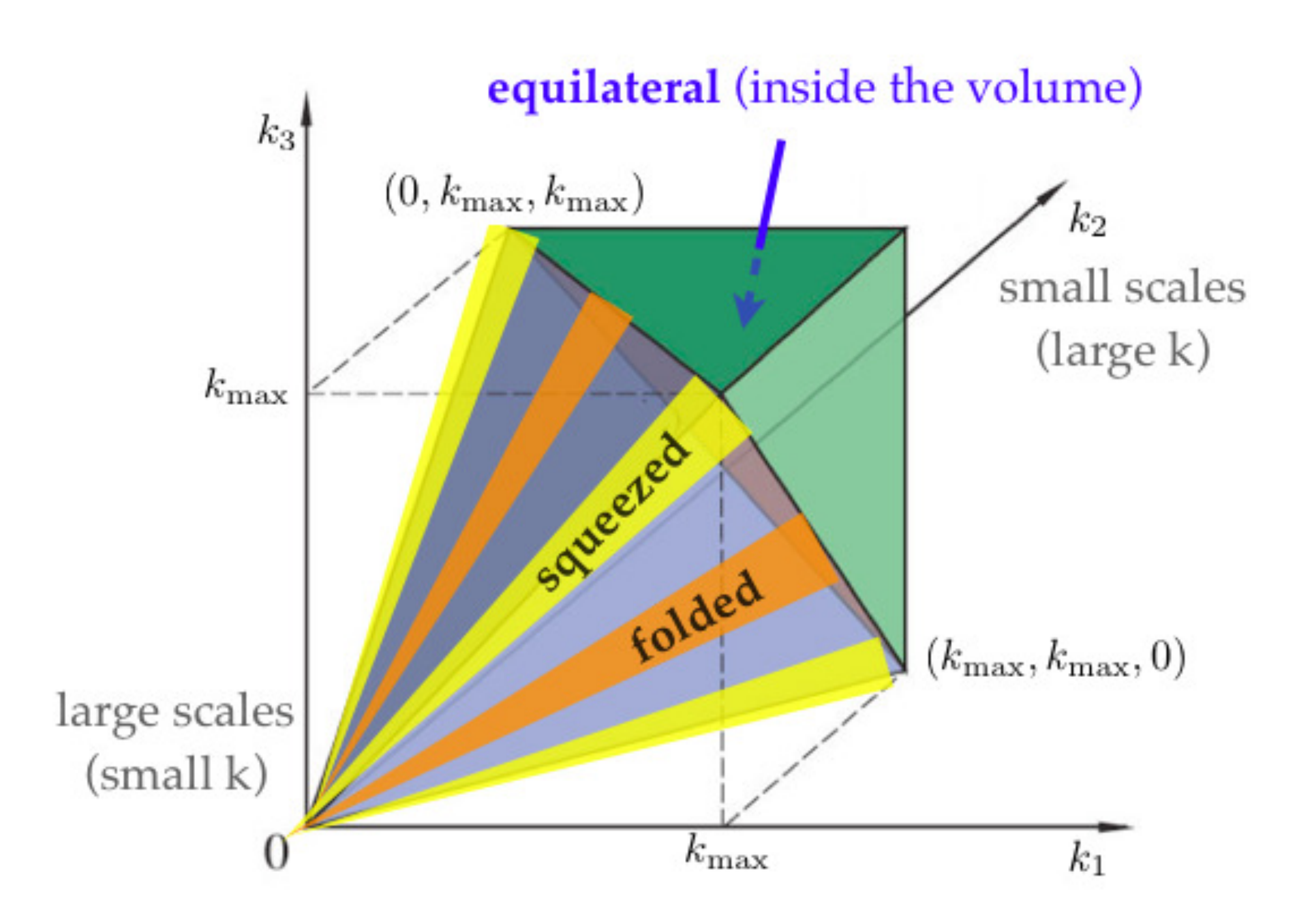}
\caption{Space of triangles with sides $k_1,k_2,k_3$,
  i.e.~each point inside the tetrapyd volume corresponds to a
  triangle configuration. Squeezed, folded and
  equilateral configurations are highlighted.  }
\label{fig:tetrapyd}
\end{figure}

At late times, the model of a Gaussian pdf for the density
perturbation is over-simplified because nonlinear gravitional collapse
will lead to higher order correlations.  To study deviations from
Gaussianity, it is useful to consider the \emph{bispectrum} $B_\delta$
(or three-point correlator transform), which is defined by
\begin{align}
  \label{eq:bispectrum-defn}
  \langle \delta(\bk_1)\delta(\bk_2)\delta(\bk_3) \rangle=(2\pi)^3 \delta_D\left(\Sigma_i \bk_i\right) B_\delta(k_1,k_2,k_3),
\end{align}
assuming again statistical homogeneity and isotropy. The Dirac delta
function imposes a triangle constraint on the bispectrum arguments
$k_1,k_2,k_3$. The space of triangle configurations is sometimes
called a tetrapyd \cite{shellard0912} and is shown in \fig{tetrapyd}.
Additionally to the triangle constraint, we impose a cut off at
$k_\mathrm{max}$, corresponding to the smallest scale under
consideration.  The bispectrum shape, i.e.~its dependence on the
wavenumbers $k_i$, gives important information about the physics which
induced the bispectrum.  Not only does it help to disentangle late
time non-Gaussianity, e.g.~induced by nonlinear gravitational
collapse, from primordial inflationary non-Gaussianity, but it also
opens the intriguing possibility of distinguishing between different
inflationary models.  To see this explictly we will review the
perturbative treatment of such non-Gaussianities in the following
sections.

\subsection{Tree-level gravitational matter bispectrum}

The equations of motion for a perfect pressureless fluid in a
homogeneous and isotropic universe contain nonlinearities, which
induce a non-vanishing bispectrum (see
e.g. \cite{colombi-review,scocci9612} for details). Specifically these are the
$\nabla \cdot (\delta\mathbf{v})$ term in the continuity equation and
the $(\mathbf{v}\cdot \nabla)\mathbf{v}$ term in the Euler equation,
where $\mathbf{v}$ is the peculiar velocity, which describes
deviations from the background Hubble flow.  To see how these terms
induce a non-Gaussian density perturbation one makes a perturbative
ansatz for the density perturbation (see \cite{colombi-review} for a review)
\begin{align}
  \label{eq:density-pert-expansion}
  \delta(\mathbf{k},t)=\sum_{n=1}^\infty a(t)^n
\delta_n(\mathbf{k})
\end{align}
and similarly for the peculiar velocity, which is assumed to be
described completely by its divergence, $\theta\equiv \nabla\cdot \mathbf{v}$. 
Then the equations of motion imply (see \cite{colombi-review} and
references therein)
\begin{align}
  \label{eq:density-Fn}
  \delta_n(\mathbf{k}) = \int  d^3\mathbf{q}_1\cdots \int d^3\mathbf{q}_n
\delta_D(\mathbf{k}-\mathbf{q}_1-\cdots
-\mathbf{q}_n)\nonumber\\
F_n^{(s)}(\mathbf{q}_1,\dots,\mathbf{q}_n)\delta_1(\mathbf{q}_1)\cdots \delta_1(\mathbf{q}_n),
\end{align}
and determine the kernels $F_n^{(s)}$, e.g. 
\begin{equation}
  \label{eq:F2-long-intro}
  F_2^{(s)}(\bk_1,\bk_2) = \frac{5}{7} + \frac{1}{2}\,
  \frac{\bk_1\cdot\bk_2}{k_1k_2} \(\frac{k_1}{k_2}+\frac{k_2}{k_1}\)
+ \frac{2}{7}\(\frac{\bk_1\cdot\bk_2}{k_1k_2} \)^2.
\end{equation}
Here the first and second terms come, respectively, from $\delta\;\nabla\cdot\mathbf{v}$
and $\mathbf{v}\cdot\nabla \delta$ in the continuity equation, while the last term comes from
$(\mathbf{v}\cdot \nabla)\mathbf{v}$ in Euler's equation
\cite{scocci9612}.

  At high redshift, the
expansion \eqref{eq:density-pert-expansion} is dominated by
 $\delta_1$ and $\delta_2\sim F_2^{(s)}\delta_1^2$.  If we assume $\delta_1$ to be Gaussian,
 the leading order contribution to
 the gravitational bispectrum is  given by 
\cite{fry84}
\begin{align}
  \label{eq:Bgrav}
    B_\delta^\mathrm{grav}(k_1,k_2,k_3) = 2P_\delta^L(k_1)P_\delta^L(k_2)F_2^{(s)} (\bk_1,\bk_2) +\mbox{2 perms},
\end{align}
where $P_\delta^L$ is the power spectrum of the linear perturbation
$\delta_1$. 

\begin{figure*}[htp]
\centering
\hspace{-0.9cm}
\subfloat[][]{
\includegraphics[height=0.22\textheight]{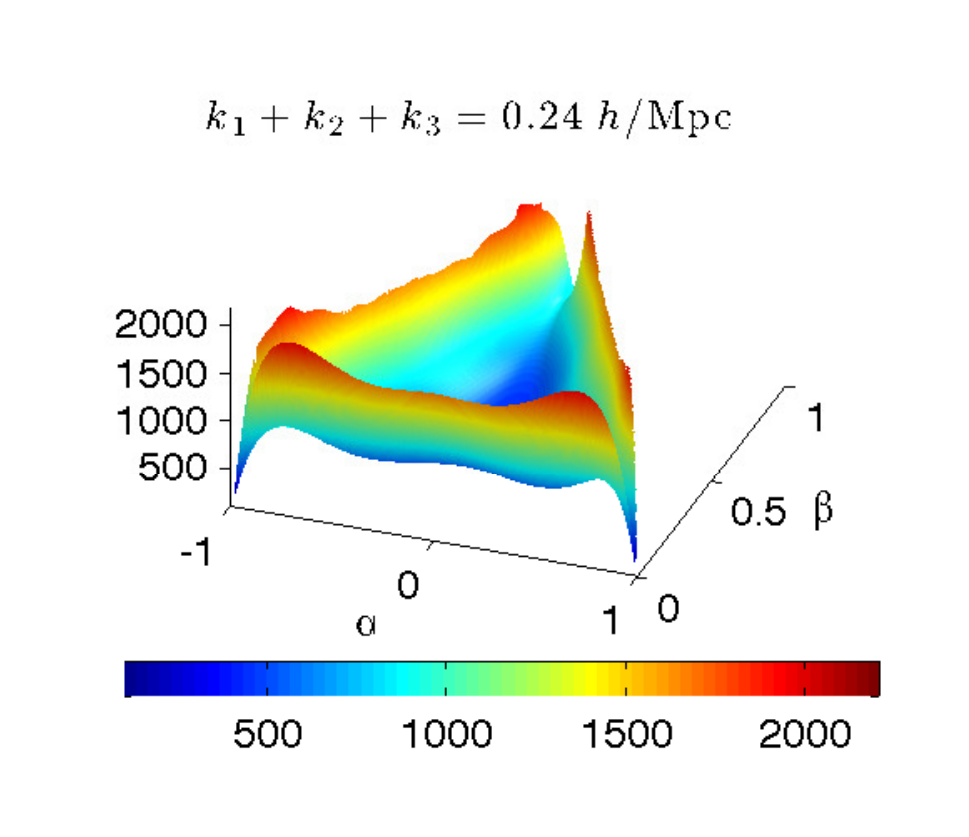}
\label{fig:Bgrav_2dslice}}
\subfloat[][]{
\includegraphics[height=0.22\textheight]{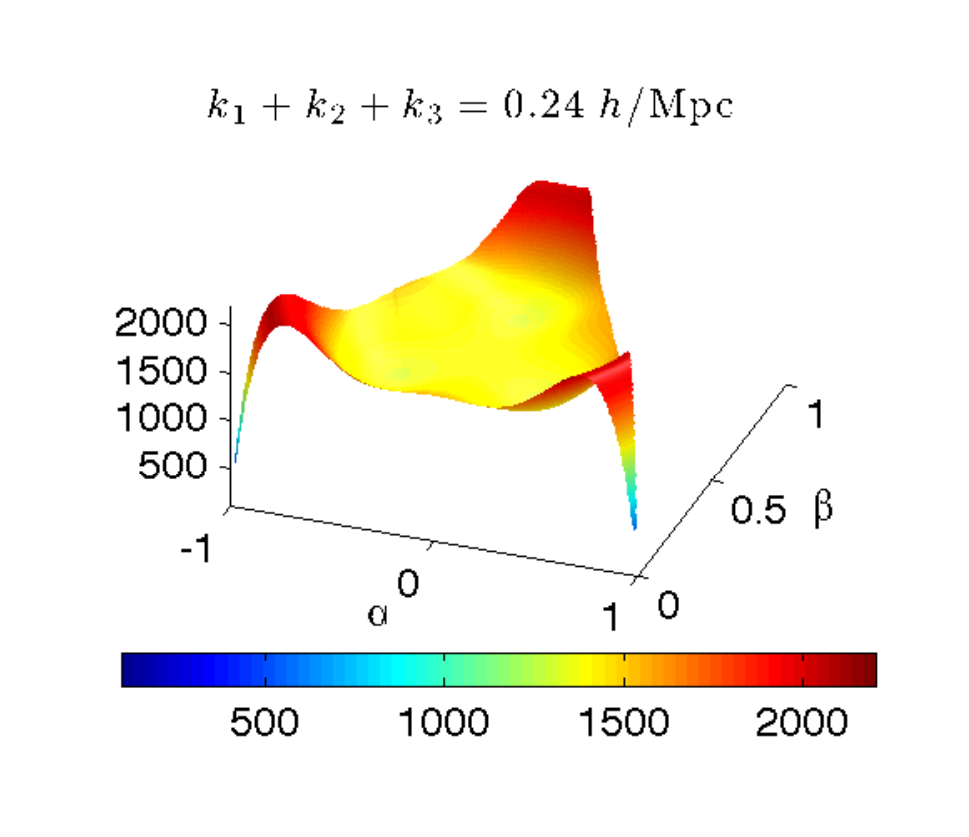}
\label{fig:Bgrav_constantF2}}
\subfloat[][]{
\includegraphics[height=0.22\textheight]{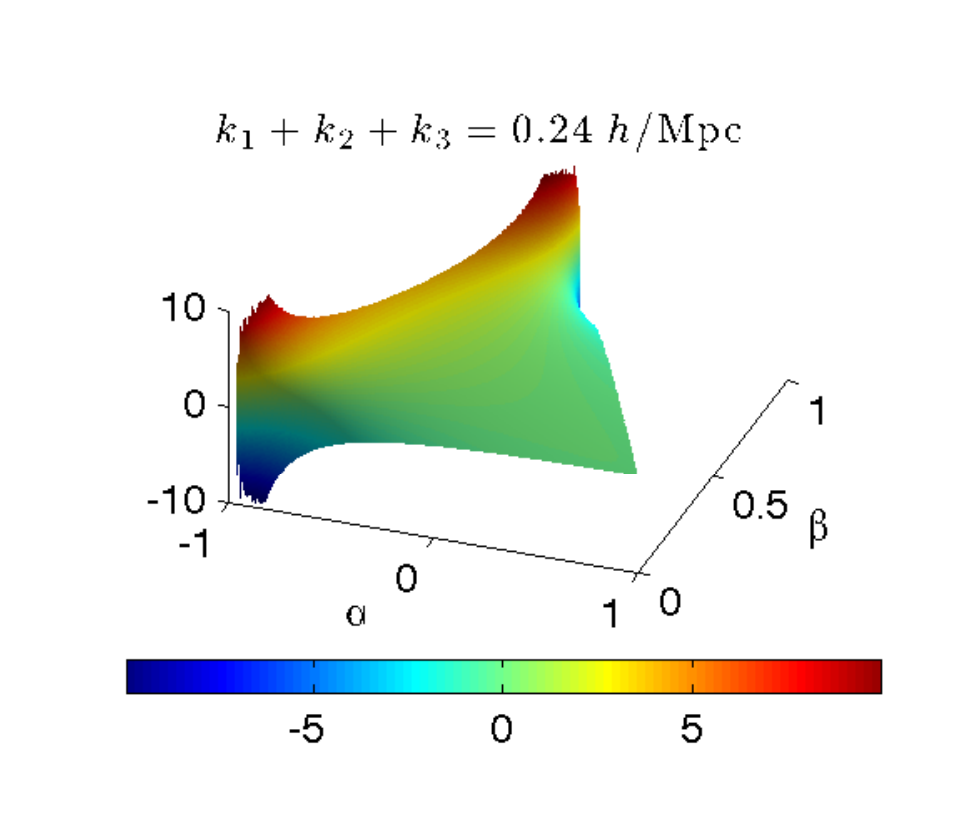}
\label{fig:Bgrav_F2only}} \mycaption{Two-dimensional slices of (a) the
leading order gravitational bispectrum $B_\delta^\mathrm{grav}$ from
\eqref{eq:Bgrav}, (b) the expression
$2P^L_\delta(k_1)P^L_\delta(k_2)+\mathrm{perms}$ and (c) the kernel
$F_2^{(s)}$ defined in \eqref{eq:F2-long-intro} (with cut-offs in the squeezed limit where the kernel diverges).  All slices are at
fixed $k_1+k_2+k_3=0.24h/\mathrm{Mpc}$ and we restrict the plot to
$k_i\ge 0.0013h/\mathrm{Mpc}$, corresponding to a $5000\mathrm{Mpc}/h$
box.  The linear power spectra are evaluated at $z=30$. }
\label{fig:bgrav_tree_discussion_all}
\end{figure*}

To discuss the shape of this gravitational bispectrum let us consider
two-dimensional slices through the tetrapyd shown in \fig{tetrapyd},
with $k_1+k_2+k_3=\mathrm{const}$. We denote one side of these
two-dimensional triangular slices as $\alpha$ and parameterise the
other direction with $\beta$.\footnote{Details can be found in
  \cite{shellard0812}. The slice parameters $\alpha,\beta$ should not
  be confused with the expansion coefficients $\alpha^{\{Q,R\}}_n$ and
  $\beta^{\{Q,R\}}_n$ to be defined later.}  A slice through the
gravitational bispectrum \eqref{eq:Bgrav} is shown in
\fig{Bgrav_2dslice} (see \cite{jeong09} for more slices at slightly
different length scales).  The bispectrum is maximal at the edges of
the plot, corresponding to flattened triangle configurations, where
$k_1+k_2=k_3$ or permutations thereof, i.e.~where the wavevectors
$\vk_i$ are parallel or anti-parallel to each other. However there is
a suppression in the corners of the plot, corresponding to squeezed
triangle configurations with $k_1\ll k_2\approx k_3$ or permutations
thereof.  Non-flattened and equilateral triangle configurations in the
centre of the plot are also suppressed.  This tree level gravitational
bispectrum is also illustrated in \fig{tet3dplots_show_expansions_grav_loc_eq_orth} as a function over the full tetrapyd.

To understand the basic shape of the gravitational bispectrum \eqref{eq:Bgrav} we plot in
\fig{Bgrav_constantF2} the expression
$2P_\delta^L(k_1)P_\delta^L(k_2)+2~\mathrm{perms}$, which corresponds to
replacing $F_2^{(s)}$ in \eqref{eq:Bgrav} by a constant.  Comparing
\fig{Bgrav_constantF2} with \fig{Bgrav_2dslice} shows that the
configuration dependence of the kernel \eqref{eq:F2-long-intro}, which is
induced by the terms containing scalar products, leads to an
enhancement of flattened and particularly folded configurations, where
two of the wavevectors $\vk_i$ equal each other. Non-flattened
configurations are relatively suppressed.  As we go along the edge of
the plot in \fig{Bgrav_2dslice}, from folded ($k_1=k_2=k_3/2$ or
permutations) to squeezed configurations ($k_1\ll k_2\approx k_3$ or
permutations), the bispectrum shape reflects the shape of the power
spectrum $P_\delta^L$, which peaks at $k_\mathrm{eq}\approx
0.02h/\mathrm{Mpc}$ and then decreases with decreasing $k_1$ because
$P_\delta^L(k_1)\propto k_1$ on large scales. 

Further discussion is required in the squeezed limit, where $\vk_2\approx -\vk_3$.  Let us
consider the regime where $k_1<k_\mathrm{eq}$ and
$k_2,k_3>k_\mathrm{eq}$.  The term with $P_\delta^L(k_2)P_\delta^L(k_3)$ in
\eqref{eq:Bgrav} is small since the small scale power spectra
 decrease rapidly with increasing $k_2, k_3$ and
$F_2^{(s)}(\vk_2,\vk_3)$ vanishes for $\vk_2=-\vk_3$.  The other two
permutations in \eqref{eq:Bgrav} depend on the angle between the large-scale wavevector $\vk_1$ and the small-scale wavevectors $\vk_2,\vk_3$.
First consider the case where the large scale is approximately
perpendicular to the two small scales, i.e.~$\vk_1\cdot \vk_2\approx
-\vk_1\cdot\vk_3\approx 0$. Then $F_2^{(s)}(\vk_1,\vk_2)\approx
F_2^{(s)}(\vk_1,\vk_3) \approx 5/7$ and
$P_\delta^L(k_1)P_\delta^L(k_2)$ and $P_\delta^L(k_1)P_\delta^L(k_3)$
decrease as we approach more squeezed triangles, implying a suppressed
bispectrum in the squeezed limit. In the other limit, the large-scale
wavevector $\vk_1$ is not orthogonal to the small scale wavevectors,
and so is aligned with one or other of $\vk_2$,  $\vk_3$.
Then the squeezed limit $k_1\to 0$ implies
$F_2^{(s)}(\vk_1,\vk_2)\propto k_1^{-1}\to \infty$ and
$F_2^{(s)}(\vk_1,\vk_3)\propto -k_1^{-1}\to
-\infty$, which can be seen in \fig{Bgrav_F2only}.  However, in the sum
\begin{align}
P_\delta^L(k_1)P_\delta^L(k_2) F_2^{(s)}(\vk_1,\vk_2) +
P_\delta^L(k_1)P_\delta^L(k_3) F_2^{(s)}(\vk_1,\vk_3),
\nonumber
\end{align}
 the two terms containing
$k_1^{-1}$ divergences in the kernel approximately cancel, because
$k_2\approx k_3$ and $\vk_1\cdot\vk_2\approx -\vk_1\cdot \vk_3$.
Also in the limit $k_1\to 0$, the divergences of the kernels
are regulated by the large scale power spectrum because
$k_1^{-1}P_\delta^L(k_1)=\const$ on very large scales.
\fig{Bgrav_2dslice}
shows that the divergences are indeed cancelled in the total
bispectrum \eqref{eq:Bgrav} and the squeezed limit is suppressed if
the large-scale wavenumber satisfies $k_1<k_\mathrm{eq}$.

\subsection{Gravitational matter bispectrum beyond tree level}

\subsubsection{Loop corrections}
The tree level prediction for the gravitational matter bispectrum
\eqref{eq:Bgrav} is only a good approximation on large scales and can
be improved by including so-called loop corrections, which were
derived for Gaussian initial conditions in \cite{scocci9612} and
extended to include non-Gaussian initial conditions in
\cite{sefusatti09}.  Important loop corrections can be included simply
by replacing the linear power spectrum $P_\delta^L$ by the nonlinear
power spectrum $P_\delta$ in the tree level expression
\eqref{eq:Bgrav} for Gaussian initial conditions, that is,
\begin{align}
  \label{eq:Bgrav_Pnonlinear}
  B_{\delta,\mathrm{NL}}^\mathrm{grav}\equiv 2P_{\delta}(k_1)P_{\delta}(k_2)F_2^{(s)}(\vk_1,\vk_2)
  + 2\mbox{ perms}\,.
\end{align}
We will find the expression \eqref{eq:Bgrav_Pnonlinear} to be a key
approximation to the full gravitational bispectrum as we probe beyond
the mildly nonlinear regime later.  Of course, this omits several loop
corrections containing powers of $F_2^{(s)}$ or higher order kernels,
but it is easy to evaluate with the nonlinear power spectrum from CAMB
\cite{camb}.  We will discuss further loop corrections much more
quantitatively in a forthcoming publication \cite{Reganetal2012}.

\subsubsection{Halo models}
At sufficiently small scales, well probed by our simulations, the
perturbative treatment breaks down and simulations and
phenomenological models must be used.  In the strongly nonlinear
regime the halo model can be used as a phenomenological model for the
dark matter distribution (see \cite{cooray-sheth2002} for a
review). Recently in Ref.~\cite{valageas1102} it was demonstrated that
combining 1-loop perturbation theory with the halo model describes the
matter bispectrum in simulations at the $\mathcal{O}(10\%)$ level on
nonlinear scales for equilateral and isosceles bispectrum
slices. Similar results were obtained in Ref.~\cite{figueroa1205},
where local non-Gaussian initial conditions were also considered and
compared with simulations for $k\leq 0.3h/\mathrm{Mpc}$ at $z\leq
1$. In the mildly nonlinear regime, when the approximation that all
dark matter particles are inside halos is no longer valid, the halo
model becomes less accurate than in the strongly nonlinear regime. It
should also be noted that important ingredients of the halo model like
the halo density profile and halo mass function cannot be derived
analytically but are obtained from fits to $N$-body simulations.

In detail, the halo model prediction for the dark matter bispectrum is
given by three contributions corresponding to the three cases that the
three points of the three point function lie in one, two or three
halos (see e.g.~\cite{cooray-sheth2002,valageas1102,figueroa1205}):
\begin{align}
  \label{eq:halomodel_sum}
  B_\mathrm{HM}=B_\mathrm{1H}+B_\mathrm{2H}+B_\mathrm{3H},
\end{align}
where
\begin{align}
  \label{eq:halomodel_1H}
  B_\mathrm{1H}(k_1,k_2,k_3) &= \frac{1}{\bar{\rho}^3}\int dm\, n(m)
  \prod_{i=1}^3 \hat{\rho}(k_i,m),
\end{align}
\begin{align}
  B_\mathrm{2H}(k_1,k_2,k_3) &= \frac{1}{\bar{\rho}^3}\int dm_1\, n(m_1)
\hat{\rho}(k_1,m_1) \int dm_2\nonumber\\
  \label{eq:halomodel_2H}
n(m_2)\hat{\rho}(k_2,&m_2)\hat{\rho}(k_3,m_2)
P_h(k_1,m_1,m_2) + \mathrm{perms},
\end{align}
\begin{align}
  B_\mathrm{3H}(k_1,k_2,k_3) &=
  \frac{1}{\bar{\rho}^3}\left[\prod_{i=1}^3\int dm_i\,
    n(m_i)\hat{\rho}(k_i,m_i)\right] \nonumber\\
  \label{eq:halomodel_3H}
&\quad \;\; B_h(k_1,m_1;k_2,m_2;k_3,m_3).
\end{align}
Here $n(m)$ is the mass function and $\hat{\rho}(k,m)$ is the Fourier
transform of the density profile of a halo of mass $m$, i.e.~for
spherically symmetric density profiles (like the commonly used NFW profile \cite{NFW_profile})
\begin{align}
  \label{eq:HM_rhohat}
  \hat \rho(k,m)=4\pi \int dr\,r^2 \rho(r,m)\frac{\sin(kr)}{kr}.
\end{align}
$P_h$ and $B_h$ denote the halo power spectrum and bispectrum, which
can be related to the tree level dark matter power spectrum and
bispectrum on large scales using bias relations.

\subsubsection{Constant  `halo'  model}

As we consider the more strongly 
nonlinear regime, $k\ge 1h/\mathrm{Mpc}$ at $z=0$, the halo model
bispectrum is dominated by the 1-halo contribution $B_\mathrm{1H}$.
Neglecting the overall bispectrum amplitude, we find that in this
regime the 1-halo shape for Gaussian initial conditions can be
approximated by a simple constant bispectrum,
\begin{align}
  \label{eq:Bconst_grav}
B_{\delta,\mathrm{const}}^\mathrm{grav}(k_1,k_2,k_3)| _{z}\equiv c_1 \, \bar
D^{n_h}(z)\,(k_1+k_2+k_3)^{\nu}  \,.
\end{align}
Here, `constant'  refers to constancy on
$k_1+k_2+k_3=\mbox{const.}$ slices (see \cite{shellard0812}), noting that this is a bispectrum
typical of isolated point-like objects (e.g.~particles randomly distributed in a box have a completely constant shot noise bispectrum).  The expression
\eqref{eq:Bconst_grav} has an overall wavenumber
scale-dependence with exponent $\nu$ and a time-dependence on the
linear growth function $\bar D(z)$ with halo exponent $n_h$. The wavenumber
scaling $\nu\approx-1.7$ is chosen such that it approximately reflects
the scaling for equilateral triangle configurations in this regime as
measured in our simulations (and which is approximately predicted by the halo
model for Gaussian initial conditions
\cite{valageas1102,figueroa1205}).  The exponent $n_h$ is similarly
defined by the fast growth factor appropriate for the halo model for
the scales under study $0.2h/\mathrm{Mpc}\lesssim k\leq
2h/\mathrm{Mpc}$, typically with $n_h \approx 6$--$8$.  For
$k_\mathrm{max}=2h/\mathrm{Mpc}$ and $z=0$, this simple model achieves
a shape correlation of $99.7\%$ with the 1-halo contribution
$B_\mathrm{1H}$ and $99.3\%$ with the full halo model bispectrum
\eqref{eq:halomodel_sum}.    

In later sections, we will note that \eqref{eq:Bconst_grav}  provides an excellent approximation to
the late-time bispectrum 
 in the nonlinear regime when we use it
in a simple fitting formula together with the modified tree-level gravitational bispectrum \eqref{eq:Bgrav_Pnonlinear}.
We shall investigate the other halo contributions in more detail
elsewhere \cite{Reganetal2012}.

As we shall see, on smaller nonlinear scales (large $k>1 h/\mathrm{Mpc}$) the fast bispectrum 
growth begins to slow down by the present day $z=0$.   In this case, we should really replace
the power law growth  $D^{n_h}(z)$ for the constant mode using a more general growth factor ${\cal T}(\tilde k, z, z_\mathrm{i})$, where
 the `slice' or `average' wavenumber 
\begin{align}
 \tilde k = (k_1+k_2+k_3)/3\,.
\end{align}
For future reference, it is convenient to use this growth rate in  a  more general integral form of the `constant' model  \eqref{eq:Bconst_grav}, that is, 
\begin{align}
  \label{eq:Bconst_gravnew}
B_{\delta,\mathrm{const}}^\mathrm{grav}(k_1,k_2,k_3)| _{z}&=  {\cal T}(\tilde k, z, z_\mathrm{i})\;B^\mathrm{init}_\mathrm{const}(\tilde k,z_\mathrm{i}) \\
\nn&\equiv  \int^z_{z_\mathrm{i}} {\cal G}(\tilde k, z)\, d z \;B^\mathrm{init}_\mathrm{const}(\tilde k,z_\mathrm{i})\,,
\end{align}
where $z_\mathrm{i}$ is the redshift at which this rapid `halo' growth
takes hold (it has an implicit $\tilde k$ dependence).  The quantity
$B^\mathrm{init}_\mathrm{const}(\tilde k,z_\mathrm{i})$ represents the
initial condition at $z=z_\mathrm{i}$ for the constant part of the
bispectrum.  Naively, we might take this to be
$B^\mathrm{init}_\mathrm{const}(\tilde
k,z_\mathrm{i})=B^\mathrm{grav}_\delta (\tilde k, \tilde k, \tilde
k)|_{z=z_\mathrm{i}}$, that is, the equilateral or constant part of
the tree-level gravitational bispectrum \eqref{eq:Bgrav} at
$z=z_\mathrm{i}$.  However, to date, determining the amplitude of this
initial constant bispectrum has relied on simulations.
 We will use this simple model to characterise both
the time- and scale-dependence of the gravitational bispectrum, as
well as primordial non-Gaussianity as we approach the strongly
nonlinear regime.

\subsubsection{Alternative phenomenological fit}
An alternative description of the gravitational matter bispectrum in
the non-perturbative regime was proposed in
Ref.~\cite{scoccimarro-couchman01}, who constructed a fitting formula
which interpolates between the perturbative prediction on large scales
and a local-type bispectrum on small scales, which was suggested by
early simulations.  In their terminology, the small-scale bispectrum
was denoted as a `constant reduced' bispectrum, which implies that it
takes the same form as the local shape $B_\delta\sim
P_\delta(k_1)P_\delta(k_2) + \hbox {perms}$, in contrast to the
constant shape \eqref{eq:Bconst_grav} above.  Recently
Ref.~\cite{verde1111} extended this fitting formula with updated
simulations into mildly nonlinear scales $0.03h/\mathrm{Mpc}\le k\le
0.4h/\mathrm{Mpc}$ at $0\le z\le 1.5$.

We review for the sake of completeness this phenomenological
$9$-parameter fit, for which we will test the regime of validity.  The
linear power spectrum in \eqref{eq:Bgrav} is replaced by the nonlinear
one $P_\delta$, and the symmetrised kernel $F_2^{(s)}$ is replaced by
\begin{align} \label{eq:gaussLoop}
F_2^{(s)\,\mrm{eff}}(\bk_1,\bk_2)&=\frac{5}{7}a(n_1,k_1)a(n_2,k_2)\nonumber\\
&+\frac{1}{2}\,
  \frac{\bk_1\cdot\bk_2}{k_1k_2} \(\frac{k_1}{k_2}+\frac{k_2}{k_1}\)b(n_1,k_1)b(n_2,k_2)\nonumber\\
  &+ \frac{2}{7}\(\frac{\bk_1\cdot\bk_2}{k_1k_2} \)^2 c(n_1,k_1)c(n_2,k_2)\,,
\end{align}
where 
\begin{align}
a(n,k)&=\frac{1+\sigma_8^{a_6}(z)\sqrt{0.7Q_3(n)} (a_1 q)^{a_2+n} }{1+(a_1 q)^{a_2+n}}\,,\\
b(n,k)&=\frac{1+0.2 a_3 (n+3) (qa_7)^{n+3+a_8}}{1+(q a_7)^{3.5+n+a_8}}\,,\\
c(n,k)&=\frac{1+4.5a_4/[1.5+(n+3)^4] (qa_5)^{n+3+a_9}}{1+(q a_5)^{3.5+n+a_9}}\,.
\end{align}
In these formulae, $n$ represents the slope of the linear power
spectrum at $k$, i.e.~$n=d\ln P^L_\delta(k)/d \ln k $ (with an
additional spline interpolation as described in \cite{verde1111}),
$q=k/k_{\rm{nl}}$ with $k_{\rm{nl}}$ defined by $k_{\rm{nl}}^3
P^L_\delta(k_{\rm{nl}})/(2\pi^2) =1$, and the function $Q_3(n)$ is
given by
\begin{align}
Q_3(n)=\frac{4-2^n}{1+2^{n+1}}\,.
\end{align}
The best-fit values for the free parameters $a_1 - a_9$ were found by
simulations \cite{verde1111}
\begin{align}
a_1&=0.484,\quad a_2=3.740,\qquad a_3=-0.849,\nonumber\\
a_4&=0.392,\quad a_5=1.013,\qquad a_6=-0.575,\nonumber\\
a_7&=0.128,\quad a_8=-0.722,\quad\,\, a_9=-0.926\,.
\end{align}

\begin{figure*}[p]
\centering
\includegraphics[height=0.9\textheight,angle=0]{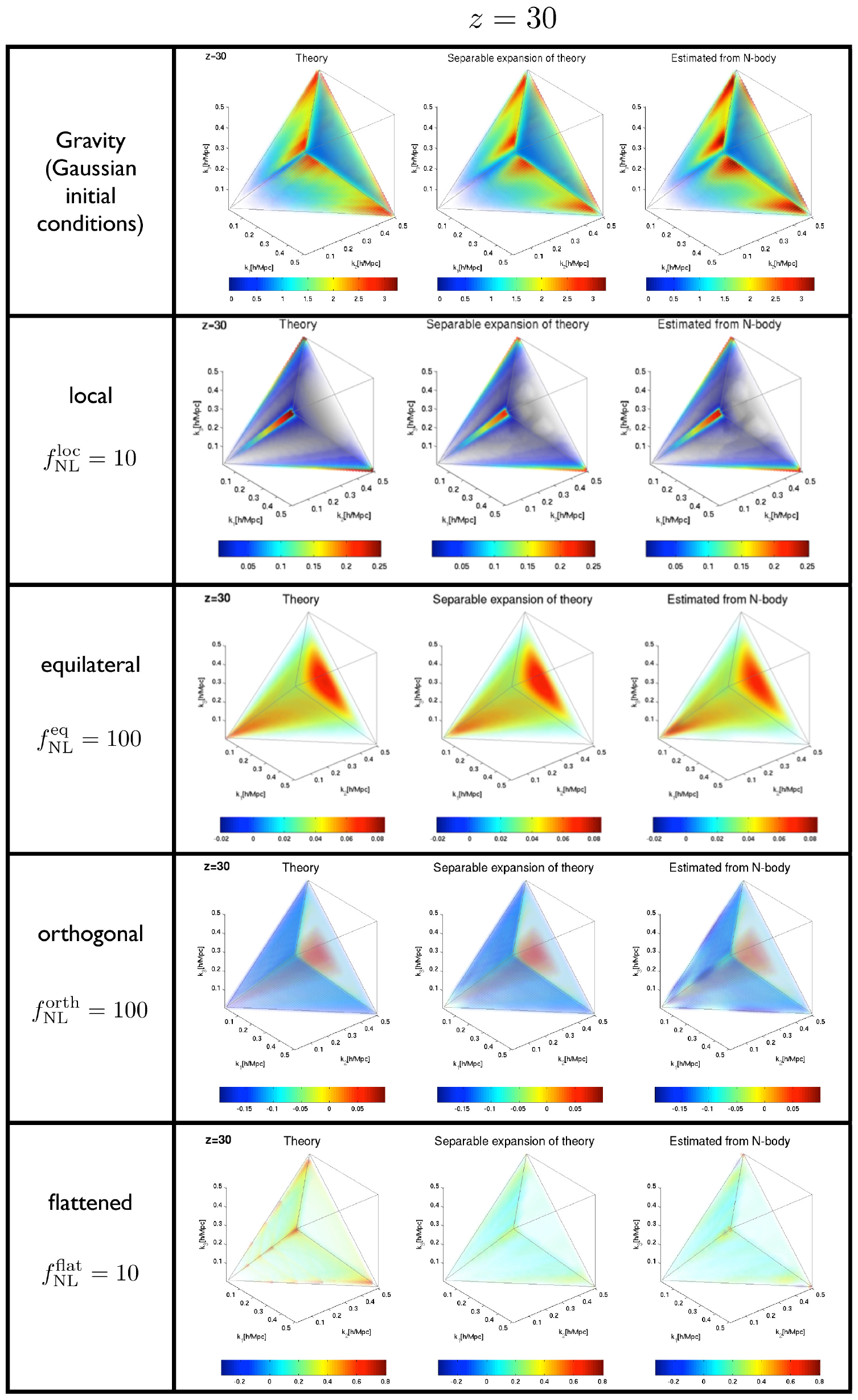}
\mycaption{Illustration of separable expansion \eqref{eq:bispDec2} for
  some theoretical tree level bispectra.  Each black cell of $3$ plots
  contains from left to right: Theoretical tree level prediction
  \eqref{eq:Bgrav} or \eqref{eq:bprim-tree}, separable expansion of
  the tree level theory \eqref{eq:bispDec2} and estimated bispectrum
  from $N$-body simulations \eqref{eq:bisp-from-betas} (for
  simulations G512b, Loc10, Eq100, Orth100 and Flat10 from top to
  bottom). 
 The plots show signal to noise weighted bispectra (as on the left hand side of 
\eqref{eq:bispDec2})
on the tetrapyd domain in \fig{tetrapyd}, evaluated at
  redshift $z=30$ for $512^3$ particles in a box with 
  $L=1600\mathrm{Mpc}/h$.  The opaqueness of the points reflects the
  absolute value of the signal to noise weighted bispectrum (values
  close to $0$ are completely transparent).  Colors and transparency
  are consistent within each black cell but differ across different
  black cells. The plot axes are $k_1, k_2, k_3 \le
  0.5h/\mathrm{Mpc}$.
 For better visibility we do not plot points with
  $k_1+k_2+k_3>2k_\mathrm{max}$ (corresponding to the green region in
  \fig{tetrapyd}).
}
\label{fig:tet3dplots_show_expansions_grav_loc_eq_orth}
\end{figure*}

\section{III. Primordial Non-Gaussianity}
Additional to the contribution $B_\delta^\mathrm{grav}$ from nonlinear
gravity, the matter bispectrum can have primordial contributions
$B_\delta^\mathrm{prim}$ from inflation or some other early universe
model such as cosmic defects.  While the simple model of single field
slow roll inflation gives only a small primordial bispectrum,
$f_\mathrm{NL}\sim\mathcal{O}(10^{-2})$, other models can yield large
non-Gaussianities with $f_\mathrm{NL}>1$ (see
e.g.~\cite{chen1002,shellard-review10,bartolo04,10036097,Yadav2010}
for reviews).  Such models can be distinguished if they induce
different bispectrum shapes, i.e.~different dependencies of the
bispectra on the momenta $k_1,k_2,k_3$ as illustrated in \fig{tet3dplots_show_expansions_grav_loc_eq_orth}. However, before reviewing
primordial bispectrum shapes we describe how primordial
non-Gaussianity changes the dark matter bispectrum.

\subsection{Primordial contribution to the matter bispectrum}
Let us assume that an inflationary model produces a primordial
bispectrum $B_\Phi$ with nonlinear amplitude $f_\mathrm{NL}$, i.e.\footnote{Non-linear corrections to the mapping between $\mathcal{R}$ and $\Phi$, which are not taken into account here, induce an additional bispectrum of $\Phi$, which is however smaller than the primordial contributions considered here \cite{bartolo04}.}
\begin{align}
  \label{eq:fnl}
\langle \Phi(\bk_1)\Phi(\bk_2)\Phi(\bk_3) \rangle&=(2\pi)^3 \delta_D(\Sigma_i \bk_i)f_\mathrm{NL} B_\Phi(k_1,k_2,k_3).
\end{align}
From the linear Poisson equation \eqref{eq:poisson} we see that the
primordial contribution to the matter bispectrum is given at leading
order by
\begin{align}
  \label{eq:bprim-tree}
  B_\delta^\mathrm{prim}&(k_1,k_2,k_3)| _{z}=\nonumber\\
  &M(k_1,z)M(k_2,z)M(k_3,z)f_\mathrm{NL}B_\Phi(k_1,k_2,k_3),
\end{align}
which is valid on large scales.

A simple improvement to the tree level shape which incorporates some
loop corrections can be obtained with the nonlinear power spectrum
$P_\delta$:
\begin{align}
  \label{eq:bprim-nonlinear-evolved}
  B_{\delta,\mathrm{NL}}^\mathrm{prim}(k_1,k_2,k_3)\equiv \sqrt{\frac{P_\delta(k_1)P_\delta(k_2)P_\delta(k_3)}{P_\Phi(k_1)P_\Phi(k_2)P_\Phi(k_3)}}B_\Phi(k_1,k_2,k_3)\,,
\end{align}
where $P_\Phi(k_i)$ refers to the primordial power spectrum at some
early time in matter domination.  As will be demonstrated later in
this paper, this shape can be used to obtain simple fitting formulae
for the primordial contribution to the matter bispectrum.  A more
systematic but also more cumbersome approach is to include loop
corrections, which become important on small scales
\cite{sefusatti09,sefusatti1003}.  We shall calculate more of these
contributions in a subsequent paper and test their correspondence to
the $N$-body simulations \cite{Reganetal2012}.

We note that, while the time dependence of the tree level
gravitational bispectrum \eqref{eq:Bgrav} is given by
$B_\delta^\mathrm{grav}\propto D^4(z)$, the tree level primordial
contribution \eqref{eq:bprim-tree} only grows like
$B_\delta^\mathrm{prim}\propto D^3(z)$, implying that it is easier to
extract the primordial contribution to the dark matter bispectrum at
early times.  The simulations and fitting formulae discussed later
will show that the gravitational bispectrum also grows faster than the
primordial contribution in the strongly nonlinear regime (also by a
factor of roughly $D(z)$).

\subsection{Non-Gaussianity as a halo model time-shift}
At sufficiently small
scales the perturbative treatment breaks down and simulations and
phenomenological models must be used. The phenomenological halo model
prediction for local non-Gaussian initial conditions was computed in
\cite{figueroa1205} by incorporating modified expressions for the halo
model ingredients in presence of local non-Gaussianity. While first
tests by \cite{figueroa1205} demonstrate that this approach works well
for some one-dimensional bispectrum slices in the mildly nonlinear
regime, $k\leq 0.3h/\mathrm{Mpc}$ and $z\leq 1$, comparisons to
simulations in the strongly nonlinear regime have not been undertaken
to date.

An alternative phenomenological model we propose here is to note that
the primordial bispectrum can also contribute to the halo model
bispectrum as an initial offset or time-shift.  As we have seen the
1-halo model is highly correlated with the simple `constant'
bispectrum we described earlier \eqref{eq:Bconst_grav}.  The key
point is that this constant halo contribution grows much more rapidly
than both the tree-level primordial bispectrum \eqref{eq:bprim-tree}
or the tree-level gravitational bispectrum \eqref{eq:Bgrav}.
Consequently, if the primordial bispectrum has a significant positive
(or negative) constant component, then this can act as an initial
condition for the halo bispectrum; the faster halo amplification
growth will start earlier (or later) by a time or redshift offset
$\Delta z$.  The constant part of the primordial signal, then, will
participate in the halo bispectrum growth and can be amplified much
faster than expected from the tree-level result
\eqref{eq:bprim-tree} or even with loop corrections
\eqref{eq:bprim-nonlinear-evolved}.  If this physical picture is correct, 
then the primordial constant contribution can be described
perturbatively around the constant halo bispectrum
\eqref{eq:Bconst_grav} by expanding the growth factor 
\begin{align}
  \label{eq:timeshift_perturbed_D}
\bar D^{n_h}(z+\Delta z) \approx \bar D^{n_h}(z) +n_h
\bar D^{n_h-1}(z)\frac{d\bar D(z)}{dz}\Delta z.  
\end{align}
We determine $\Delta z$ by matching
the projection of the total measured bispectrum $\hat B_\delta$ on the constant
shape $B_{\delta,\mathrm{const}}^\mathrm{grav}$ in the Gaussian and non-Gaussian
simulations:
\begin{align}
  \label{eq:timeshift_matching_fnl_const}
  (\hat f_\mathrm{NL}^\mathrm{const})_\mathrm{Gauss}(z+\Delta z) = 
  (\hat f_\mathrm{NL}^\mathrm{const})_\mathrm{NG}(z).
\end{align}
From simulations, the time shift $\Delta z$ and the corresponding shift of the growth function,
$\Delta \bar D=\frac{d\bar D(z)}{dz}\Delta z$, are both  found to vary over
time at most by a factor of $2$ for redshifts $10\gtrsim z \gtrsim 1$
for local, equilateral and flattened initial conditions (to the extent
to which this can be tested by linearly interpolating the limited
number of output redshifts at which the bispectrum was measured).
Provided $\Delta \bar D$ is time-independent
this perturbation of the simple `constant' model \eqref{eq:Bconst_grav} means we should be able to model the halo
contribution from the primordial perturbation as 
\begin{align}\label{eq:Bconst_prim}
  B_{\delta, \mathrm{const}}^\mathrm{prim}(k_1,k_2,k_3)\equiv c_2 \,
  \bar D^{n_h^\mathrm{prim}}(z)\,
 \left(k_1+k_2+k_3\right)^{\nu} \,, 
\end{align}
where we expect $n_h^\mathrm{prim}=n_h-1$ from
\eqref{eq:timeshift_perturbed_D} and again $\nu \approx -1.7$. 
The fitting parameter $c_2$ will be related to the correlation between the primordial
shape\footnote{We refer here to the excess bispectrum compared to
  Gaussian initial conditions as measured in $N$-body simulations at
  times when the constant contribution to the bispectrum is not
  negligible compared to the partially loop-corrected tree level
  contribution \eqref{eq:bprim-nonlinear-evolved}.  Therefore $c_2$ is
  a fitting parameter to be determined from $N$-body simulations.}
and the constant model at
the time at which halos form for the length scale under consideration.  
Different non-Gaussian bispectrum
models should show consistent behaviour in the nonlinear regime
depending on the relative magnitude of their constant component.  We
shall define these quantities more precisely after introducing the
bispectrum shape correlator in the next section, but
\eqref{eq:Bconst_prim} will be an important component in our later
fitting formulae for primordial non-Gaussianity.

The simple power law in the linear growth function $\bar D(z)$ used
to model the time dependence of the constant halo bispectrum can of
course be extended to more general functions of time \eqref{eq:Bconst_gravnew}, whose time
derivative would then enter in the expansion
\eqref{eq:timeshift_perturbed_D}.  The results presented
later show that the overall normalisation of the simple fits can be
improved by a more general modeling of the time dependence (reducing
the overall growth rate in the strongly nonlinear regime).  In this case, we 
can consider a time-shift for the more general `constant' model \eqref{eq:Bconst_gravnew}
\begin{align}\label{eq:Bconst_primnew}
 \nn & B_{\delta, \mathrm{const}}^\mathrm{prim}(k_1,k_2,k_3)| _{z} =   B_{\delta,\mathrm{const}}^\mathrm{grav}(k_1,k_2,k_3)| _{z+\Delta z}\\
 \nn &\qquad\qquad \qquad\qquad\quad  - B_{\delta,\mathrm{const}}^\mathrm{grav}(k_1,k_2,k_3)| _{z}\\
  &\qquad = \frac{d\mathcal{T}(\tilde k,z,z_\mathrm{i})}{dz} \;B^\mathrm{init}_\mathrm{const}(\tilde k,z_\mathrm{i})\Delta z + ... \,,
\end{align}
where we neglect higher order terms assuming them to be subdominant.   We can determine the time-shift $\Delta z$ by determining the magnitude of the constant part of the primordial bispectrum at $z=z_\mathrm{i}$.

While more
sophisticated models of the time evolution are left for future work, we note that the
correlation with the measured bispectrum shape cannot be improved much, 
because simulations are so well described by a combination of the
 'constant' shape \eqref{eq:Bconst_prim} and the modified tree-level gravitational 
 shape \eqref{eq:bprim-nonlinear-evolved} used in the fitting formulae presented later.

\subsection{Primordial bispectrum shapes from inflation}

\subsubsection{Local shape}
The fiducial $\fnl$ model of primordial non-Gaussianity is the local
model \cite{SalopekBond}, which is described in this way because it can be generated
simply by squaring a Gaussian field $\Phi_G$ in real space,
\begin{align}
  \label{eq:local-phi-square}
  \Phi(\vx) = \Phi_G(\vx)+f_\mathrm{NL}[  \Phi^2_G(\vx)-\<\Phi_G^2\>],
\end{align}
where $\<\Phi_G^2\>$ denotes an average over $\vx$ space and ensures
that the average perturbation is zero.    The resulting bispectrum takes the form
\begin{align}
\label{eq:Blocal}
  B_\Phi^\mathrm{loc}(k_1,k_2,k_3) =
  2\[P_\Phi(k_1)P_\Phi(k_2)+2 \;\mathrm{ perms}\],
\end{align}
which is illustrated in \fig{tet3dplots_show_expansions_grav_loc_eq_orth} and peaks at squeezed triangle configurations, where one wavenumber
is much smaller than the other two.  Multiple field inflation models
are one potential source of this shape (for a
review see \cite{chen1002}). If a bispectrum signal in the squeezed
limit is detected, then this will rule out all canonical single field
models of inflation \cite{maldacena03,creminelli0407,cheung0709}. 

\subsubsection{Equilateral shape}
Higher derivative operators in the inflationary action, arising
e.g.~in DBI inflation \cite{tong04} and in effective field theory
approaches \cite{cheung0709,baumann1102}, produce a shape that can be
approximated by the separable equilateral template
\cite{creminelli0509,chen1002,wagner-verde1102}
\begin{align}  \label{eq:equilateral-bisp}
\nonumber
  B_\Phi^\mathrm{eq}= 6\big[ & 
-(P_\Phi(k_1)P_\Phi(k_2)+2\mbox{ perms})  \\ 
& - 2(P_\Phi(k_1)P_\Phi(k_2)P_\Phi(k_3))^{2/3}\\
& +(P_\Phi^{1/3}(k_1)P_\Phi^{2/3}(k_2)P_\Phi(k_3)+5\mbox{ perms})
\big],\nonumber
\end{align}
which peaks in the equilateral limit, $k_1=k_2=k_3$.  For equilateral
triangles with $k_1=k_2=k_3$ summing up the corresponding three plane
waves, $\sum_j\Phi(\vk_j)e^{i\vk_j\vx}$, gives filamentary
overdensities, i.e.~overdense cylinders along the direction
perpendicular to the plane of the triangle of $(\vk_1,\vk_2,\vk_3)$,
surrounded by underdensities (as motivated in Ref.~\cite{lewis1107}).
A primordial equilateral shape from higher derivatives, therefore,
must be distinguished carefully from the stronger equilateral and
(highly correlated) constant contribution produced at late times by
nonlinear gravitational collapse and the emergence of filamentary and
point-like structures (see e.g.~\cite{fry84,colombi-review}).
We will confirm this expectation by measuring the bispectrum in $N$-body
simulations.

\subsubsection{Orthogonal shape}
Another shape that can arise from single field inflation is the orthogonal shape \cite{Senatore:2009gt}, which
is roughly orthogonal to the equilateral and the local shape and peaks
(with opposite sign) for both equilateral and flattened/elongated
triangle configurations ($k_3=k_1+k_2$). It can be approximated by the
separable template \cite{Senatore:2009gt} 
\begin{align}\label{eq:Borth}
  \nonumber
  B_\Phi^\mathrm{orth} = 6\big[ &
3(P_\Phi^{1/3}(k_1)P_\Phi^{2/3}(k_2)P_\Phi(k_3)+5\mbox{ perms}) \\ 
& -\frac{3}{2}B_\Phi^\mathrm{loc} -8 (P_\Phi(k_1)P_\Phi(k_2)P_\Phi(k_3))^{2/3}
\big].
\end{align}
Being orthogonal to the equilateral shape, the constant mode is
relatively suppressed, a fact which will be important in later
discussions.  We note that all the shapes above
\eqref{eq:Blocal}--\eqref{eq:Borth} are written explicitly as the sum
of separable functions of the wavenumbers $k_1,\, k_2, \, k_3$.

\subsubsection{Flattened shape}

A non-Bunch-Davies vacuum leads to shapes which peak for
flattened/elongated ($k_1+k_2=k_3$) and folded ($k_1=2k_2=2k_3$)
triangle configurations\footnote{Sometimes other names are used for
  these triangles, we follow \cite{jeong09}. }
\cite{chen06_foldedshape,HolmanTolley2008,MeerburgVanDerSchaarCorasaniti2009,chen1002}.
This flattened shape depends on terms such as $1/(k_1+k_2-k_3)$
\cite{chen06_foldedshape}. Therefore it provides
an important example of a bispectrum which is inherently
non-separable, so it is computationally difficult to generate initial
conditions for this shape and to extract its amplitude from data.  We
will use an expansion in separable modes to solve both problems
\cite{shellard0812,shellard0912,shellard1008,shellard1108}.  Of
course, there is a divergence for elongated triangles, $k_3=k_1+k_2$,
which must be removed with a physically motivated cutoff.  For
definiteness we use the same shape and cutoff as was used for the CMB
in \cite{shellard0812,shellard1006}, setting $B=0$ for
$k_1+k_2-k_3<0.03(k_1+k_2+k_3)$ and permutations, and then smoothing
on $k_1+k_2+k_3=\const$ slices with a Gaussian filter with FWHM of
$0.03/(k_1+k_2+k_3)$.

It is worth noting that the equilateral and
orthogonal templates are separable approximations of  inflationary
shapes, with good agreement where the primordial bispectrum signal
peaks. However late time observables may be sensitive to suppressed
triangle configurations for which template and physical shape may
differ significantly (e.g.~scale dependent halo bias mainly depends on
the squeezed limit of the bispectrum
\cite{schmidt1008,wagner-verde1102}). The methods described below
enable us to simulate and estimate physical non-separable bispectra
without using such separable templates. However we do not expect
significant differences for the dark matter bispectrum, because its
peaks are well approximated by the separable templates. For this
reason and for easier comparability with other simulations, we use the
equilateral and orthogonal templates instead of the physical
shapes. Future work, especially on the power spectrum and bispectrum
of halos, will instead focus on the physical shapes.

\section{IV. Bispectrum Estimation Methodology}
\subsection{$f_\mathrm{NL}$ estimator}
If our theoretical model is that the density perturbation $\delta$ has
power spectrum $P_\delta(k)$ and bispectrum $f^\th_\mathrm{NL}\Btheo$,
then the maximum likelihood estimator for the amplitude of this
bispectrum in the limit of weak non-Gaussianity is given by
\cite{shellard1008,shellard1108}\footnote{In the denominator we have
  assumed that $\langle \delta \delta\rangle$ is diagonal which is
  valid for a statistically homogeneous density perturbation.  The linear term
  $\langle\delta\delta \rangle\delta$ in the numerator is written
  out for completeness but it is not used in
  our simulations because it vanishes for a statistically homogeneous
  field (because $\delta_{\vk=0}=0$).}
\begin{align}
  \label{eq:fnl_esti}
  \hatfnl^\th =& \frac{(2\pi)^3}{N_{\th}} \int \frac{\Pi_{i=1}^3
    d^3\bk_i}{(2\pi)^9}(2\pi)^3\delta_D(\bk_1+\bk_2+\bk_3)\nonumber\\
&\times\frac{
  \Btheo(k_1,k_2,k_3)[\delta_{\bk_1}\delta_{\bk_2}\delta_{\bk_3}-3\langle
\delta_{\bk_1}\delta_{\bk_2}\rangle \delta_{\bk_3}]}{P_\delta(k_1)P_\delta(k_2)P_\delta(k_3)}
  \end{align}
where $\delta$ is the observed density perturbation. If this has a
bispectrum $\Bobs$, i.e. 
\begin{align}
  \label{eq:bobs}
  \langle \delta_{\bk_1}\delta_{\bk_2}\delta_{\bk_3} \rangle&=(2\pi)^3 \delta_D\left(\Sigma_i \bk_i\right) \Bobs(k_1,k_2,k_3),
\end{align}
then the expectation value of \eqref{eq:fnl_esti} is given
by\footnote{Performing the angular integrals shows for arbitrary $F(k_1,k_2,k_3)$
  \begin{align*}
  \int\frac{\Pi_{i=1}^3d^3\vk_i}{(2\pi)^9}    
  (2\pi)^6\delta^2_D(\Sigma_{j=1}^3\vk_j)F
=
\frac{V}{8\pi^4}\int_{\mathcal{V}_B}dk_1dk_2dk_3\,k_1k_2k_3F
,
  \end{align*}
 which corrects for a factor of $(2\pi)^3$ missing in
  \cite{shellard1008,shellard1108}. }
\begin{align}
\langle \hatfnl^\th \rangle=&\frac{1}{N_\th}\,\frac{V}{\pi}\int_{\mathcal{V}_B}dV_k\, k_1 k_2 k_3 \nonumber\\
\label{eq:avg_fnl_esti}
&\times \frac{ \Bobs(k_1,k_2,k_3)\Btheo(k_1,k_2,k_3)}{P_\delta(k_1)P_\delta(k_2)P_\delta(k_3)},
\end{align}
where $dV_k\equiv dk_1dk_2dk_3$, $\mathcal{V}_B$ is the tetrahedral domain allowed by the
triangle condition on the wavenumbers $k_i$,  and $V$ is a volume
factor given by $V=(2\pi)^3\delta_D (\mathbf{0})=L^3$.
Demanding $\langle \hatfnl^\th
\rangle=1$ for $\Btheo=\Bobs$ fixes the normalisation such that
\begin{align}
  \label{eq:Nfnl}
  N_\th = \frac{V}{\pi}\int_{\mathcal{V}_B}dV_k\, \frac{k_1 k_2 k_3 [\Btheo(k_1,k_2,k_3)]^2}{P_\delta(k_1)P_\delta(k_2)P_\delta(k_3)}.
\end{align}

\subsection{Shape and size comparisons}
Eq. \eqref{eq:avg_fnl_esti} motivates the definition of a scalar product between two bispectrum shapes \cite{shellard1008,babich-shape-correlator},
\begin{equation}
  \label{eq:inner-prod-from-avg-fnl}
  \< B_i, B_j\> \equiv 
\frac{V}{\pi}\int_{\mathcal{V}_B}dV_k\, \frac{k_1k_2k_3
    B_i(k_1,k_2,k_3)B_j(k_1,k_2,k_3)}{ P_\delta(k_1)P_\delta(k_2)P_\delta(k_3)},
\end{equation}
which can be normalised to a number between $-1$ and $1$ by defining
the cosine or shape correlation
\begin{equation}
  \label{eq:shape-correl}
  \mathcal{C}(B_i,B_j) \equiv \frac{\<B_i,B_j\>}{\sqrt{\<B_i,B_i\>\<B_j,B_j\>}}.
\end{equation}
If two theoretical bispectra $B_1$ and $B_2$ have a small shape
correlation, $|\mathcal{C}(B_1,B_2)|\ll 1$, the optimal estimator for
$f_\mathrm{NL}^{B_1}$ will perform badly in detecting
$f^{B_2}_\mathrm{NL}$ and vice versa. Therefore the shape correlation
is a useful measure of the similarity of bispectrum shapes.

 To measure the total integrated size of a
bispectrum we define the squared norm as the cumulative signal to
noise squared of the bispectrum \cite{shellard1006,creminelli0509},
\begin{align}
  \label{eq:Fnl_norm}
\| B \|^2 \equiv 
\frac{V}{(2\pi)^3}\int_{\mathcal{V}_B}dV_k\,\frac{s_B}{6} \frac{B^2(k_1,k_2,k_3)}{\mathrm{var } \,B(k_1,k_2,k_3)}
=\frac{\langle B,B\rangle}{6 (2\pi)^3},
\end{align}
which equals the quantity $(S/N)_B^2$ in \cite{Sef2011}. The symmetry factor $s_B$ is $6$ if
$k_1=k_2=k_3$, $2$ if only two $k_i$ equal each other and $1$ if all
$k_i$ are different from each other.  To obtain the expression on the right hand side
of \eqref{eq:Fnl_norm} we assumed that different Fourier modes are
uncorrelated, $\langle \delta_{\vk}\delta_{\vk'}\rangle\propto
\delta_D(\vk+\vk')$, and we only took the Gaussian contribution to the
bispectrum noise into account, i.e.\footnote{We need an additional
  factor of $(2\pi)^3$ compared to \cite{Sef2011} because
  $P_\mathrm{them}= (2\pi)^3P_\mathrm{us}$ and $B_\mathrm{them}=
  (2\pi)^3B_\mathrm{us}$.  } 
\begin{align}
  \label{eq:B_gaussian_variance}
\mathrm{var}\,B(k_1,k_2,k_3)= (2\pi)^3\frac{s_BP_\delta(k_1)P_\delta(k_2)P_\delta(k_3)}{8\pi^2\,k_1k_2k_3}.
\end{align}

It is also convenient to normalise the total integrated bispectrum
size with respect to the linearly evolved local bispectrum  defined in
\eqref{eq:Blocal} and \eqref{eq:bprim-tree} (see \cite{shellard1006} for
a similar quantity in the CMB context):
\begin{align}
  \label{eq:FNL_localnorm}
  \bar F_\mathrm{NL}^B \equiv \frac{\|B\|}{\|B_\delta^\mathrm{prim,loc}\|}.
\end{align}
We use the \emph{linearly} evolved local bispectrum without loop corrections
to obtain an expression which can be easily evaluated.

Using definitions \eqref{eq:shape-correl} and \eqref{eq:Fnl_norm}, the expectation value \eqref{eq:avg_fnl_esti} can
be rewritten in the intuitive form 
\begin{align}
  \label{eq:avg_fnl_esti_correls}
 \quad \langle \hat{f}_\mathrm{NL}^{\th}\rangle \;=\;
\mathcal{C}(B_\delta^\mathrm{th},B_\delta^\mathrm{obs}) \; \frac{\|
 B_\delta^\mathrm{obs}\|}{\| B_\delta^\mathrm{th} \|}.
\end{align}
Therefore the bispectrum amplitude $f_\mathrm{NL}$ can be interpreted
as the projection of the observed bispectrum shape on the theoretical
bispectrum shape, which is given by the cosine of the shapes times the
ratio of their norms.

\subsection{Separable mode expansion for fast $f_\mathrm{NL}$ estimation}
Evaluating the estimator \eqref{eq:fnl_esti} is computationally very
expensive because it requires $\mathcal{O}(N^6)$ operations for a grid
with $N$ points per dimension (typically $N=\mathcal{O}(10^3)$).  For
an efficient evaluation we expand the signal to noise weighted
theoretical bispectrum in the separable form
\begin{align}\label{eq:bispDec2}
 \frac{\sqrt{k_1 k_2 k_3} \Btheo(k_1,k_2,k_3)}{\sqrt{P_\delta(k_1)P_\delta(k_2)P_\delta(k_3)}}=\sum_{n=0}^{n_\mathrm{max}-1} {\alpha}^{Q}_n q_{\{r}(k_1) q_s(k_2) q_{t\}}(k_3)\,,
\end{align}
where $\{r s t\}$ denotes a symmetrisation over the indices, and a partial ordering has been introduced to enumerate the modes (for a fuller description see \cite{shellard0912}). The expressions $q_r$ could be any set of independent one-dimensional basis
functions. 
For definiteness we choose $q_r$ to be polynomials of order
$r$ defined in \cite{shellard0912} and choose the slice ordering
defined in Eq.~(58) of \cite{shellard0912}. We truncate the modal
expansion \eqref{eq:bispDec2} after $n_\mathrm{max}=\mathcal{O}(50)$
modes, which provides an accurate representation for the shapes we are
considering as illustrated in
\fig{tet3dplots_show_expansions_grav_loc_eq_orth}, which compares
theoretical bispectra on the left with truncated expansions in the
centre (and bispectra measured in N-body simulations on the
right). More quantitative discussions of the error induced by the
truncation of the expansions will be provided later in terms of shape
correlations.

 The separable expansion
\eqref{eq:bispDec2} allows us to write the estimator and its
expectation value in the form
\begin{align}\label{eq:bispEst}
  \hatfnl^\th
=&\frac{(2\pi)^3}{N_\th}  \sum_n  {\alpha}^{Q}_n \int d^3 \bx [M_r(\bx)M_s(\bx)M_t(\bx)\nonumber\\
&\hspace{20mm}-3\langle M_{\{r}(\bx)M_{s}(\bx)\rangle M_{t\}}(\bx)],\\
\label{eq:bispEst-avg}
\langle \hatfnl^\th \rangle=&\frac{1}{N_\th}\sum_{n m}  {\alpha}^{Q}_n {\alpha}^{Q}_m \gamma_{nm},
\end{align}
where
\begin{align}
\gamma_{n m}&=\frac{V}{\pi}\int_{\mathcal{V}_B}dV_k\, Q_n(k_1,k_2,k_3)Q_m(k_1,k_2,k_3),\nonumber\\
M_r(\bx)&=\int \frac{d^3\bk}{(2\pi)^3} \frac{\delta_{\bk} q_r(k)}{\sqrt{k P_\delta(k)}}e^{i\bk.\bx}\label{eq:Mrx}
\end{align}
and 
\begin{align}
  \label{eq:Qn_defn}
Q_n(k,k',k'')\equiv q_{\{r}(k) q_s(k') q_{t\}}(k'')\,.
\end{align}
 This form of  
the estimator can be evaluated efficiently because it involves only
Fourier transforms and one three-dimensional integral over position
space, leading to $\mathcal{O}(N^3)$ operations.

To motivate the choice of the prefactor in the expansion
\eqref{eq:bispDec2} let us transform  
to basis functions $R_n(k_1,k_2,k_3)$ that are orthonormal on the tetrapyd domain,
$\langle R_n,R_m\rangle_\mathrm{noweight}=\delta_{nm}$, with respect to the unweighted scalar
product
\begin{align}
  \label{eq:inner-prod-unit-weight}
   \< B_i, B_j\>_\mathrm{noweight} \equiv 
\frac{V}{\pi}\int_{\mathcal{V}_B}dV_k\,
    B_i(k_1,k_2,k_3)B_j(k_1,k_2,k_3).
\end{align}
From $\sum_n {\alpha}^Q_nQ_n=\sum_n {\alpha}^R_nR_n$ and
Eq. \eqref{eq:bispDec2} we find that 
\begin{align}
  \label{eq:Btheo_expansion_BRn}
   \Btheo(k_1,k_2,k_3) = \sum_n {\alpha}^R_nB^R_n(k_1,k_2,k_3),
\end{align}
where the contributions
\begin{align}
  \label{eq:B-n}
 B^R_n(k_1,k_2,k_3)\equiv \sqrt{\frac{P_\delta(k_1)P_\delta(k_2)P_\delta(k_3)}{k_1 k_2 k_3}}R_n(k_1,k_2,k_3)
\end{align}
to the full bispectrum are orthonormal with respect to
\eqref{eq:inner-prod-from-avg-fnl}. Thus the expansion
coefficients ${\alpha}^R_n$ measure the size of 
contributions to the bispectrum which are orthonormal with respect to the signal to noise weighted scalar product \eqref{eq:inner-prod-from-avg-fnl}.  

As a side remark, note that in contrast to the signal to noise
weighted scalar product \eqref{eq:inner-prod-from-avg-fnl} induced by
the estimator expectation value, the weight of the scalar product
\eqref{eq:inner-prod-unit-weight} is time-independent and does not
depend on the range of scales under consideration. Therefore we can
use the same set of orthonormal polynomials $R_n$ at any time and for
all length scales. If the expansion \eqref{eq:bispDec2} does not
converge as well as for the applications considered here, different
(separable) scalar product weights can be used to orthonormalise the $R_n$'s.

\subsection{Fast modal bispectrum estimator}
Importantly the separable mode expansion allows us not only to measure
$f^\th_\mathrm{NL}$ efficiently for any given theoretical bispectrum, but it also
allows us to reconstruct the full bispectrum in a model-independent
manner by measuring a set of independent $f_\mathrm{NL}$'s and summing
up the corresponding contributions to the bispectrum \cite{shellard1008,shellard1108}. To see this,
note that the transformation $R_n=\sum_m\lambda_{nm}Q_m$ implies
${\alpha}^Q_n=\sum_p \lambda_{pn}{\alpha}_p^R$ and from $\langle
R_n,R_m\rangle_\mathrm{noweight}=\delta_{nm}$ we get $\gamma= \lambda^{-1}
(\lambda^{-1})^T$. Plugging this into \eqref{eq:bispEst} gives
\begin{align}
\label{eq:bispEst2}
  \hatfnl^\th
=&\frac{1}{N_\th}  \sum_m  {\alpha}^{R}_m \beta^R_m,
\end{align}
with
\begin{align}
\label{eq:betaR}
\beta^R_m\equiv& (2\pi)^3 \sum_n \lambda_{mn}  \int d^3 \bx [M_rM_sM_t-3\langle M_{\{r}M_{s}\rangle M_{t\}}],
\end{align}
where the index $n$ labels the combination $r,s,t$. 
If $\Btheo=\langle B_\delta^\mathrm{obs}\rangle$ then \eqref{eq:bispEst-avg} becomes
\begin{align}
\label{eq:bispEst-avg2}
\langle \hatfnl^\th \rangle=&\frac{1}{N_\th}\sum_{m}
{\alpha}^{R}_m {\alpha}^{R}_m.
\end{align}
Equations \eqref{eq:bispEst2} and \eqref{eq:bispEst-avg2}  imply
\begin{align}
  \label{eq:betaIsAlpha}
\langle \beta^R_n\rangle = {\alpha}^R_n.  
\end{align}
Therefore we can estimate the full bispectrum with \cite{shellard1008,shellard1108}
\begin{align}
  \label{eq:bisp-from-betas}
 \hat B_\delta(k_1,k_2,k_3) = \sum_{n=0}^{n_\mathrm{max}-1}
\beta_n^R B^R_n(k_1,k_2,k_3),
\end{align}
where $B^R_n$ is given by \eqref{eq:B-n}. Intuitively the coefficients
$\beta_n^R$ measure the components of the orthonormal basis bispectra
$B_n^R$ in the data and $\hat B_\delta$ measures the projection of the
bispectrum on the subspace of bispectra spanned by the $B^R_n$. 

This approach is extremely efficient with the calculation of each $\beta_n^R$ coefficient being equivalent to a single $3$D integration over products of fast Fourier transforms. While theoretically a complete basis would require $N^3$ modes in practice far fewer modes ($\mathcal{O}(50)$) are necessary to reconstruct the bispectrum accurately.

The primordial
contribution to the matter bispectrum will be extracted by measuring
the difference of the bispectrum between non-Gaussian
and Gaussian initial conditions,
\begin{align}
  \label{eq:B_ng_meas}
  {\hat B}_\mathrm{NG} = \sum_{n=0}^{n_\mathrm{max}-1} \left[ \beta^R_n
    - (\beta^R_n)_\mathrm{Gauss}\right] B_n^R.
\end{align}
As will be
discussed in more detail later, the difference is computed seed by seed to reduce error bars.

The time-dependence of the expansion coefficients ${\alpha_n^R}$ and the
$\beta^R_n$ coefficients can be
read off from \eqref{eq:bispDec2} to be $({\alpha}^R_n)_\mathrm{grav}
\propto D(z)$ for the gravitational bispectrum and
$({\alpha}^R_n)_\mathrm{NG} \propto D(z)^0$ for primordial bispectra,
assuming tree level perturbation theory where $P_\delta\propto
D^2(z)$.  In practice we must use the nonlinear power in
\eqref{eq:bispDec2} and \eqref{eq:Mrx} implying slightly different
time dependences for ${\alpha_n^R}$ in the nonlinear regime (see e.g.~the middle panel of
\fig{bgrav_18BBfromglass}, which will be discussed in more detail later).

\subsection{Fast modal bispectrum correlations}
To analyse the reconstructed bispectrum and to control the accuracy of
the separable method, the following shape correlations are computed using \eqref{eq:shape-correl}:
\begin{align}
  \label{eq:shape-correl-alphatheo}
\mathcal{C}_{\alpha,\mathrm{th}}\equiv &\, \mathcal{C}\left(\sum_n{\alpha}_n^RB^R_n,B^\mathrm{th}_\delta\right) = 
  \sqrt{\frac{\sum_n({\alpha}^R_n)^2}{\langle
      B^\mathrm{th}_\delta,B^\mathrm{th}_\delta\rangle}},\\
  \label{eq:shape-correl-alphabeta}
\mathcal{C}_{\beta,\alpha} \equiv &\,  \mathcal{C}\left(\langle \hat B_\delta\rangle_\mrm{sim},\sum_n{\alpha}_n^RB^R_n\right) = \nonumber\\
& \quad \qquad \qquad \qquad
   \frac{\sum_n{\alpha}^R_n\langle\beta^R_n\rangle_\mrm{sim} }{\sqrt{\sum_m({\alpha}^R_m)^2
       \sum_p \langle \beta^R_p\rangle_\mrm{sim}^2 }},\\
  \label{eq:shape-betatheo}
\mathcal{C}_{\beta,\mathrm{th}}\equiv &\, \mathcal{C}\left(\langle \hat B_\delta\rangle_\mrm{sim},B^\mathrm{th}_\delta\right) =
\mathcal{C}_{\beta,\alpha}\mathcal{C}_{\alpha,\mathrm{th}}.
\end{align}
Here $\langle \hat B_\delta\rangle_\mrm{sim}$
denotes
the average of \eqref{eq:bisp-from-betas} over independent simulations
 and we truncate all sums appearing in the above expressions and in
\eqref{eq:bisp-from-betas} after $n_\mathrm{max}$ modes.
The first shape correlation  measures how well the separable expansion
\eqref{eq:bispDec2} of the theoretical bispectrum approximates the
theoretical bispectrum. The second shape correlation quantifies how
well the estimated bispectrum $\langle \hat B_\delta\rangle_\mrm{sim}$ agrees with the
separable expansion of the theoretical bispectrum. The product of
these two shape correlations gives the shape correlation between the
reconstructed bispectrum and the theoretical bispectrum.

\subsection{Cumulative measures of non-Gaussianity}
Additionally to these shape correlations we will also compare the
projections \eqref{eq:bispEst2} of the measured bispectrum on theory
bispectra, as well as the cumulative signal to noise of the average
reconstructed bispectrum, which can be expressed in terms of the
measured average $\langle\beta^R_n\rangle_\mrm{sim}$
as\footnote{Alternatively using $\langle
  \sum_n(\beta^R_n)^2\rangle_\mrm{sim}$
would add undesirable contributions from
  the variances of the $\beta^R_n$, which become relevant if the
  measurements are not signal-dominated.}
\begin{align} 
  \label{eq:FNL_reconstructed}
\| \langle \hat B_\delta\rangle_\mrm{sim} \| 
=
\sqrt{ \frac{\sum_n \langle\beta^R_n\rangle_\mrm{sim}^2}{6 (2\pi)^3}},\quad
{{\bar F}}_\mathrm{NL}^{\mrm{sim}} = \sqrt{\frac{\sum_n \langle\beta^R_n\rangle_\mrm{sim}^2}{N_{B_\delta^\mathrm{prim,loc}}}}
.
\end{align}

The norm was defined in \eqref{eq:Fnl_norm} using the cumulative
signal to noise squared of the bispectrum.  As a consistency check we
check if \eqref{eq:FNL_reconstructed} also measures the cumulative
signal to noise squared of the $\beta^R_n$ coefficients. The
theoretical covariance of $\beta^R_n$ involves the
$6$-point function of the density perturbation. Taking only the
Gaussian contribution into account and assuming that different Fourier
modes are uncorrelated we find
\begin{align}
  \label{eq:beta-variance}
\langle  \beta^R_n\beta^R_m\rangle = 6\,(2\pi)^3 \, \delta_{nm}.
\end{align}
This predicts $\sigma_{\beta^R_n}=\sqrt{6 (2\pi)^3}$ for Gaussian
simulations and therefore confirms that \eqref{eq:FNL_reconstructed}
measures the cumulative signal to noise of the reconstructed
bispectrum.  We confirmed the prediction for $\sigma_{\beta^R_n}$
quantitatively in Gaussian simulations, but we have not assessed
non-Gaussian corrections to the predicted noise (see e.g.~\cite{FS3}).
Note that our plots show $1\sigma$ sample standard deviations obtained
by running realisations with different random number seeds.

\subsection{Towards experimental setups}
For the bispectrum measurements presented below we will drop the
second term in the square brackets in \eqref{eq:betaR}, i.e.~we assume
that different modes of the density perturbation are uncorrelated,
$\langle\delta_\bk\delta_{\bk'}\rangle\propto \delta_D(\bk+\bk')$,
which is true for a statistically homogeneous field.
In experimental setups, which are not considered here, off-diagonal
mode couplings from inhomogeneous noise can be incorporated in
\eqref{eq:betaR} with the linear term $\langle MM\rangle M$, without
affecting the efficiency of the bispectrum estimator
\eqref{eq:bisp-from-betas}.  Ref.~\cite{FS3} explores how
modifications to the denominator of the estimator \eqref{eq:fnl_esti}
in presence of off-diagonal mode couplings can be incorporated
efficiently with modal expansions and an implementation for CMB experiments has been
developed successfully. 

\subsection{Comparison to other bispectrum estimators}
An alternative way to reconstruct the bispectrum is to estimate the
three-point correlation function directly, taking a subset of all
possible triangle configurations that yields the desired accuracy of
the bispectrum \cite{sefusatti1003,verde1111,scocci1108}.  In contrast
our estimator takes all possible triangle configurations into account
in a very efficient way, so that the bispectrum can be measured on the
complete range of scales that were included in the $N$-body
simulation.
For example, with our method the extraction of the bispectrum of a
$1024^3$ grid takes about one hour on $6$ cores, which is only a small
fraction of the time required to run an $N$-body simulation of this
size.  Moreover our estimator compresses the information about the
shape of the bispectrum, which is a function of the whole
three-dimensional tetrahedral domain allowed by the triangle
condition, to $n_\mathrm{max}$ numbers $\beta^R_n$, where
$n_\mathrm{max}$ is the number of basis functions used in the
expansion \eqref{eq:bispDec2}. This simplifies further processing of
the bispectrum, e.g.~for comparisons with theoretical models and for obtaining fitting formulae.  Once $\beta^R_n$
is measured from the data, we can not only obtain the full bispectrum
with \eqref{eq:bisp-from-betas}, but we can also calculate full
three-dimensional shape correlations, the cumulative signal to noise
and the nonlinearity parameter $f^\th_\mathrm{NL}$ associated to any
theoretical bispectrum $\Btheo$ without additional computational cost,
using \eqref{eq:shape-correl-alphabeta}, \eqref{eq:FNL_reconstructed}
and \eqref{eq:bispEst2}, respectively.  In presence of inhomogeneous
noise we can in principle include off-diagonal covariance elements
$\langle\delta_\bk\delta_{\bk'}\rangle$ in a straightforward way.
Finally our approach allows us to estimate the trispectrum efficiently
\cite{shellard1008}, which has been shown for a class of trispectrum
shapes in \cite{shellard1108} (see also \cite{RSF10,FRS2}), but we
leave the application to $N$-body simulations for future work.

\subsection{Bispectrum visualisation}

\begin{figure}[htp]
\centering
\includegraphics[width=0.40\textwidth]{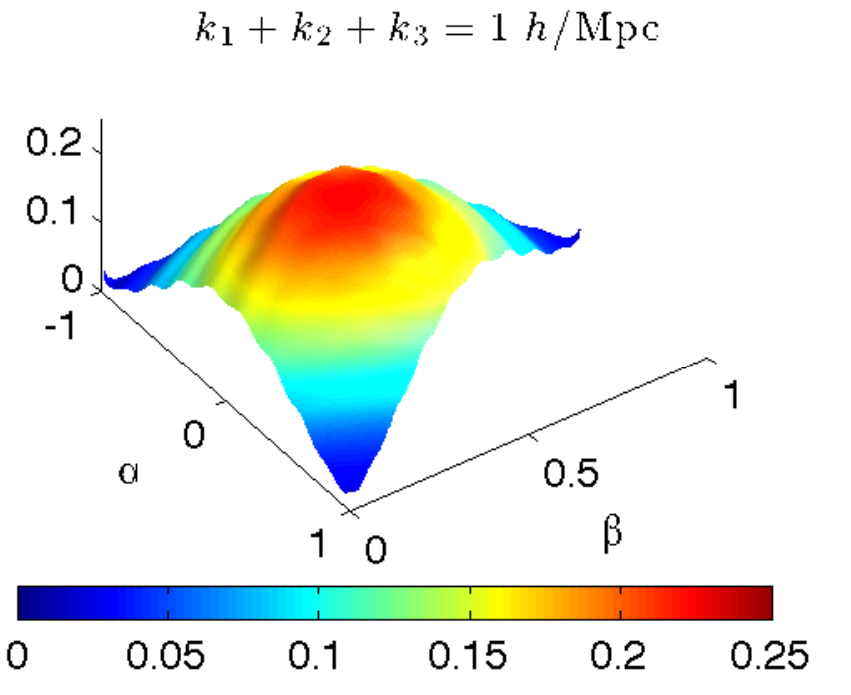}
\label{fig:PPplusperms_weight}
\includegraphics[width=0.40\textwidth]{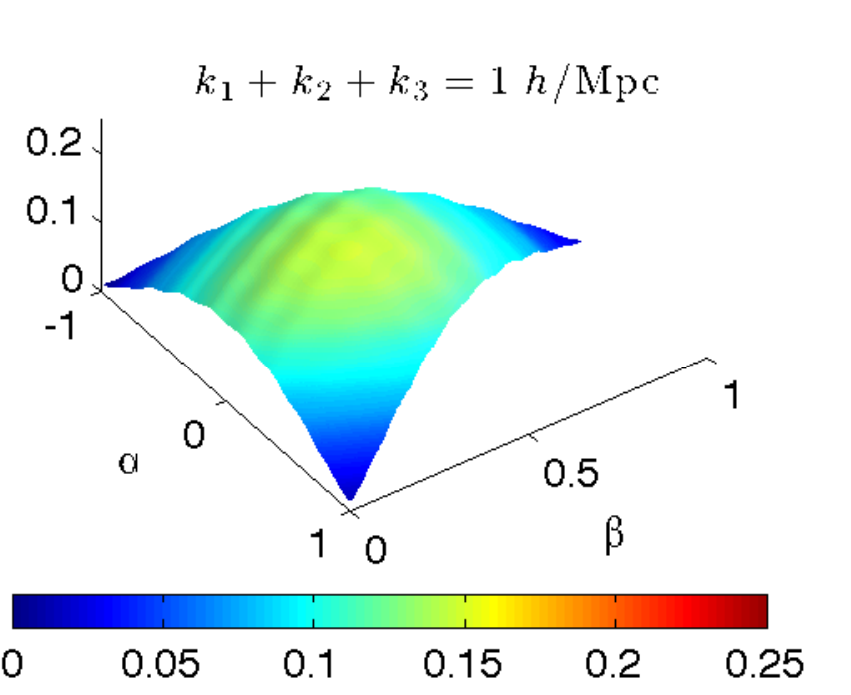}
\label{fig:SN_weight}
\mycaption{Bispectrum weights  
$[P_\delta(k_1)P_\delta(k_2)+2\;\mathrm{perms}]^{-1}$ (top) and 
$\sqrt{k_1k_2k_3/[P_\delta(k_1)P_\delta(k_2)P_\delta(k_3)]}$ (bottom)
evaluated with CAMB \cite{camb} at redshift $z=30$ on slices with
$k_1+k_2+k_3=1h/\mathrm{Mpc}$.}
\label{fig:Bweights_all}
\end{figure}

Often in the literature the bispectrum is visualised by plotting one-
or two-dimensional slices through the tetrapyd shown in
\fig{tetrapyd}. E.g.~the plots in \cite{babich-shape-correlator}
correspond to two-dimensional slices through the tetrapyd obtained by
varying $k_2$ and $k_3$ at fixed $k_1$ and removing the $k_3>k_2$ part
of the slice by symmetry. Some plots in \cite{shellard0812} and in
this paper show two-dimensional tetrapyd slices with
$k_1+k_2+k_3=\mathrm{const}$. While such slices are sufficient for
scale-invariant primordial bispectra, late time bispectra typically
have different triangle dependences at different overall scales
$k_1+k_2+k_3$. This motivates plotting late time bispectra on the full
three-dimensional tetrapyd instead of plotting particular slices
through the tetrapyd.

Instead of the unweighted bispectrum the so-called reduced bispectrum
$Q\equiv B_\delta/(P_\delta(k_1)P_\delta(k_2)+2\mathrm{ perms})$ is
often shown, because it reduces the dynamical range of the bispectrum
and for the tree level gravitational contribution it is independent of
time, overall scale and power spectrum normalisation and almost
independent of cosmology \cite{colombi-review}.  Instead we will plot
the signal to noise weighted bispectrum
$\sqrt{k_1k_2k_3/(P_\delta(k_1)P_\delta(k_2)P_\delta(k_3))}B_\delta$
because then the product of the shown functions gives the scalar
product defined in \eqref{eq:inner-prod-from-avg-fnl}
\cite{babich-shape-correlator}, visual similarity indicates a high shape
correlation \eqref{eq:shape-correl} and the dynamical range of the
unweighted bispectrum is also greatly reduced, which is of advantage
for plotting purposes. For linearly evolved primordial bispectra
\eqref{eq:bprim-tree}, the signal to noise weighted bispectrum is
time-independent in the linear regime.

Qualitatively the two weights are quite similar in the regime relevant
for most plots of this paper, see \fig{Bweights_all}. However they
differ in the squeezed limit, because for decreasing $k_1$,
$P_\delta(k_1)P_\delta(k_2)$ turns over at $k_1=k_\mathrm{eq}$,
whereas $\sqrt{P_\delta(k_1)P_\delta(k_2)P_\delta(k_3)/(k_1k_2k_3)}$
turns constant where $P(k_1)/k_1$ turns constant, which is on somewhat
larger scales than $k_\mathrm{eq}$.  While it is not straightforward
to deduce the squeezed limit of an unweighted bispectrum from a plot
of its weighted form, one can compare plots of different bispectra if
they are weighted in the same way to deduce the relative behavior in
the squeezed limit.

\section{V. Simulation Setup, Initial Conditions and Validation}

\subsection{$N$-body simulations setup}

We use the separable mode expansion method
described in \cite{shellard1008,shellard1108} to generate realisations
of the initial primordial potential $\Phi=-3\mathcal{R}/5$ during
matter domination with the desired primordial power spectrum and
bispectrum.  For $512^3$ particles this takes about $10$ minutes per
seed on one core and works for separable as well as non-separable
bispectra, therefore being more general than other proposed methods
\cite{wagner-verde1006,scocci1108}. From $\Phi$ we calculate the linear density perturbation
$\delta$ at the initial redshift of the simulation with the Poisson
equation \eqref{eq:poisson}.  We then use this initial density
perturbation to displace the initial particles from an unperturbed
distribution with the 2LPT method \cite{citeulike:9679499,scocci-transients0606}, which
also determines the initial particle velocities.  Then the $N$-body code
Gadget-3 \cite{gadget1,gadget2} simulates the time evolution until today and
we use a cloud in cell scheme to calculate the density perturbation
$\delta$ of the particle distribution on a grid at different
redshifts. After deconvolving $\delta$ with the cloud in cell kernel
we compute the power spectrum $P_\delta$ and the coefficients
$\beta^R_n$ from \eqref{eq:betaR} using $n_\textrm{max}=50$
modes. Finally we reconstruct the full bispectrum with
\eqref{eq:bisp-from-betas} and calculate its norm ${\hat{\bar
    F}}_\mathrm{NL}$ \eqref{eq:FNL_reconstructed} as well as its shape
correlation $\mathcal{C}_{\beta,\alpha}$
\eqref{eq:shape-correl-alphabeta} with theoretical bispectra and its
nonlinear amplitude \eqref{eq:bispEst2}.

Table \ref{tab:nbody} lists the parameters of the $N$-body simulations
that were performed in this work. All simulations assume a flat
$\Lambda$CDM model with the WMAP-7 \cite{wmap7} parameters
$\Omega_bh^2=0.0226,
\Omega_ch^2=0.11,\Omega_\Lambda=0.734,h=0.71,\tau=0.088,\Delta^2_\mathcal{R}(k_0)=2.43\times
10^{-9}$ and $n_s(k_0)=0.963$, where $k_0 = 0.002
\mathrm{Mpc}^{-1}$. Note that the primordial power spectrum is given by
$P_\Phi(k)=(9/25)\,(2\pi^2/k^3)\,\Delta_{\mathcal{R}}^2(k_0)(k/k_0)^{n_s-1}$. 

\begin{table}[htp]
\begin{tabular}{|p{1.55cm}|p{0.8cm}|r|c|r|r|c|c|c|}
\hline
Name	& NG shape & $f_\mathrm{NL}$ & $L[\frac{\mathrm{Mpc}}{h}]$ & $N_p$ & $z_i$ & $L_s[\frac{\mathrm{kpc}}{h}]$ & $N_r$ & glass  \\ 	\hline
G512g & -- & -- & $1600$ & $512$ & $49$ & $156$ & $3$ & yes\\
G512 & -- & -- & $1600$ & $512$ & $49$ & $156$ & $3$ & no\\
$\mathrm{G}^{512}_L$ & -- & -- & $\{400,100\}$ & $512$ & $49$ & $\{39,9.8\}$ &
$3$ & no\\
G768 & -- & -- & $2400$ & $768$ & $19$ & $90$ & $3$ & no\\
G1024 & -- & -- & $1875$ & $1024$ & $19$ & $40$ & $2$ & no\\
Loc10g & local & $10$ & $1600$ & $512$ & $49$ & $156$ & $3$ & yes\\
Loc10 & local & $10$ & $1600$ & $512$ & $49$ & $156$ & $3$ & no\\
$\mathrm{Loc10}^{512}_L$ & local & $10$ & $\{400,100\}$ & $512$ & $49$ &
$\{39,9.8\}$ & $3$ & no\\
$\mathrm{Loc10}^-$ & local & $-10$ & $1600$ & $512$ & $49$ & $156$ & $3$ & no\\
Loc20 & local & $20$ & $1600$ & $512$ & $49$ & $156$ & $3$ & no\\
Loc50 & local & $50$ & $1600$ & $512$ & $49$ & $156$ & $3$ & no\\
Eq100g & equil & $100$ & $1600$ & $512$ & $49$ & $156$ & $3$ & yes\\
Eq100 & equil & $100$ & $1600$ & $512$ & $49$ & $156$ & $3$ & no\\
$\mathrm{Eq100}^{512}_L$ & equil & $100$ & $\{400,100\}$ & $512$ & $49$ & $\{39,9.8\}$ & $3$ & no\\
$\mathrm{Eq100}^-$ & equil & $-100$ & $1600$ & $512$ & $49$ & $156$ & $3$ & no\\
Orth100g & orth & $100$ & $1600$ & $512$ & $49$ & $156$ & $3$ & yes\\
Orth100 & orth & $100$ & $1600$ & $512$ & $49$ & $156$ & $3$ & no\\
$\mathrm{Orth100}^{512}_{400}$ & orth & $100$ & $400$ & $512$ & $49$ & $39$ & $3$ & no\\
$\mathrm{Orth100^-}$ & orth & $-100$ & $1600$ & $512$ & $49$ & $156$ & $3$ & no\\
Flat10 & flat & $10$ & $1600$ & $512$ & $49$ & $156$ & $3$ & no\\
$\mathrm{Flat10}^{512}_{400}$ & flat & $100$ & $400$ & $512$ & $49$ & $39$ & $3$ & no\\
\hline
\end{tabular}
\caption{Parameters of $N$-body simulations: Non-linearity parameter $f_\mathrm{NL}$, box size $L$, number of particles per dimension $N_p$, initial redshift of the simulations $z_i$, softening length $L_s$ and number of realisations (i.e.~random seeds) $N_r$ for each parameter set. `glass' indicates if the initial particles were displaced from a regular grid or from a glass configuration. Initial conditions for non-local non-Gaussian simulations were generated with the separable method described in \cite{shellard1008,shellard1108}. All simulations use 2LPT \cite{citeulike:9679499,scocci-transients0606} to get the initial particle distribution, which is then evolved with Gadget-3 \cite{gadget1,gadget2}. }\label{tab:nbody}
\end{table}

\begin{figure*}[t]
\centering
\subfloat[][Local]{
\includegraphics[height=0.45\textheight]{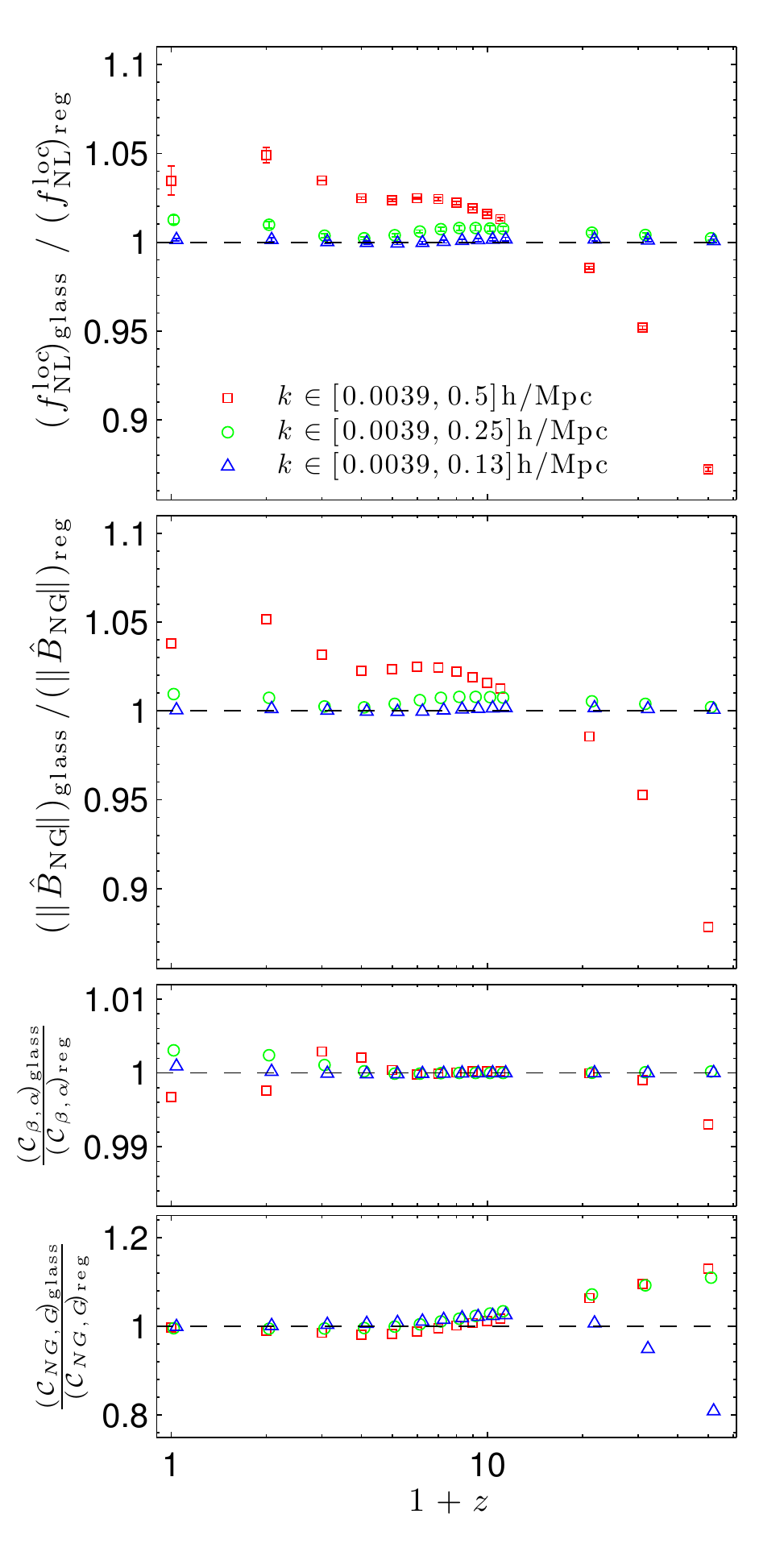}
\label{fig:scorrelz_18BBfromglass_fnl10}}
\subfloat[][Equilateral]{
\includegraphics[height=0.45\textheight]{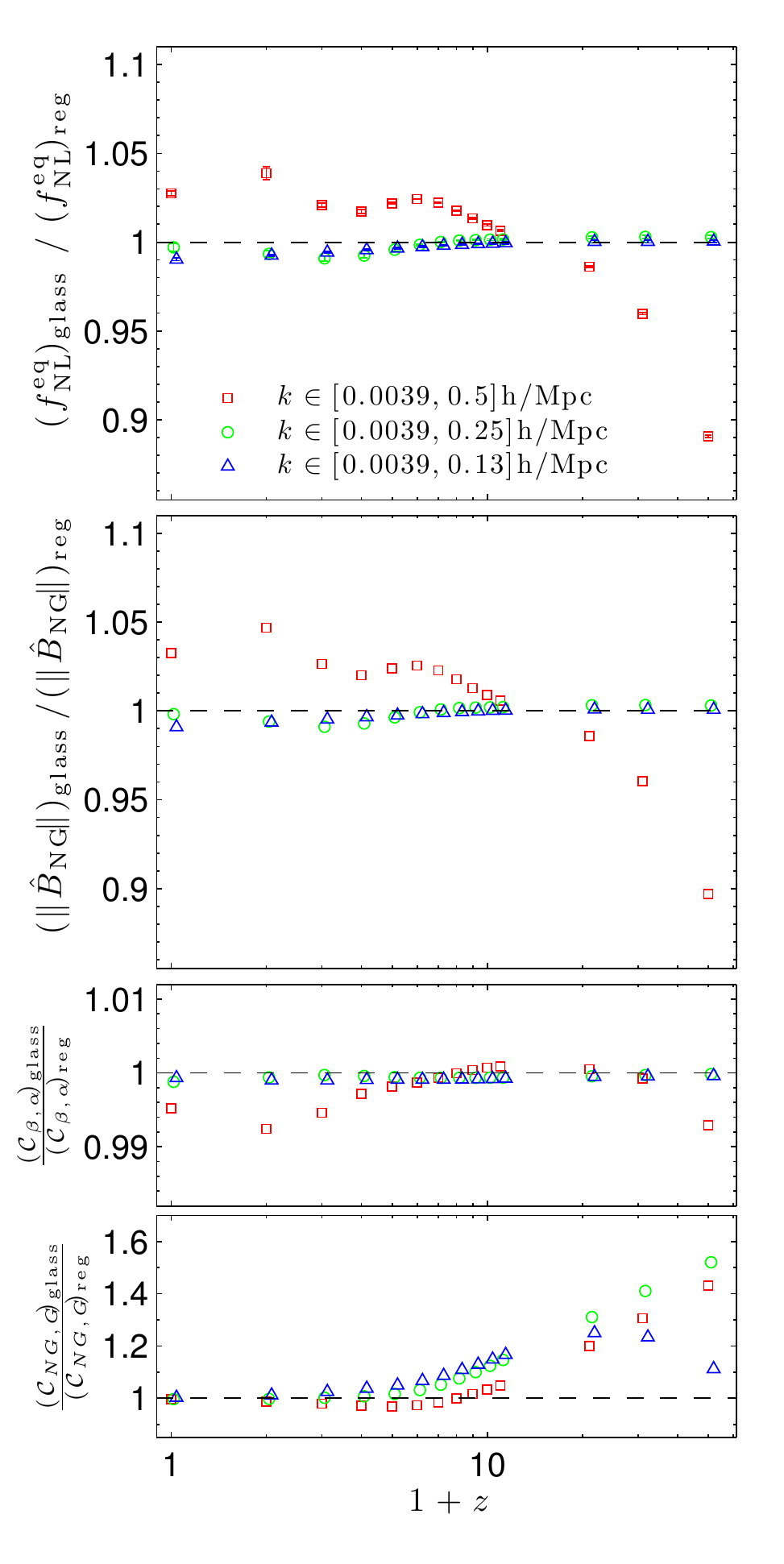}
\label{fig:scorrelz_18BBfromglass_sp6fnl100}}
\subfloat[][Orthogonal]{
\includegraphics[height=0.45\textheight]{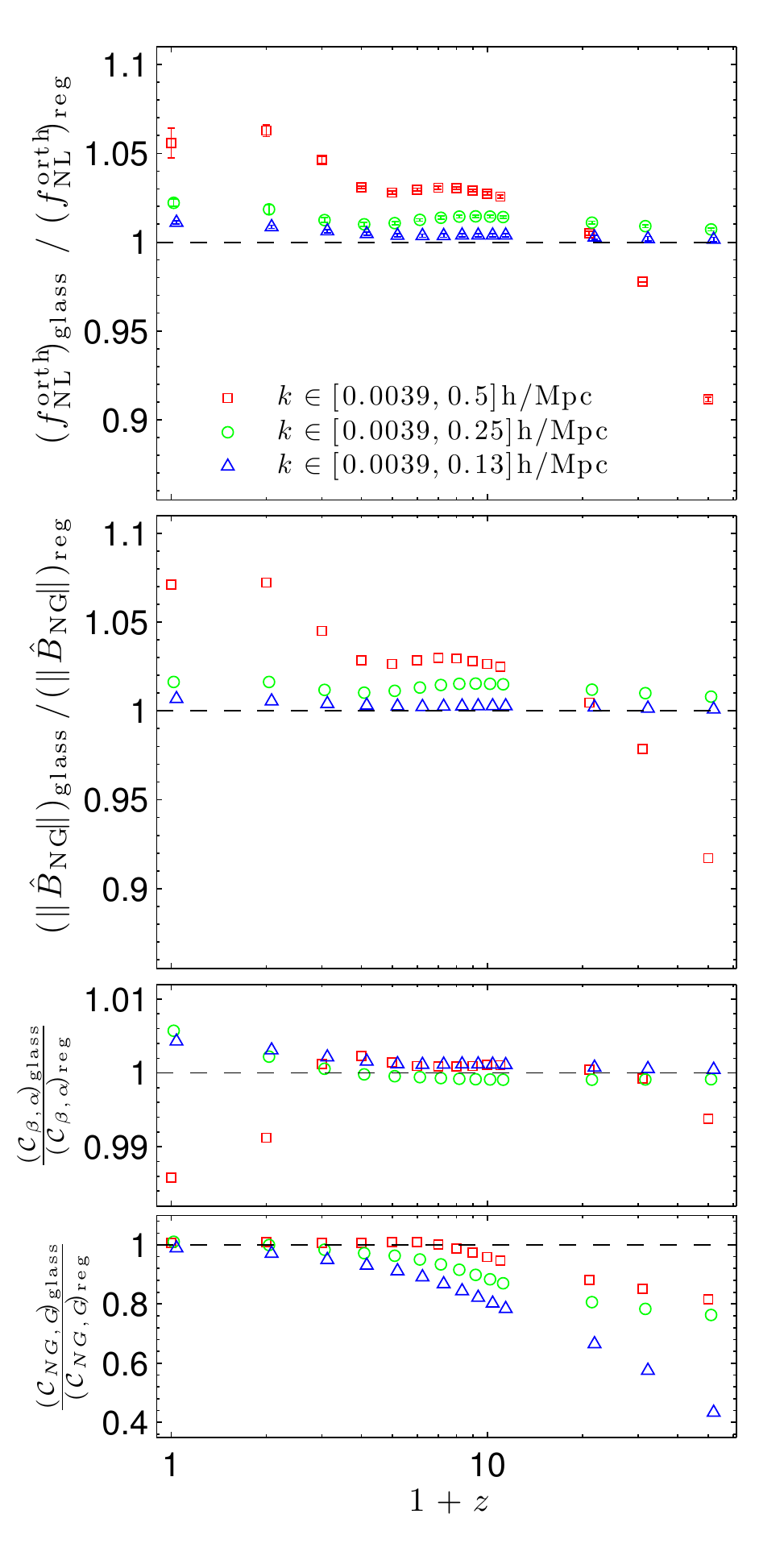}
\label{fig:scorrelz_18BBfromglass_sp22fnl100}}
\mycaption{Impact of glass initial conditions on measured non-Gaussian
  bispectra in simulations Loc10, Eq100 and Orth100. We plot the ratio
  to simulations with regular grid initial conditions.  The top panel
  contains error bars obtained by calculating the sample standard
  deviation of the ratio for each seed. We do not show error bars in
  the other panels because they depend on $\sum_n \langle
  \beta^R_n\rangle_\mathrm{sim}^2$ and are therefore more difficult to
  estimate. However they should be similar to the errors in the top
  panel because of \eqref{eq:avg_fnl_esti_correls}.  The large
  deviations in the bottom panel are mainly due to the different
  gravitational bispectra for glass and regular grid initial
  conditions as shown in \fig{bgrav_18Q_glassratio}.
}
\label{fig:scorrelz_glass_NG_all}
\end{figure*}

\subsection{Regular grid vs glass initial conditions} 

For the unperturbed particle distribution, from which initial
particles are displaced using the 2LPT method, we either use a regular
grid or a glass configuration, which is obtained by placing particles
randomly in the box and then evolving them with the sign of gravity
flipped in Gadget-3 \cite{white-glass94,gadget2}.  A disadvantage of
glass initial conditions is that they require generating a glass and
specifying when the glass configuration has converged (see
e.g.~\cite{hansen07} for a discussion).  In contrast, regular grid
initial conditions are easier to set up but break statistical
isotropy, which may have implications for structure formation.  We
will compare the different initial condition setups first for Gaussian
and then non-Gaussian initial conditions.

Measured bispectra for Gaussian $N$-body simulations will be shown in \fig{bgrav_all} in Section
VI. Simulations started from glass and regular grid initial conditions
are shown in \fig{bgrav_18BBfromglass} and \fig{bgrav_18Q},
respectively.  At early times regular grid initial conditions produce
a bispectrum which is more than three times as large as the tree level
prediction.  This significant spurious bispectrum is not present for
glass initial conditions and must therefore be caused by the breaking
of isotropy induced by the regular grid. \fig{bgrav_18Q} shows that
the spurious bispectrum decays with time as the regular grid structure
is washed out by the growing gravitational perturbations.  At late
times, $z\le 3$, the bispectra from regular grid and glass initial
conditions differ by at most $10\%$ in their cumulative signal to
noise and by less than $0.5\%$ in their shape correlation to the tree
level prediction, see \fig{bgrav_18Q_glassratio}.

We conclude that at the $10\%$ accuracy level (for Gaussian
simulations with $L=1600\mathrm{Mpc}/h$, $512^3$ particles and
$k_\mathrm{max}=0.5h/\mathrm{Mpc}$), both glass and regular grid
initial conditions can be used to extract the gravitational bispectrum
from Gaussian simulations as long as the bispectrum is measured at low
redshifts when the regular grid is washed out.  The difference between
regular grid and glass initial conditions decreases when reducing the
cutoff $k_\mathrm{max}$ used for the bispectrum estimation because
then structures of the size of the grid separation are smoothed out.

For non-Gaussian simulations, \fig{scorrelz_glass_NG_all} shows how
glass initial conditions change the non-Gaussian excess bispectrum
$\hat B_\mathrm{NG}$ \eqref{eq:B_ng_meas} compared to regular grid
initial conditions.  At $z\le 10$ the shape of $\hat B_\mathrm{NG}$ is
consistent between glass and regular grid initial conditions at the
$1.5\%$ level for the simulations Loc10, Eq100 and Orth100.  However, the total
integrated bispectrum $\bar F_\mathrm{NL}$ differs by up to $7\%$ for
$z\le 10$.  At earlier times the deviations are somewhat larger.
Compared to the full gravitational bispectrum for Gaussian initial conditions shown
in \fig{bgrav_18BBfromglass}, the impact of glass initial conditions
on $\hat B_\mathrm{NG}$ is quite small, possibly because the spurious
bispectrum due to the regular grid is partly subtracted out when
calculating $\hat B_\mathrm{NG} =\hat B-\hat B_\mathrm{Gauss}$.

Whether regular grid or glass initial conditions are
representing the statistics of the non-Gaussian density perturbation
more successfully is not unambiguous.  Regular grid initial conditions introduce a spurious
contribution to the full bispectrum $\hat B$ at early times, which can
be avoided using glass initial conditions. However glass initial
conditions represent the primordial contribution $\hat B_\mathrm{NG}$
slightly less accurately than regular grid initial conditions at the starting
redshift of the simulation. However these effects impact
$\hat B_\mathrm{NG}$ only at the $\mathcal{O}(5\%)$ level at $z\le
10$, so that both glass or regular grid initial conditions can be used
at this level of precision.  As for Gaussian simulations the two
initial condition methods agree better when $k_\mathrm{max}$ is
reduced (see \fig{scorrelz_glass_NG_all}) and grid discretization effects are 
minimized.

\subsection{Validation and convergence tests}

\begin{figure}[ht]
\hspace*{-0.5cm}
\centering
\includegraphics[width=0.53\textwidth]{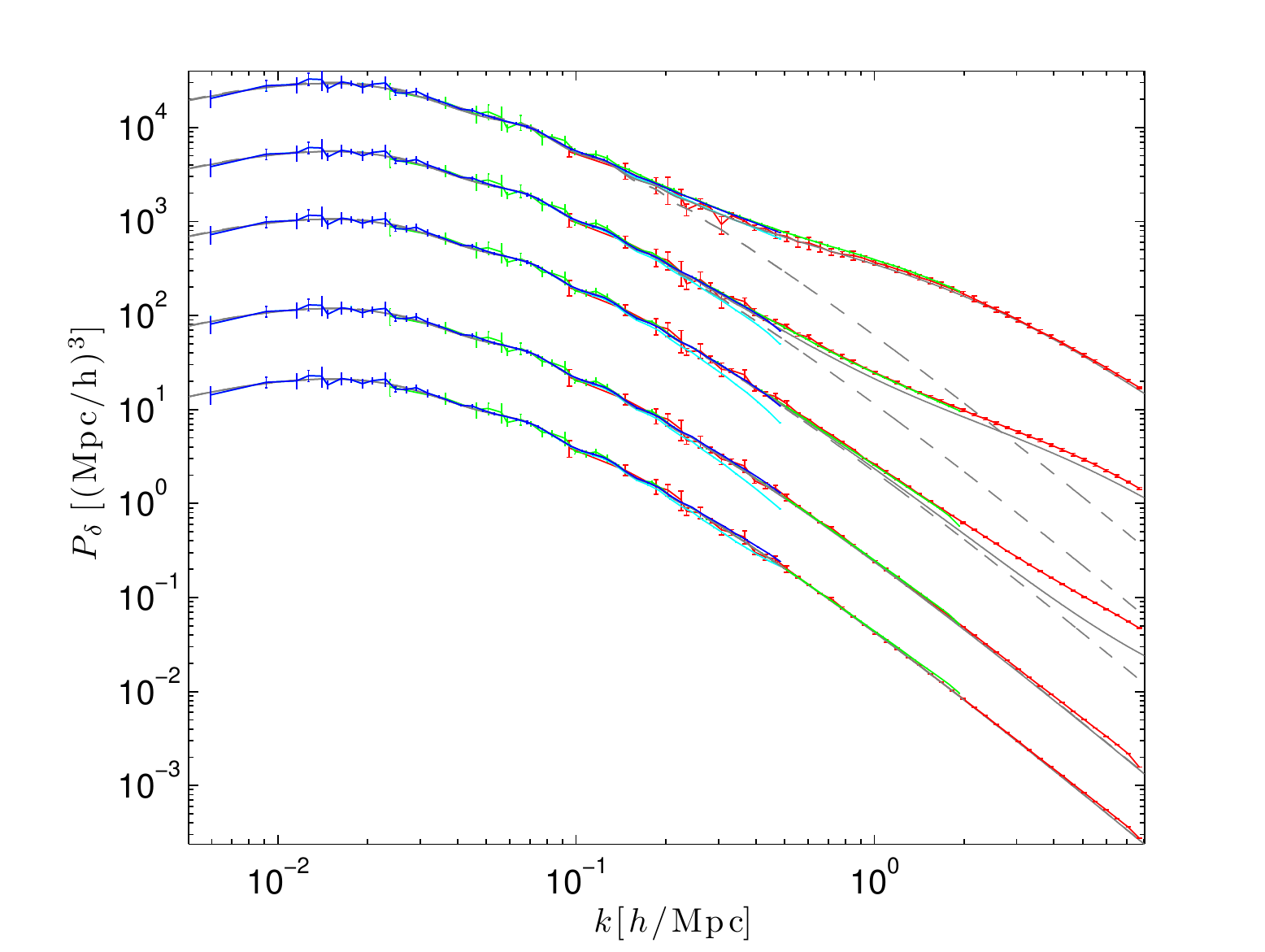}
\caption{Measured matter power spectra in simulations $\rm{G}^{512}_{100}$
  (red), $\rm{G}^{512}_{400}$ (green), $\rm{G512}$ (blue) and $\rm{G512g}$ (cyan) in
  comparison with power from linear perturbation theory (grey dashed) and CAMB (grey
  solid, \cite{camb,halofit}). See Table \ref{tab:nbody} for parameters.
  Curves from bottom to top correspond to redshifts
  $z=49,20,6,2,0$. }
\label{fig:nbody-power}
\end{figure}

\begin{figure}[ht]
\centering
\includegraphics[width=0.48\textwidth]{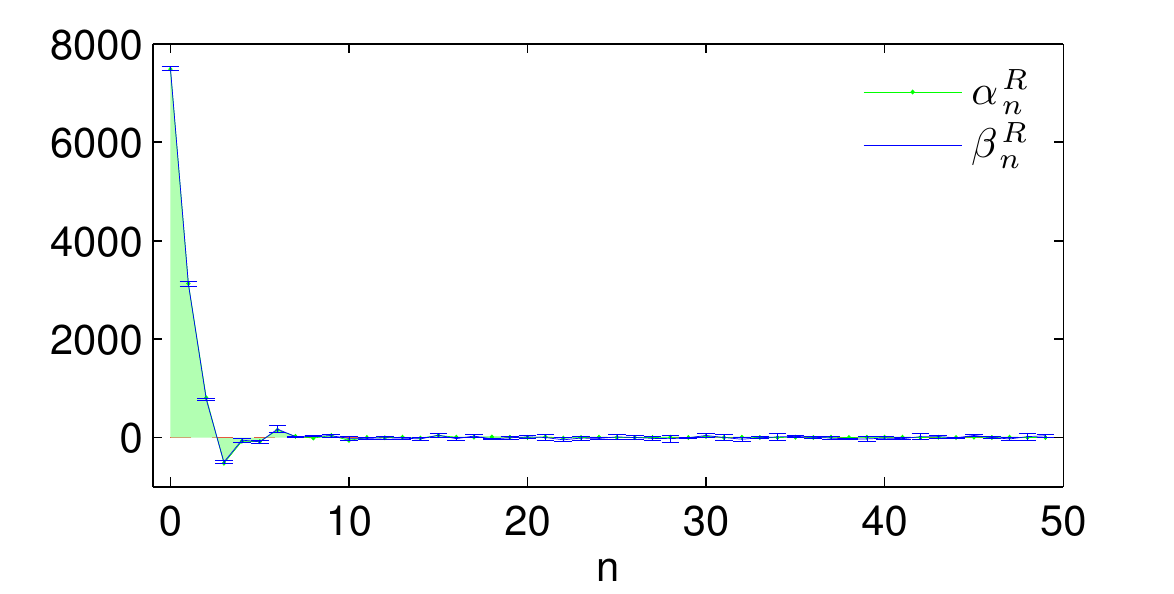}
\caption{Shot noise validation test for the separable bispectrum
  estimator: Measured bispectrum coefficients $\beta^R_n$
  \eqref{eq:betaR} (blue, averaged over $3$ seeds) for $512^3$
  particles placed randomly in a $L=1600\mathrm{Mpc}/h$ box, compared
  with the expansion coefficients ${\alpha}^R_n$ \eqref{eq:bispDec2}
  (green) for the expected pure shot noise bispectrum. The shape
  correlation is
  $\mathcal{C}_{\beta,\alpha}=0.9998$. 
  For better visibility the region under the $\alpha^R_n$ curve is
  colored green.  }
\label{fig:alphabeta_shotnoise}
\end{figure}

\begin{figure*}[t]
\centering
\subfloat[][Gaussian]{
\includegraphics[height=0.45\textheight]{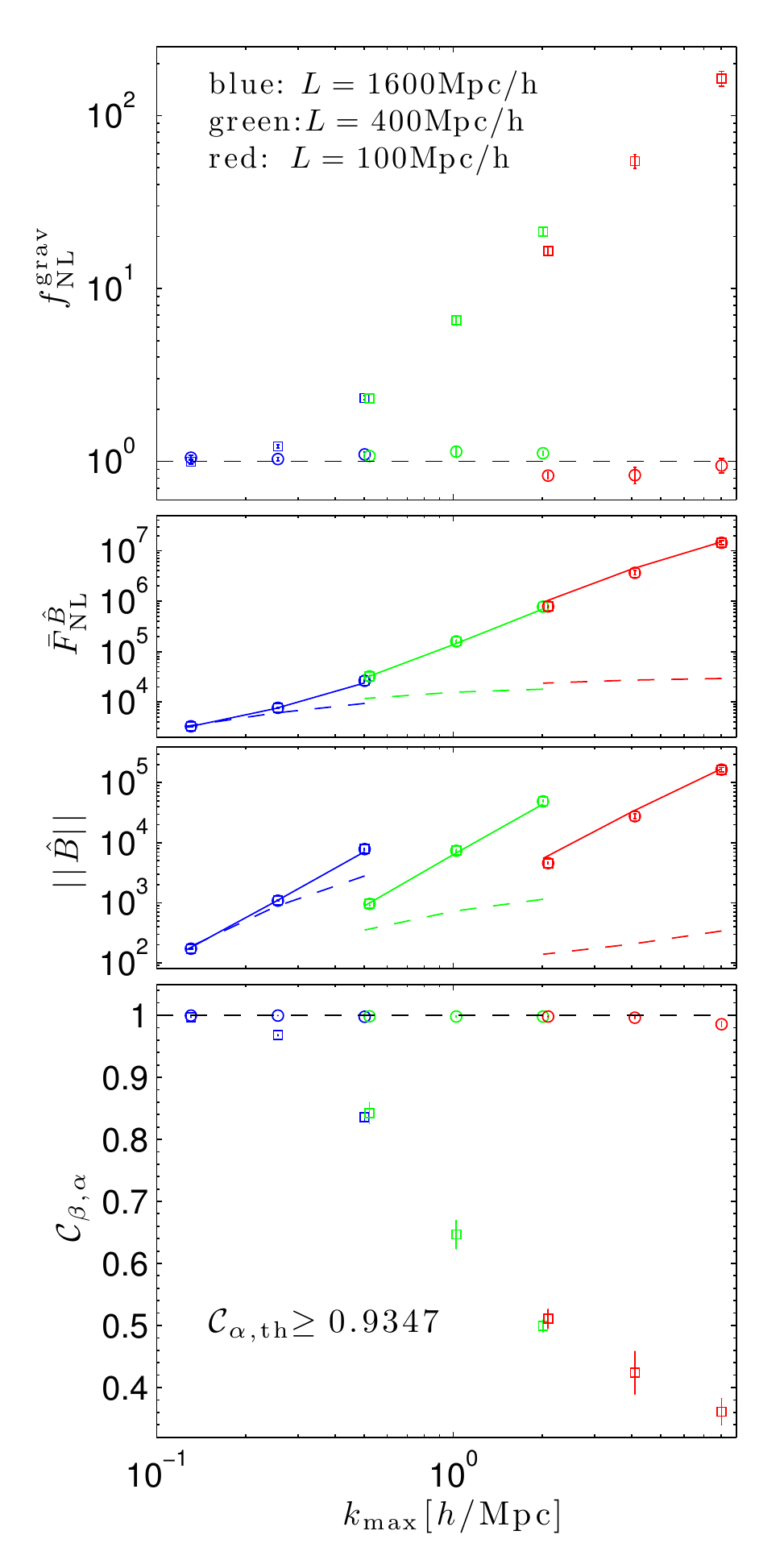}
\label{fig:scorrelK_grav}}
\subfloat[][Local]{
\includegraphics[height=0.45\textheight]{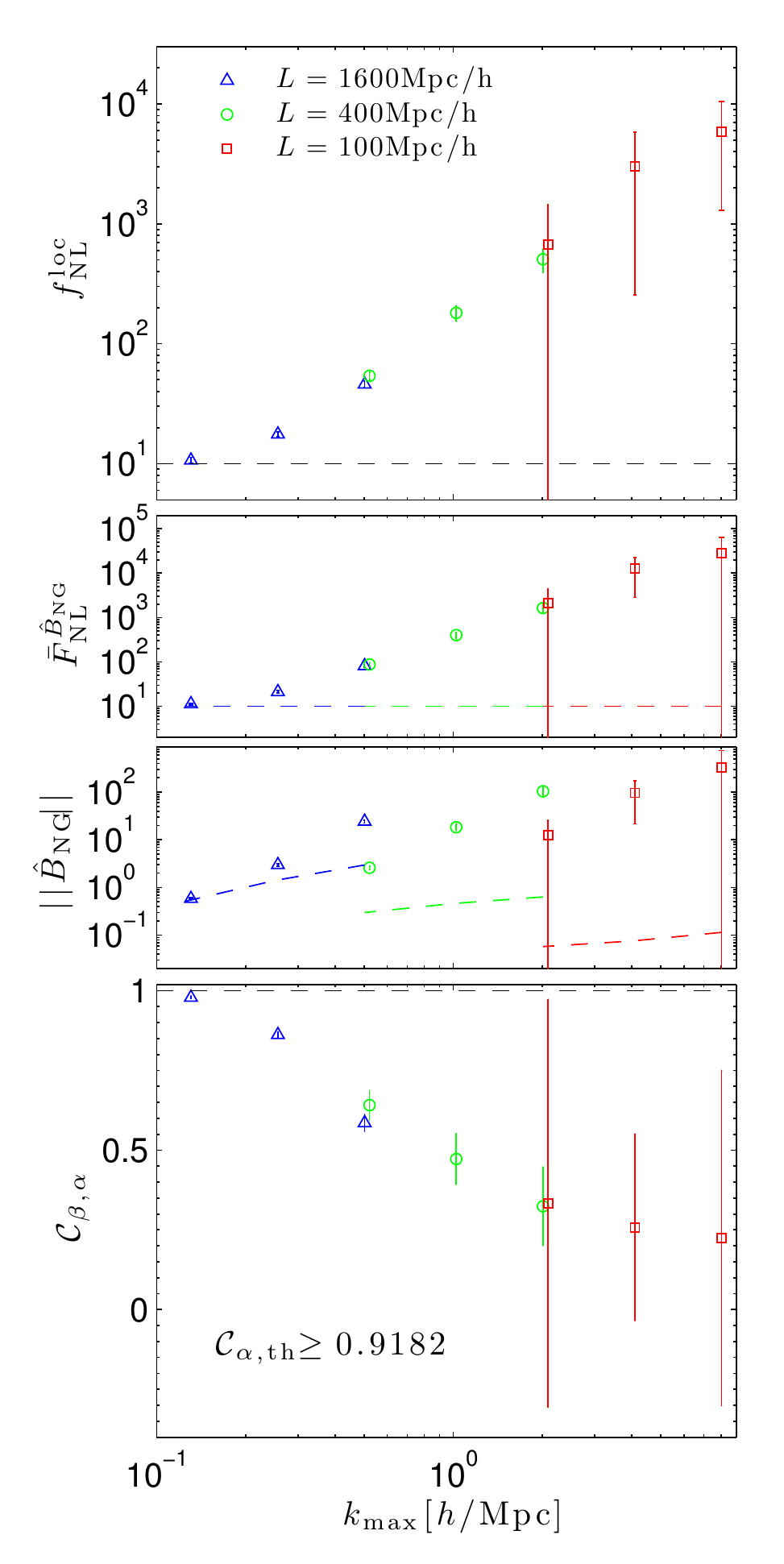}
\label{fig:scorrelK_local}}
\subfloat[][Equilateral]{
\includegraphics[height=0.45\textheight]{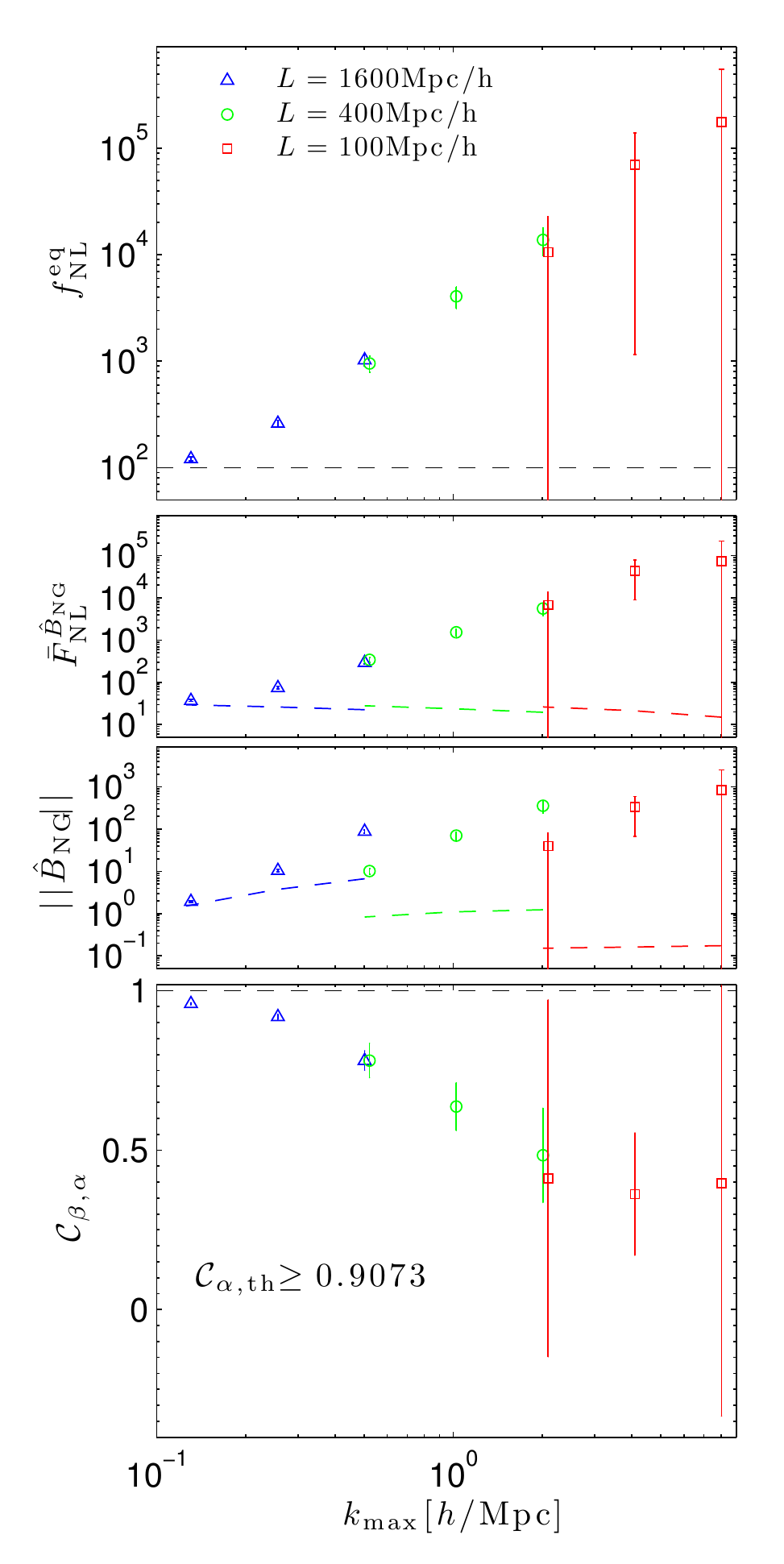}
\label{fig:scorrelK_eq}}
 \mycaption{Convergence tests: Dependence of measured
bispectra on $k_\mathrm{max}$  at $z=0$ for simulations with $512^3$
particles and box sizes
$L=\{1600,400,100\}\mathrm{Mpc}/h$
(G512, $\mathrm{G}^{512}_{\{400,100\}}$, Loc10, $\mathrm{Loc10}_{\{400,100\}}$,
Eq100 and $\mathrm{Eq100}_{\{400,100\}}$). (a) \emph{Top panel:} Projection of measured
bispectrum on the tree level prediction \eqref{eq:Bgrav} (squares) and
on the fitting formula from \cite{verde1111} (circles). \emph{Middle panels:}
Cumulative signal to noise of measured bispectrum (squares/circles),
tree level prediction \eqref{eq:Bgrav} (dashed lines) and fitting
formula from \cite{verde1111} (solid lines). \emph{Bottom panel:} Shape correlation of
measured bispectrum with tree level prediction (squares) and fitting
formula from \cite{verde1111} (circles). (b-c) Measured non-Gaussian
bispectra $\hat B_\mathrm{NG}$ \eqref{eq:B_ng_meas} in simulations
with local and equilateral initial conditions
compared to the linearly evolved primordial bispectrum
\eqref{eq:bprim-tree} (dashed lines). Note comments on the cumulative signal to noise
on small scales in the main text. }
\label{fig:scorrelK_all}
\end{figure*}

\begin{figure*}[t]
\centering
\subfloat[][]{
\includegraphics[height=0.45\textwidth]{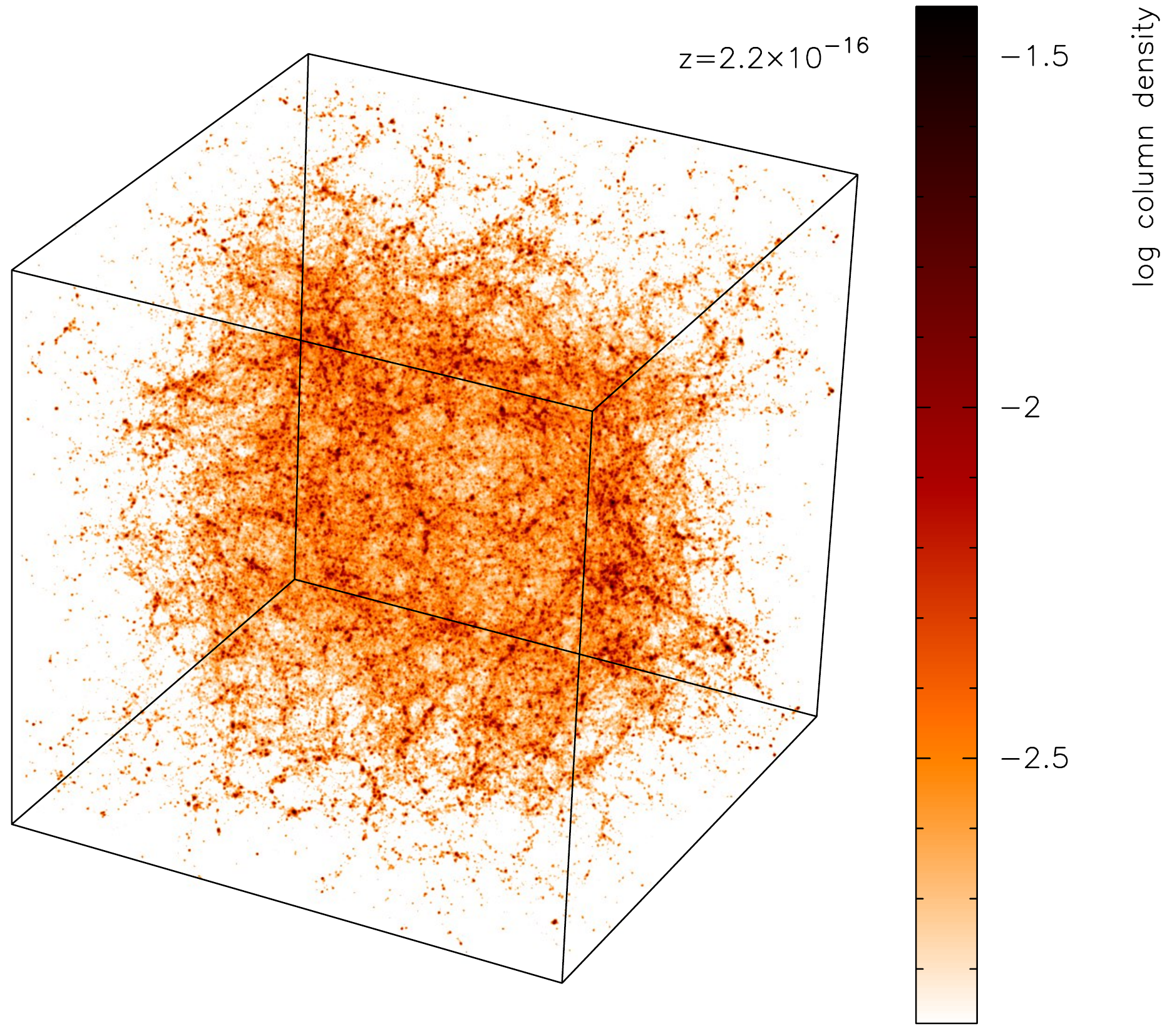}
\label{fig:splash3d_18DD_full_z0}}
\qquad
\subfloat[][]{
\includegraphics[height=0.45\textwidth]{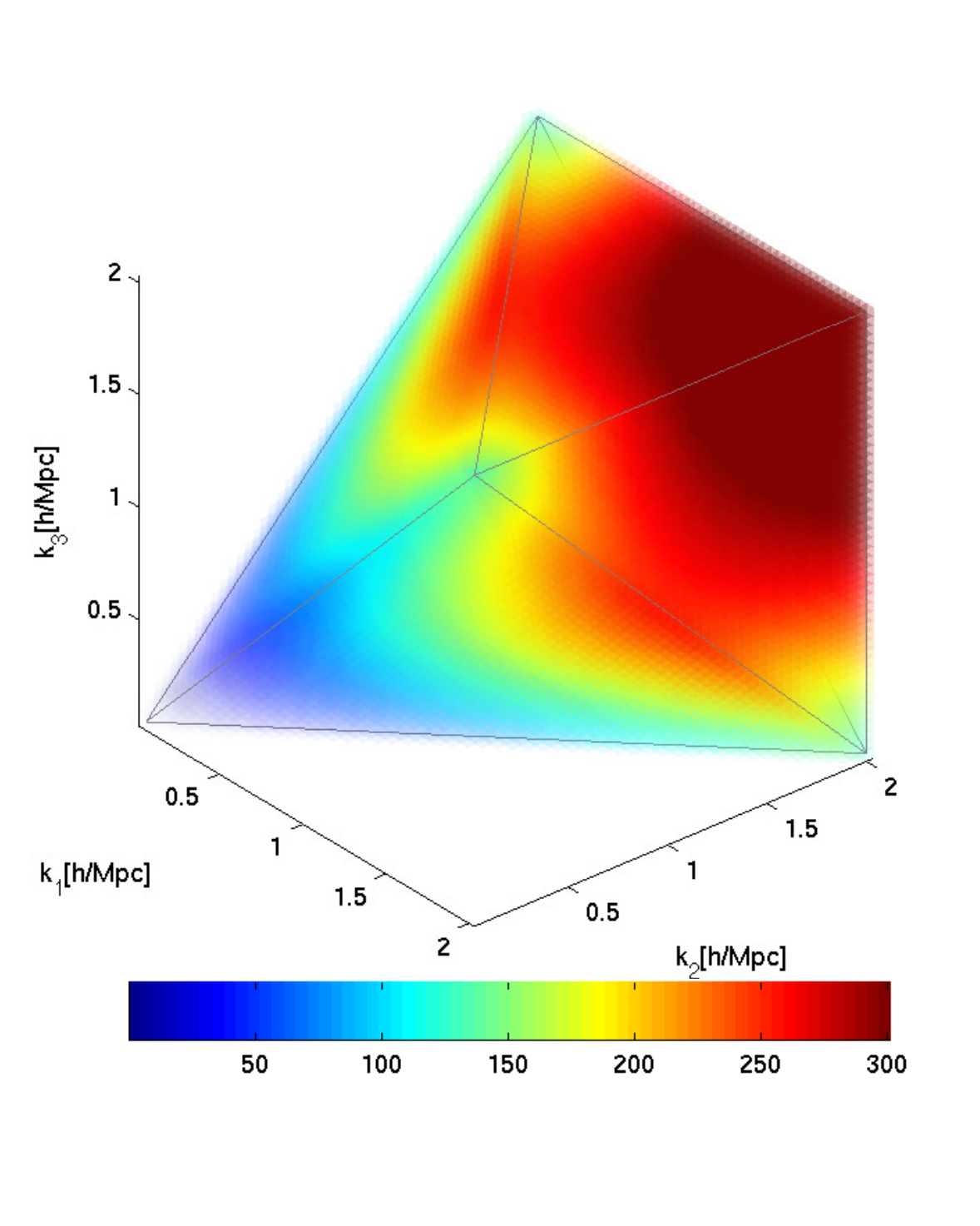}
\label{fig:bgrav_big_tetrapyd_z0}} \mycaption{(a) Dark matter
distribution in one of the $G^{512}_{400}$ simulations of $512^3$
particles in a box with $L=400\mathrm{Mpc}/h$ at redshift $z=0$. (b)
Measured (signal to noise weighted) bispectrum in the range
$0.016h/\mathrm{Mpc}\le k\le 2 h/\mathrm{Mpc}$, averaged over the
simulation on the left and two additional seeds.    The opaqueness of the plotted points
reflects the absolute value of the bispectrum. }
\label{fig:bgrav_splashred_and_bigtetra}
\end{figure*}

First we test the setup of the initial conditions and the $N$-body
simulations by comparing the measured matter power spectrum with the 
power spectrum predicted by linear theory and by CAMB
\cite{camb,halofit} in \fig{nbody-power}. Next we perform a simple
test of the bispectrum estimator by distributing particles randomly in
a box and comparing the measured bispectrum with the pure shot noise
bispectrum $B_\delta^\mathrm{shot}=L^6/N_p^6$ \cite{jeong-phd},
see \fig{alphabeta_shotnoise}.  These two tests show that the basic
setup of the simulations is correct and that the separable bispectrum
estimator works well.

To check convergence of the $N$-body simulations and the bispectrum
measurements we compare results for box sizes $L=\{1600, 400,
100\}\mathrm{Mpc}/h$ with $N_p^3=512^3$ particles.
 For each simulation we measure the bispectrum from
$k_\mathrm{min}=2\pi/L$ to $k_\mathrm{max}= \{1,0.5,0.25\}\times
(\tfrac{N_p}{4}\tfrac{2\pi}{L})$ by imposing a sharp $k$ filter on
the density perturbation $\delta_\vk$.  

The measured late time bispectra are shown in \fig{scorrelK_all} as
functions of $k_\mathrm{max}$.  Data points with the same
$k_\mathrm{max}$ measured in a low and high resolution simulation can
differ in the plots because, for these data, $k_\mathrm{min}$ differs
by a factor of $4$ and, therefore, the number of modes for the
bispectrum estimation differs by a factor of $4^3=64$, which affects
particularly the cumulative signal to noise $\|\hat B \|$.  In
contrast the quantity $\bar F_\mathrm{NL}^B$ is normalised with
respect to the linearly evolved local shape over the given range of
scales and therefore takes differences in the number of modes into
account. Thus the agreement of $\bar F_\mathrm{NL}^B$ (and
$\mathcal{C}_{\beta,\alpha}$) for the different simulations seen in
\fig{scorrelK_all} shows that the $N$-body simulations have converged
and the bispectrum estimator gives consistent results.

It should be noted that the shown cumulative signal to noise takes
only Gaussian noise into account, c.f.~\eqref{eq:B_gaussian_variance}.
 This simplification is not expected to be valid in the
strongly nonlinear regime so that the correct signal to noise may
differ significantly.  However to assess this issue in a robust manner
we need to run more realisations of the $N$-body simulations or
use larger boxes with more particles.

\subsection{Error bars}
To obtain error bars for quantities based on the primordial
contribution to the bispectrum, we use seed by seed subtracted
coefficients $\langle\beta^R_m -
(\beta^R_m)_\mathrm{Gauss}\rangle_\mathrm{sim}$ and calculate their
sample standard deviations, which are then used for standard error propagation
assuming uncorrelated $\beta^R_m$. 
 The error bars therefore measure how much the primordial
contribution to the matter bispectrum scatters for different seeds of
the initial Gaussian field $\Phi_G$ in the simulations.  Compared to
calculating the difference between the average bispectra,
$\langle\beta^R_m\rangle_\mathrm{sim} -
\langle(\beta^R_m)_\mathrm{Gauss}\rangle_\mathrm{sim}$, the seed by
seed subtraction reduces the error bars, because the late time
bispectra for Gaussian and non-Gaussian initial conditions are very
correlated due to the presence of the large gravitational bispectrum
in both cases.

In real observations we do not know the realisation of our universe
with perfectly Gaussian initial conditions and therefore error bars
will be larger if we measure $ \beta^R_m -
\langle(\beta^R_m)_\mathrm{Gauss}\rangle_\mathrm{sim}$. We do not
 discuss the interesting issue of observability of primordial
non-Gaussianity here, because instead of dark matter bispectra this
requires halo bispectra, which we leave for future work (but see
\cite{Sef2011} for the local shape).

\section{VI. Gravitational Bispectrum Results}

\subsection{Gravitational collapse and bispectrum evolution}

The cosmic web simulated by $N$-body simulations has a complex
filamentary structure, which is illustrated in
\fig{splash3d_18DD_full_z0}.  The bispectrum of this dark matter
distribution is shown in \fig{bgrav_big_tetrapyd_z0}.  In the
following we will discuss how the bispectrum can be used as a
quantitative tool to characterise the structures formed by
gravitational collapse and to study modifications due to primordial
non-Gaussianity. We will demonstrate the unbiasedness of our
bispectrum estimator by comparison with perturbation theory
 for large scales at early times.

\begin{figure*}[htp]
\centering
\subfloat[][Dark matter, $z=4$]{
\includegraphics[height=0.26\textheight]{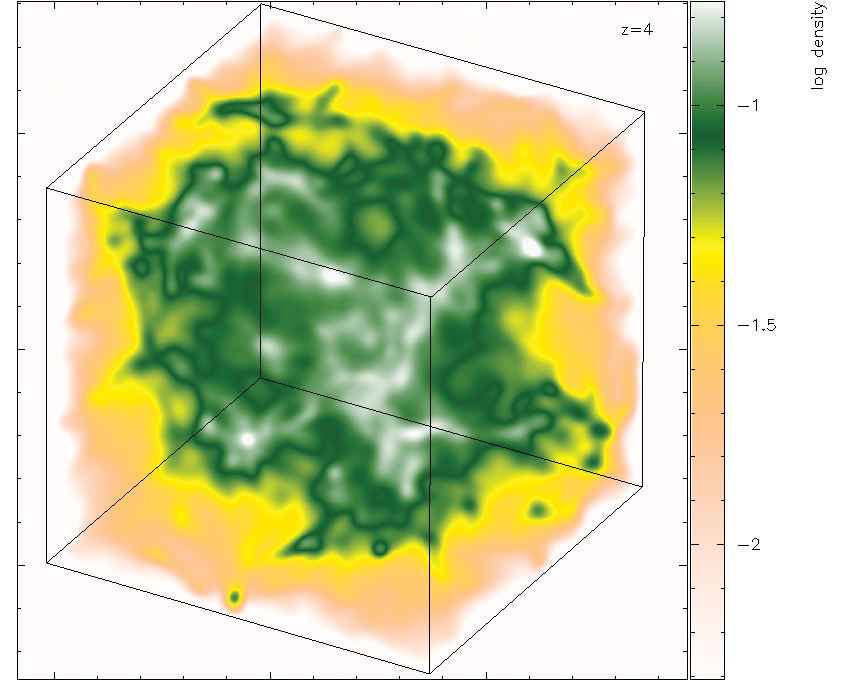}
\label{fig:splash_topology_z4}}
\qquad
\subfloat[][Bispectrum signal, $z=4$]{
\includegraphics[height=0.28\textheight]{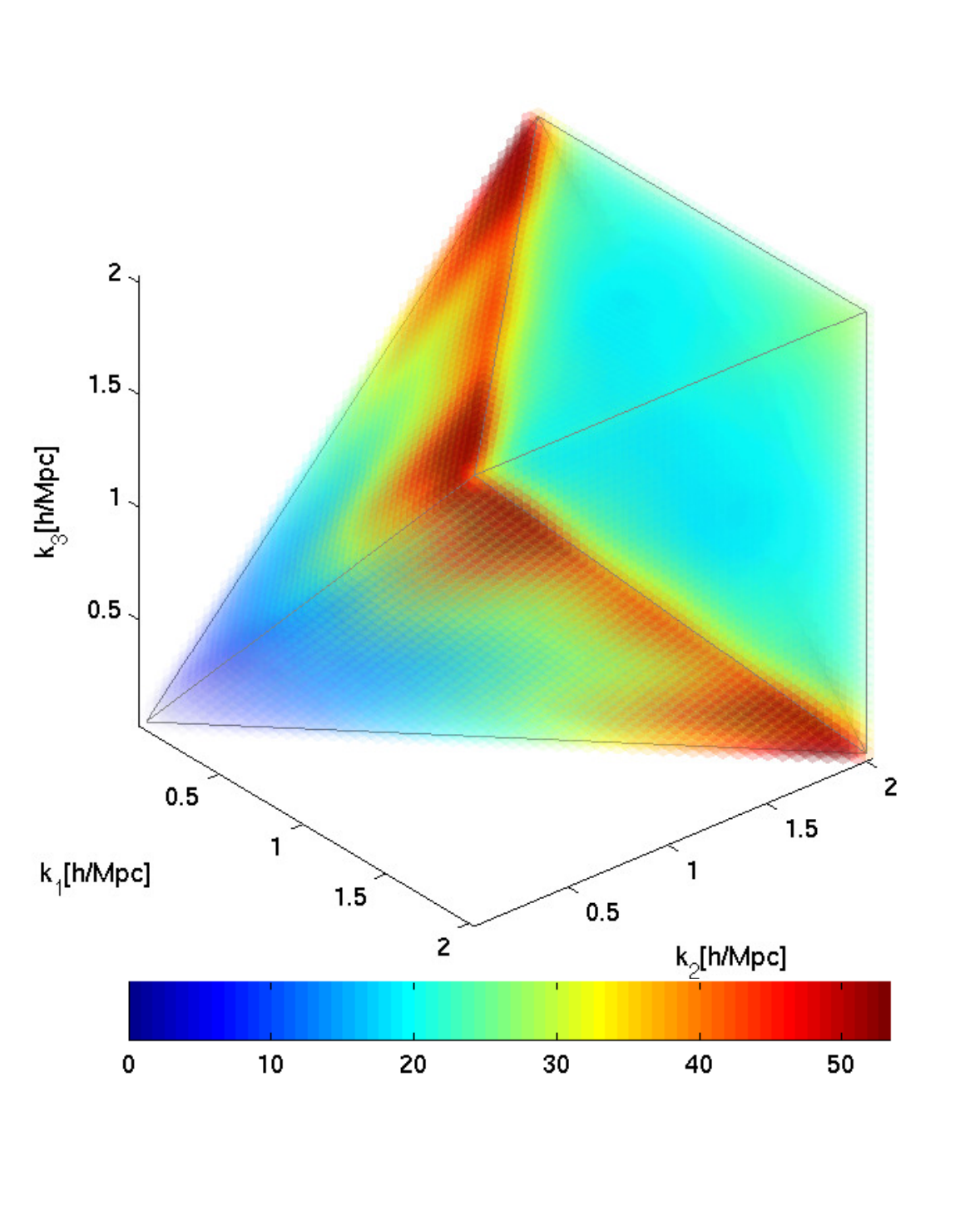}
\label{fig:tet3dwithcap_18DD_z4}}
\\
\subfloat[][Dark matter, $z=2$]{
\includegraphics[height=0.26\textheight]{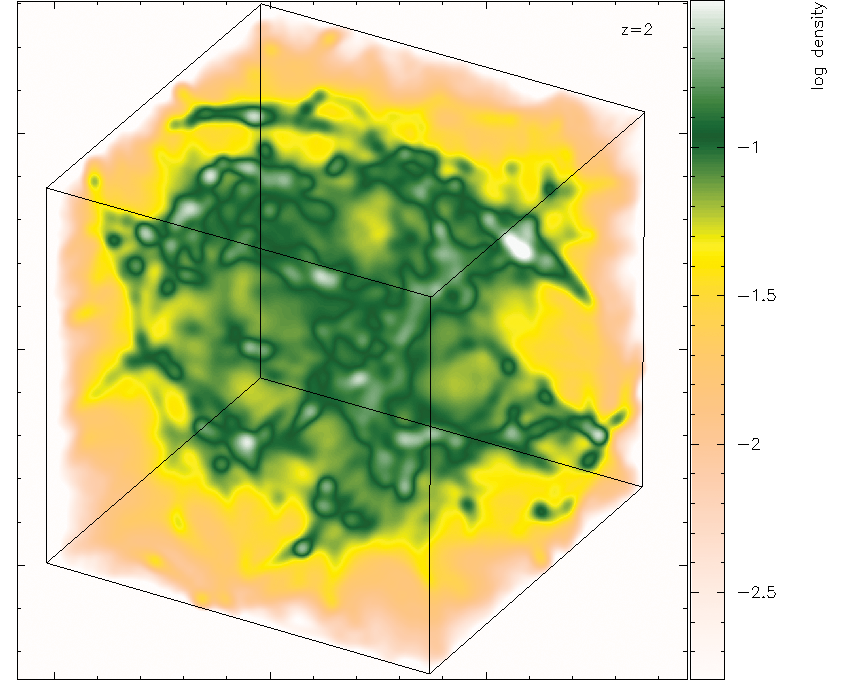}
\label{fig:splash_topology_z2}}
\qquad
\subfloat[][Bispectrum signal, $z=2$]{
\includegraphics[height=0.28\textheight]{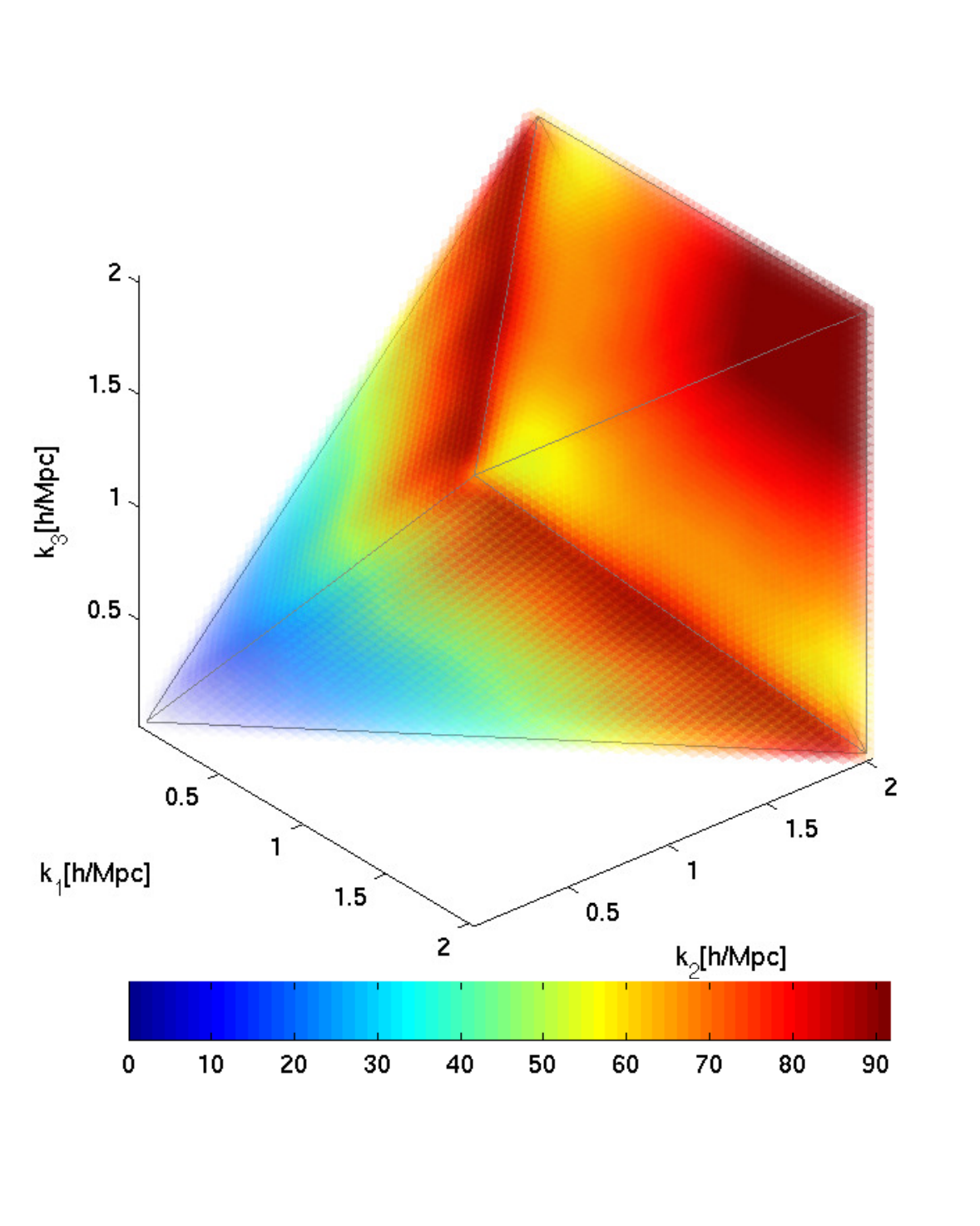}
\label{fig:tet3dwithcap_18DD_z2}}
\\
\subfloat[][Dark matter, $z=0$]{
\includegraphics[height=0.26\textheight]{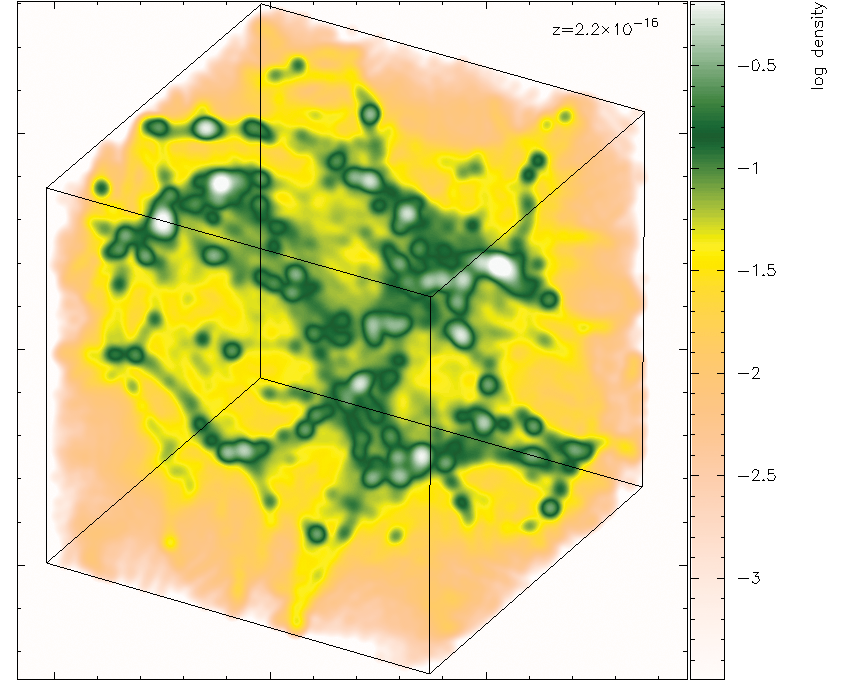}
\label{fig:splash_topology_z0}}
\qquad
\subfloat[][Bispectrum signal, $z=0$]{
\includegraphics[height=0.28\textheight]{figs_23july/tetwithcap_18DD_z0_kmax1d1_kmin1d1_prefac0_npoints50_ap200_3seeds_estpower_alphapow1_alphaamp0pt15_alphaoff0pt0025_clim361over1pt2_sett6.pdf}
\label{fig:tet3dwithcap_18DD_z0}} \mycaption{\emph{Left:} Dark matter
distribution in a $(40\mathrm{Mpc}/h)^3$ subbox of one of the
$G^{512}_{400}$ simulations at redshifts $z=4,2$ and $0$, from top to
bottom. \emph{Right:} Measured (signal to noise weighted) bispectrum
in the range $0.016h/\mathrm{Mpc}\le k\le 2 h/\mathrm{Mpc}$, averaged
over the simulation on the left and two additional seeds.  }
\label{fig:bgrav_splash_and_big_tetrapyd}
\end{figure*}

\begin{figure}[htp]
\hspace*{-0.6cm}
\subfloat[][Bispectrum signal in G512g, $z=0$]{
\includegraphics[width=0.51\textwidth]{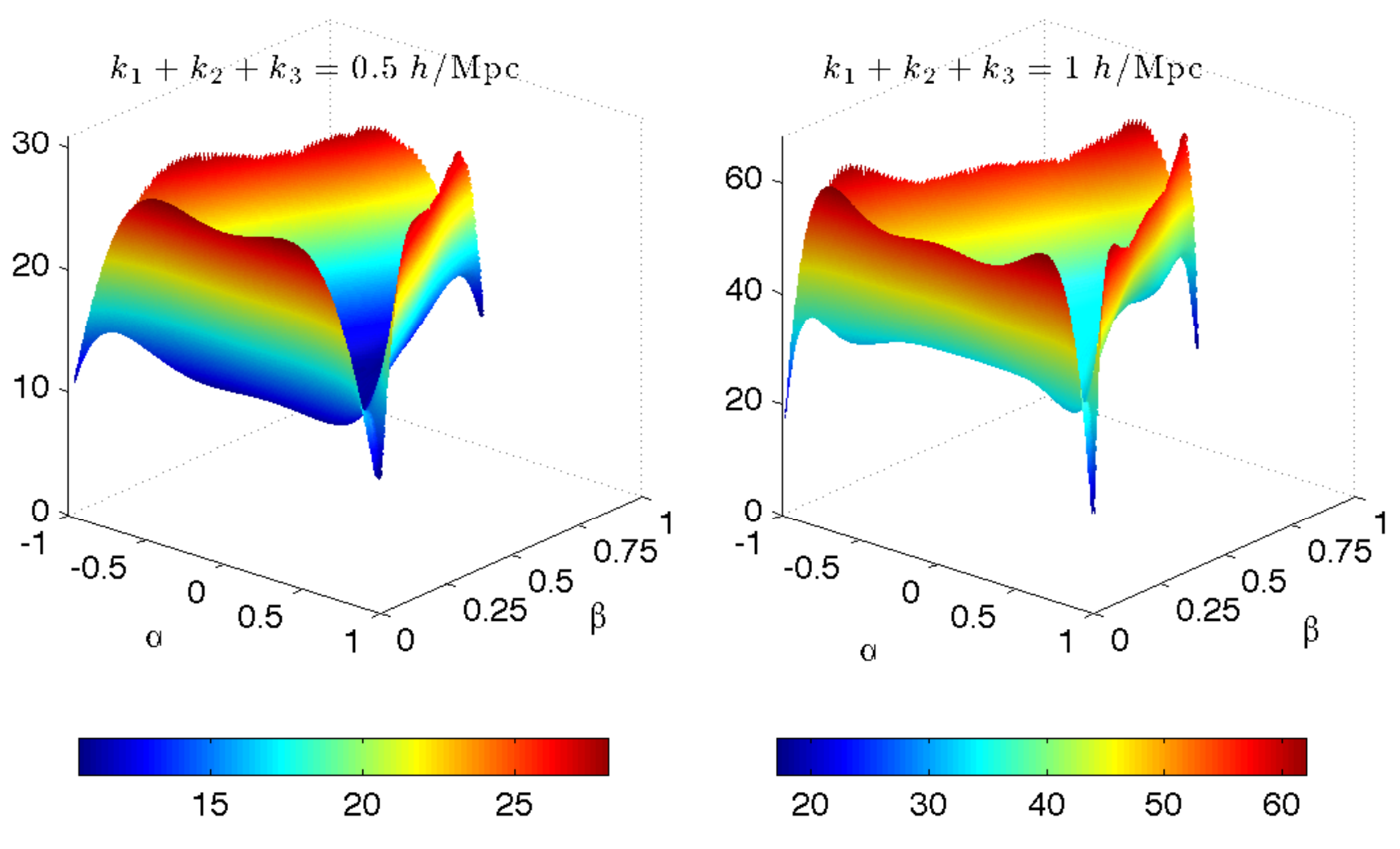}
\label{fig:tetslice_ab_18BBfromglass_z0_a}}
\\
\hspace*{-0.6cm}
\subfloat[][Bispectrum signal in $G^{512}_{400}$, $z=0$]{
\includegraphics[width=0.51\textwidth]{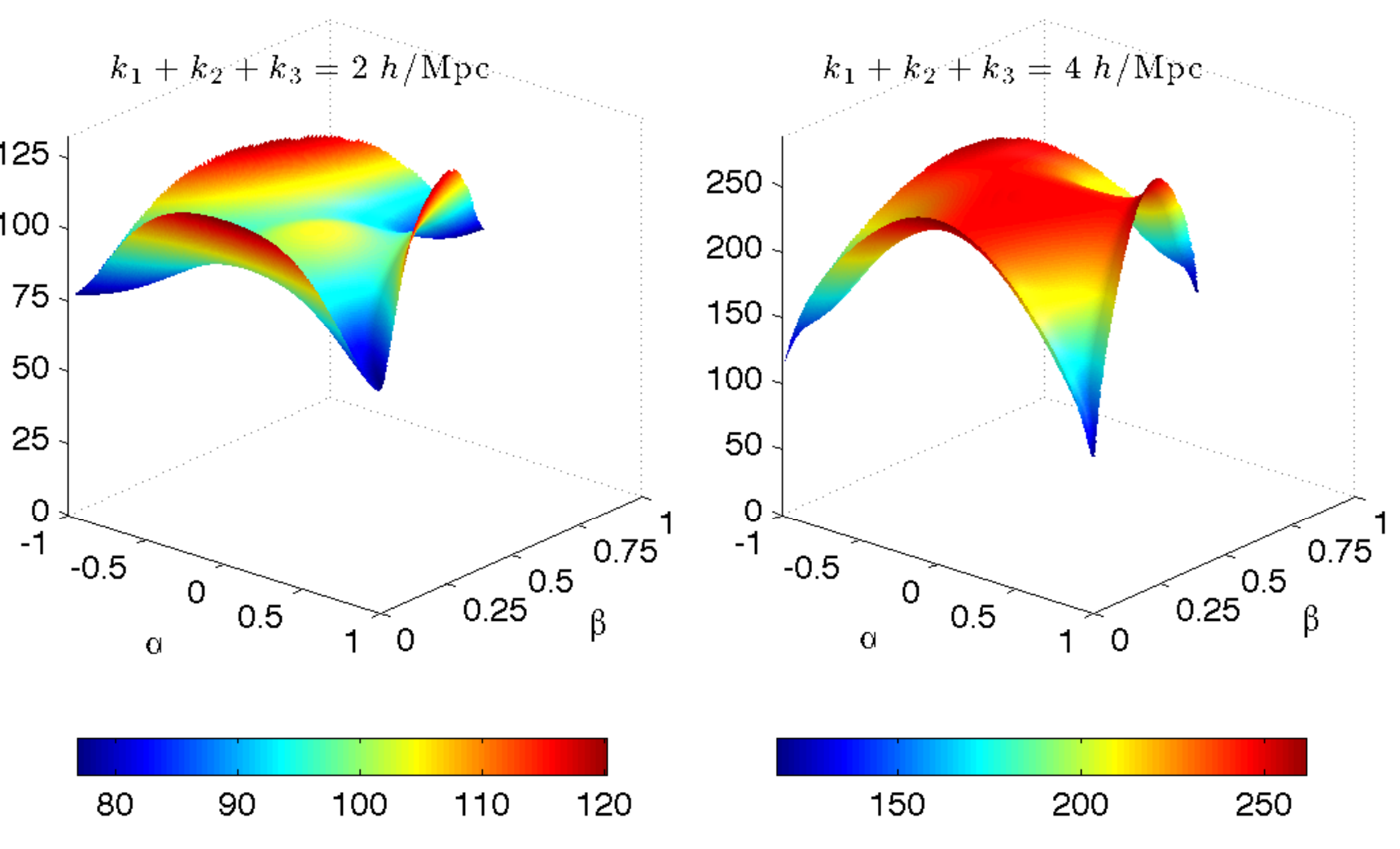}
\label{fig:tetslice_ab_18BBfromglass_z0_b}} \mycaption{Bispectrum
slices at fixed redshift $z=0$ for different overall wavenumbers
$k_\mathrm{sum}=k_1+k_2+k_3$.    For
$k_\mathrm{sum}\leq 1h/\mathrm{Mpc}$ (upper row) the bispectrum signal
dominates for flattened configurations, while an enhanced equilateral
contribution clearly emerges for $k_\mathrm{sum}=2h/\mathrm{Mpc}$
(bottom left). For $k_\mathrm{sum}=4h/\mathrm{Mpc}$ (bottom right) the
signal is large for all triangle configurations away from the squeezed
limit, that is, it is nearly constant.  The slices show  signal to
noise weighted bispectra measured in
Gaussian simulations (a) G512g and (b) $G^{512}_{400}$.}
\label{fig:tetslices_ab_z0}
\end{figure}

We can develop a qualitative and visual understanding of the bispectrum by comparing it to the form of large-scale structures. 
\fig{bgrav_splash_and_big_tetrapyd} shows snapshots of the dark matter
distribution at redshifts $z=4, 2$ and $0$ on the left and the
corresponding measured bispectrum signal on the right.  It is apparent
that the shape of the bispectrum characterises the three-dimensional
structures that have formed. The diffuse blob- and pancake-like structures at early times,
$z=4$, correspond to a flattened bispectrum, which peaks at the edges
of the tetrapyd and is predicted by leading order perturbation
theory. At later times we find that filaments and clusters induce a
bispectrum with a relatively enhanced signal for equilateral triangle
configurations, as expected from sections II and III.
As illustrated in \fig{tetslices_ab_z0}, the bispectrum signal shows a
similar transition for varying overall scale $k_1+k_2+k_3$ at fixed
time $z=0$, which indicates self-similar behaviour.

\begin{figure}[htb]
\centering
\subfloat[][Gaussian]{
\includegraphics[width=0.24\textwidth]{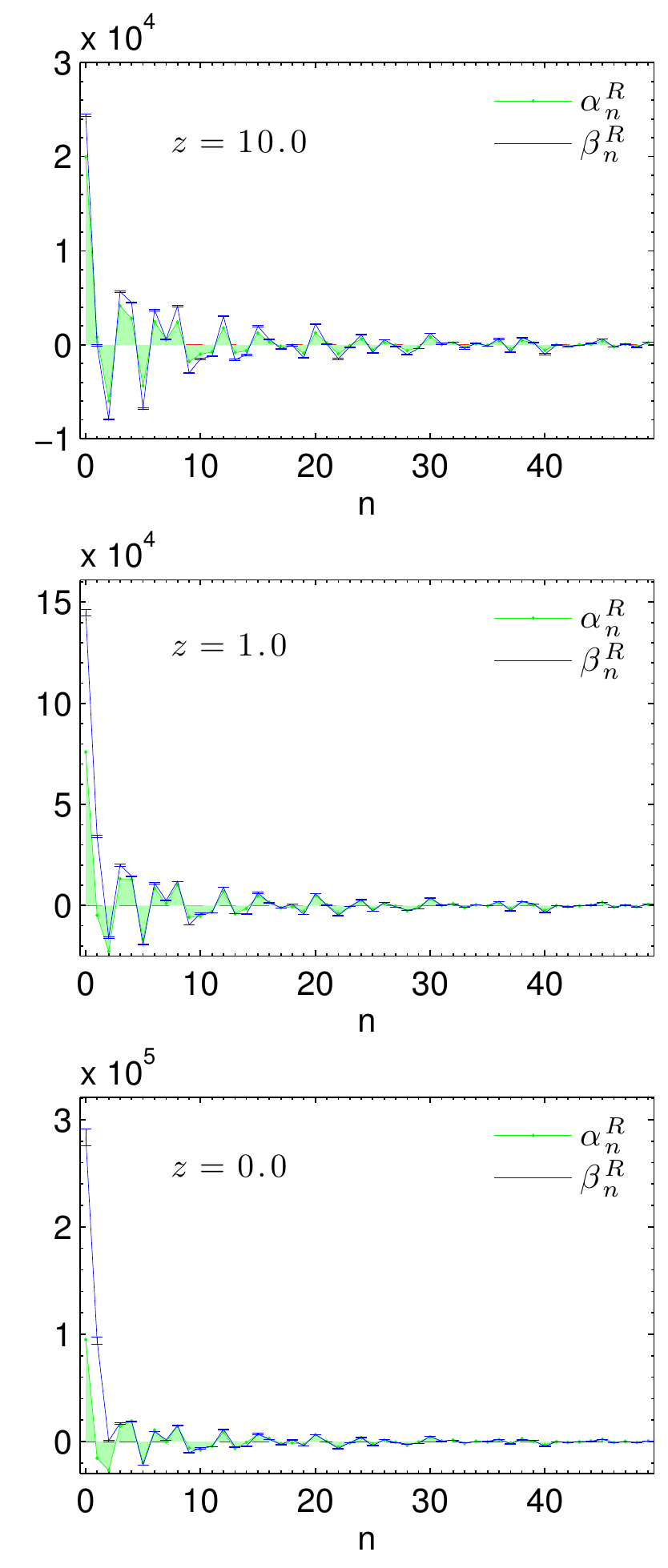}
\label{fig:alphabeta_gaussian}}
\subfloat[][Local]{
\includegraphics[width=0.24\textwidth]{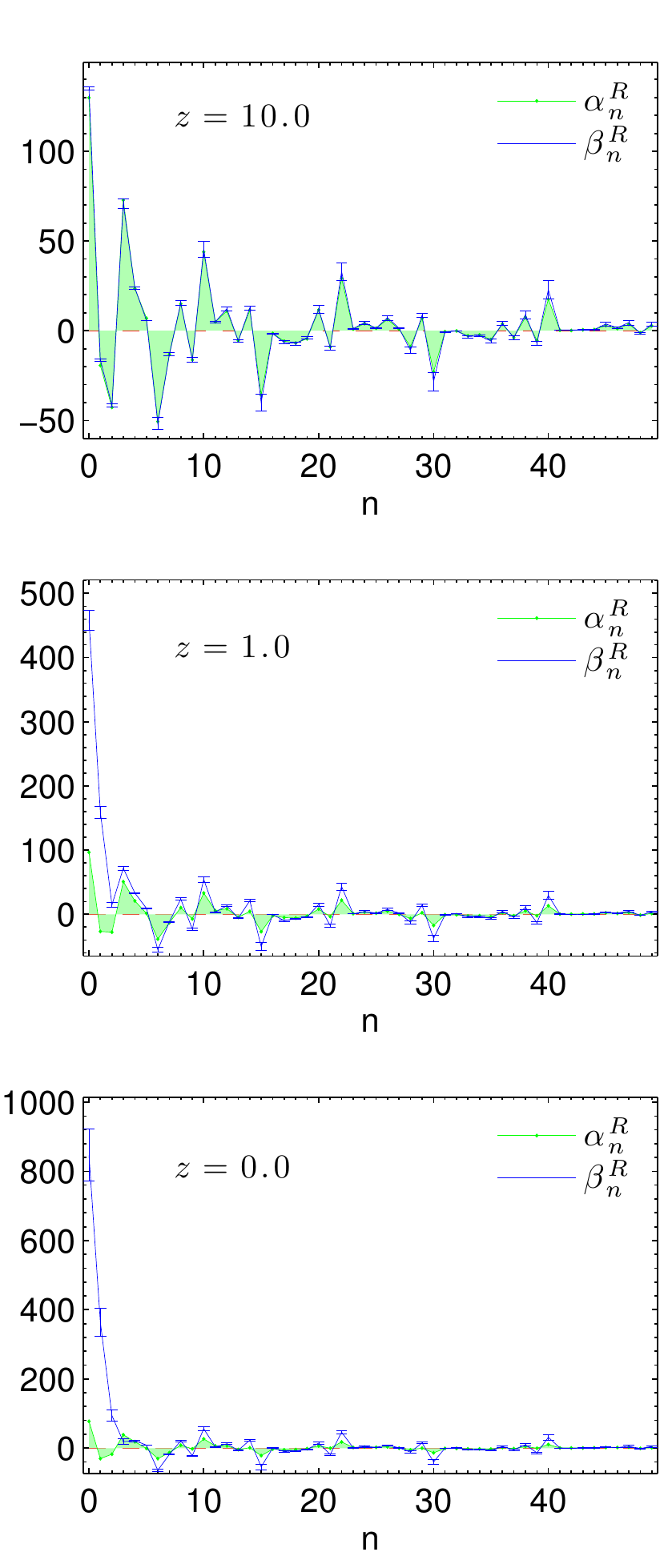}
\label{fig:alphabeta_local}}
\mycaption{(a) $\beta^R_n$ coefficients measured with \eqref{eq:betaR}
  for the Gaussian simulations G512 compared to expansion coefficients
  ${\alpha}^R_n$ of the tree level gravitational bispectrum
  \eqref{eq:Bgrav}. (b) $\beta^R_n-(\beta^R_n)_\mathrm{Gauss}$ measured in
  Loc10 simulations with $f_\mathrm{NL}^\mathrm{loc}=10$ compared to
  ${\alpha}^R_n$ coefficients of the linearly evolved local shape
  (see \eqref{eq:bprim-tree} and \eqref{eq:Blocal}). 
$\beta$'s were
 calculated using all modes $\delta_\bk$ with $0.0039h/\mathrm{Mpc} \le
 k\le 0.5h/\mathrm{Mpc}$. For better visibility the region under the $\alpha^R_n$ curves is
  colored green.}
\label{fig:18Q_alphabeta}
\end{figure}

\begin{figure*}[p]
\centering
\includegraphics[height=0.9\textheight,angle=0]{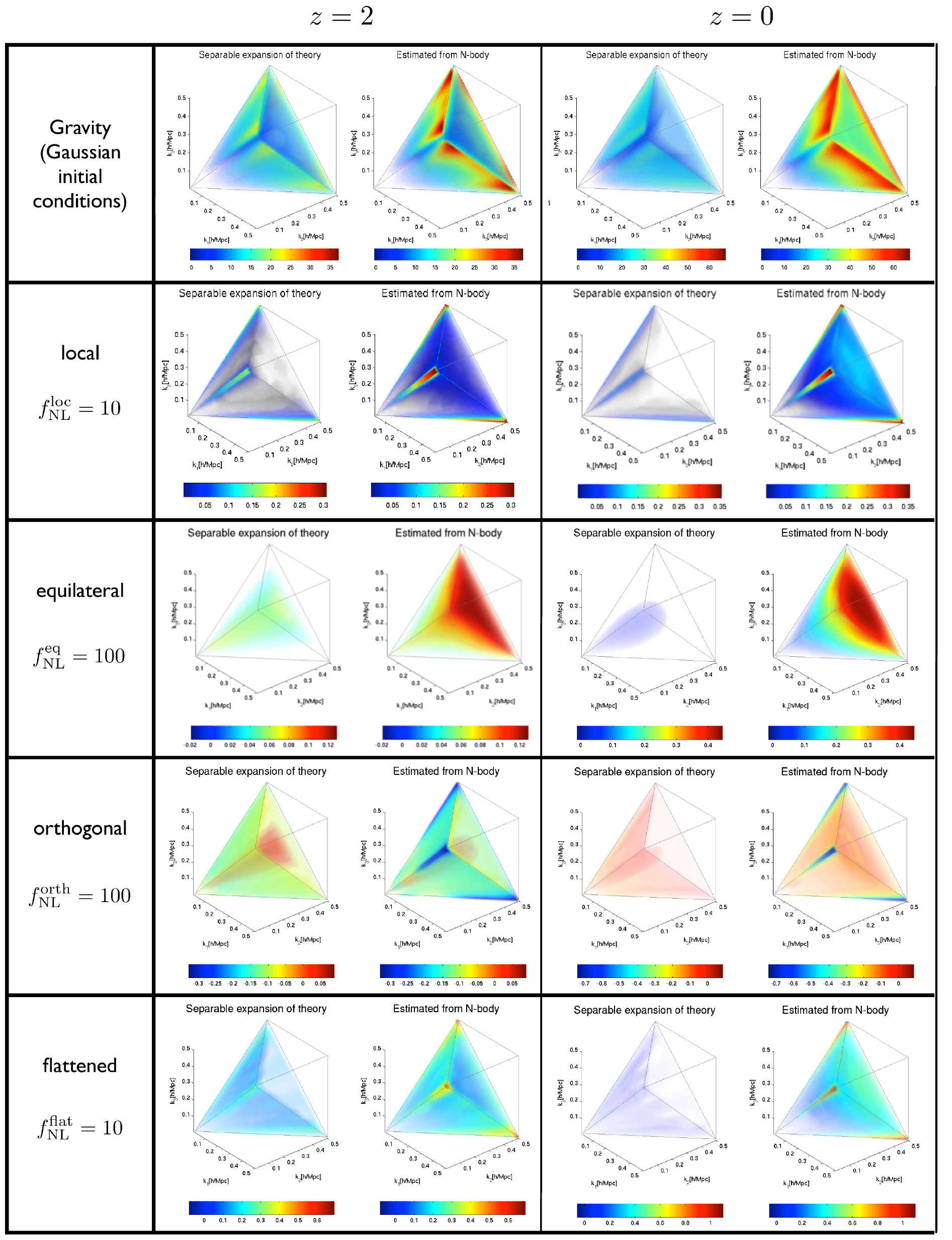}
\caption{Comparison of measured $N$-body bispectra
  \eqref{eq:bisp-from-betas} (on the right in each black cell) with
  separable expansion of tree level theory \eqref{eq:bispDec2} (on the
  left in each black cell) for simulations G512b, Loc10, Eq100,
  Orth100 and Flat10 at redshifts $z=2$ and $z=0$. The plot axes
  are $k_1, k_2, k_3\le 0.5h/\mathrm{Mpc}$ and we plot signal to noise
  weighted bispectra
  $\sqrt{k_1k_2k_3/(P_\delta(k_1)P_\delta(k_2)P_\delta(k_3))}B_\delta$
  (see \fig{tet3dplots_show_expansions_grav_loc_eq_orth} for
  comparisons at $z=30$ and further descriptions of the plots).}
\label{fig:tet3dplots_grav_loc_eq_orth}
\end{figure*}

\subsubsection{Comparison with leading order PT}

On large scales and at early times we expect the gravitational
bispectrum to agree with leading order perturbation theory
\eqref{eq:Bgrav}.  To test this \fig{18Q_alphabeta}a shows the
$\beta^R_n$ coefficients \eqref{eq:betaR} measured in the Gaussian
$N$-body simulations $\mathrm{G512}$ with $L=1600\mathrm{Mpc}/h$ in
comparison with the expansion coefficients $\alpha^R_n$
\eqref{eq:bispDec2} of the tree level gravity bispectrum 
\eqref{eq:Bgrav}.  The corresponding
theoretical and estimated bispectra \eqref{eq:bisp-from-betas} are
plotted over the full tetrapyd in
\fig{tet3dplots_show_expansions_grav_loc_eq_orth} for $z=30$ and
\fig{tet3dplots_grav_loc_eq_orth} for $z=2$ and $z=0$. At early times,
$z=30$, the bispectrum agrees with tree level perturbation theory
\eqref{eq:Bgrav}, which predicts that the signal to noise weighted
bispectrum is large for folded and elongated configurations but is
suppressed in the squeezed limit.  However at $z=0$ for $k\gtrsim
0.25h/\mathrm{Mpc}$ the $N$-body bispectrum has an enhanced amplitude
for elongated configurations and an additional equilateral
contribution not captured by tree level perturbation theory, which
breaks down on these scales as expected.

\begin{figure*}[htp]
\centering
\vspace{-0.4cm}
\subfloat[][G512g]{
\includegraphics[height=0.45\textheight]{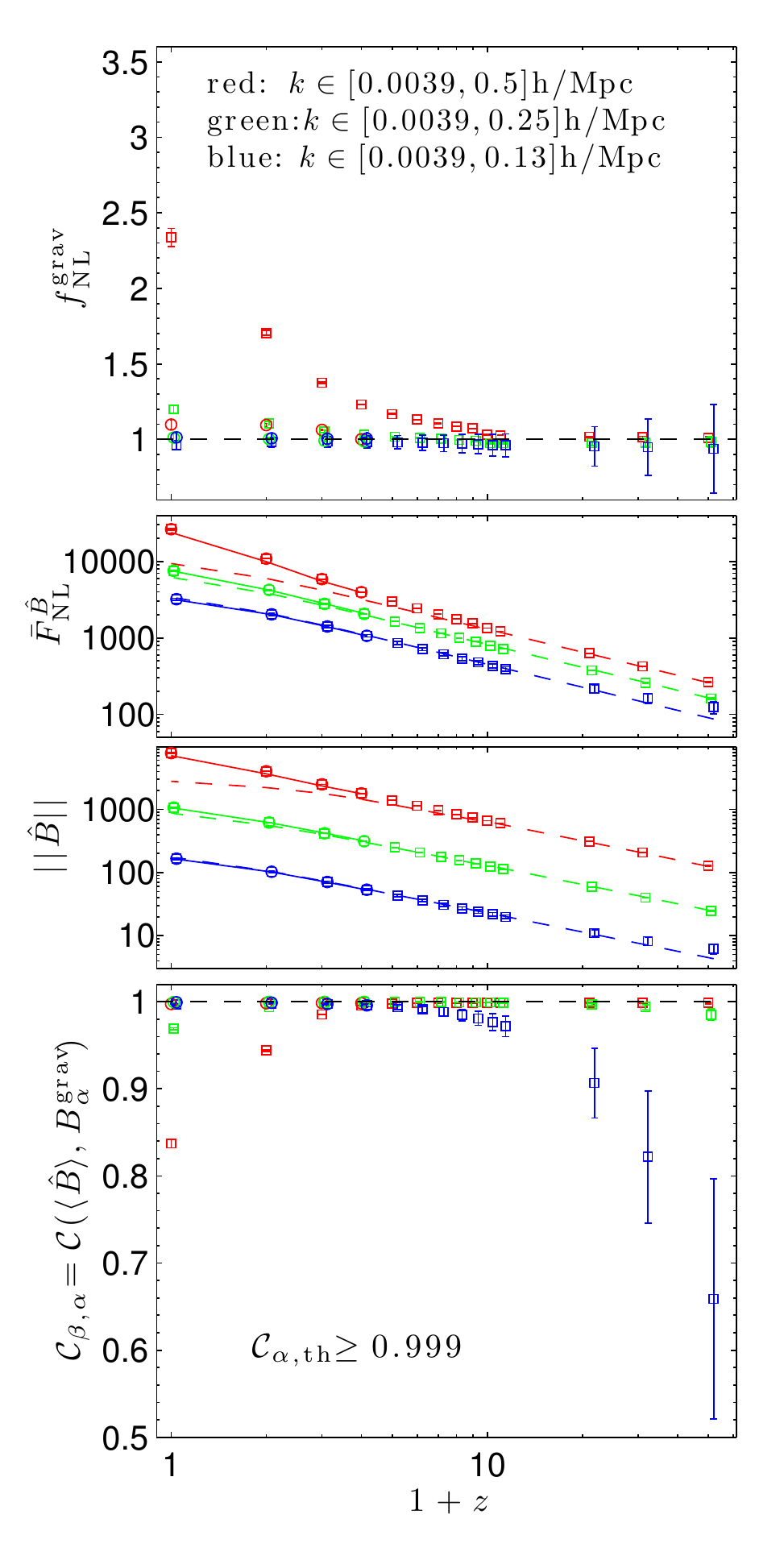}
\label{fig:bgrav_18BBfromglass}}
\subfloat[][G512]{
\includegraphics[height=0.45\textheight]{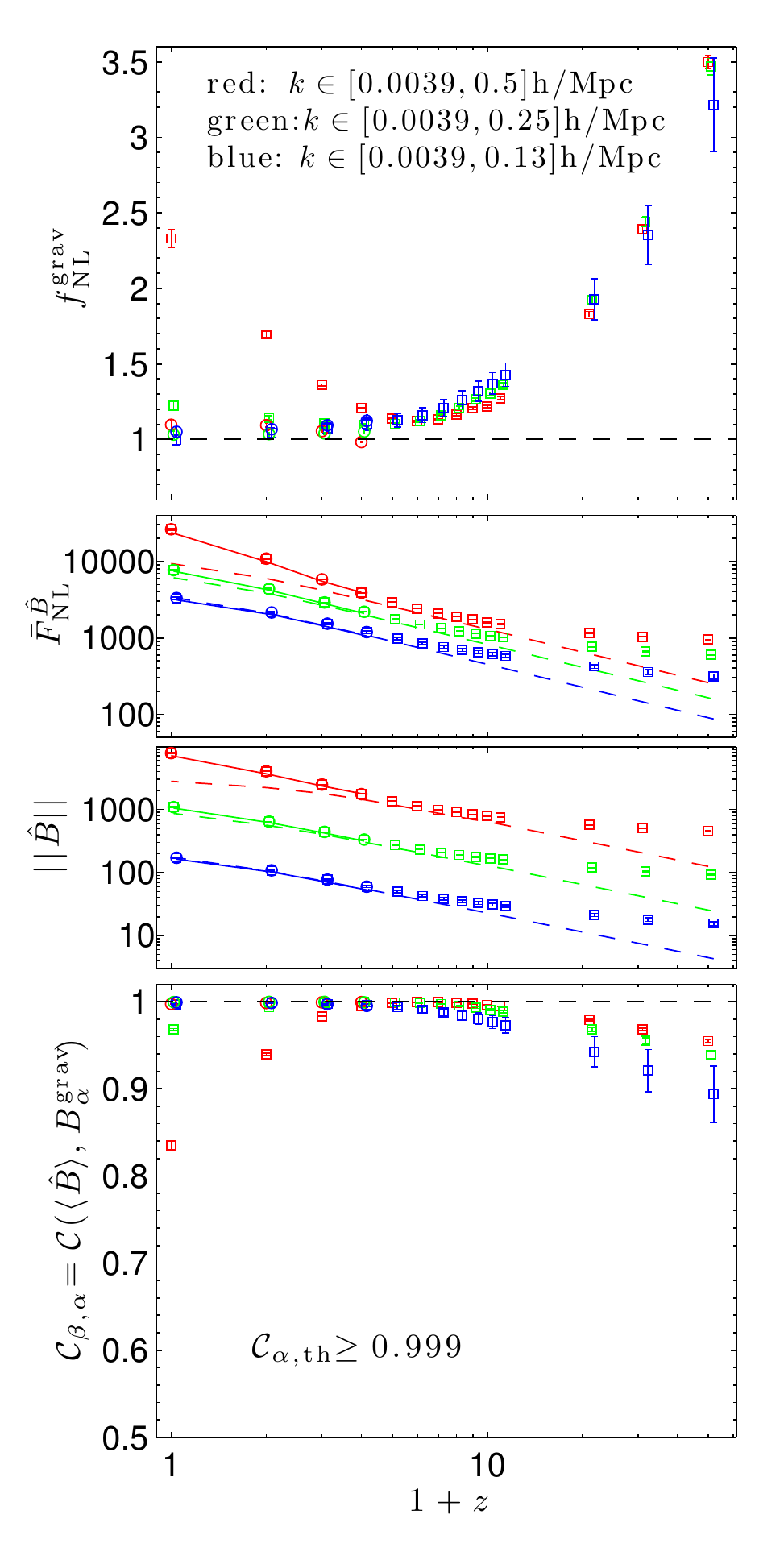}
\label{fig:bgrav_18Q}}
\subfloat[][Ratio G512g/G512]{
\includegraphics[height=0.45\textheight]{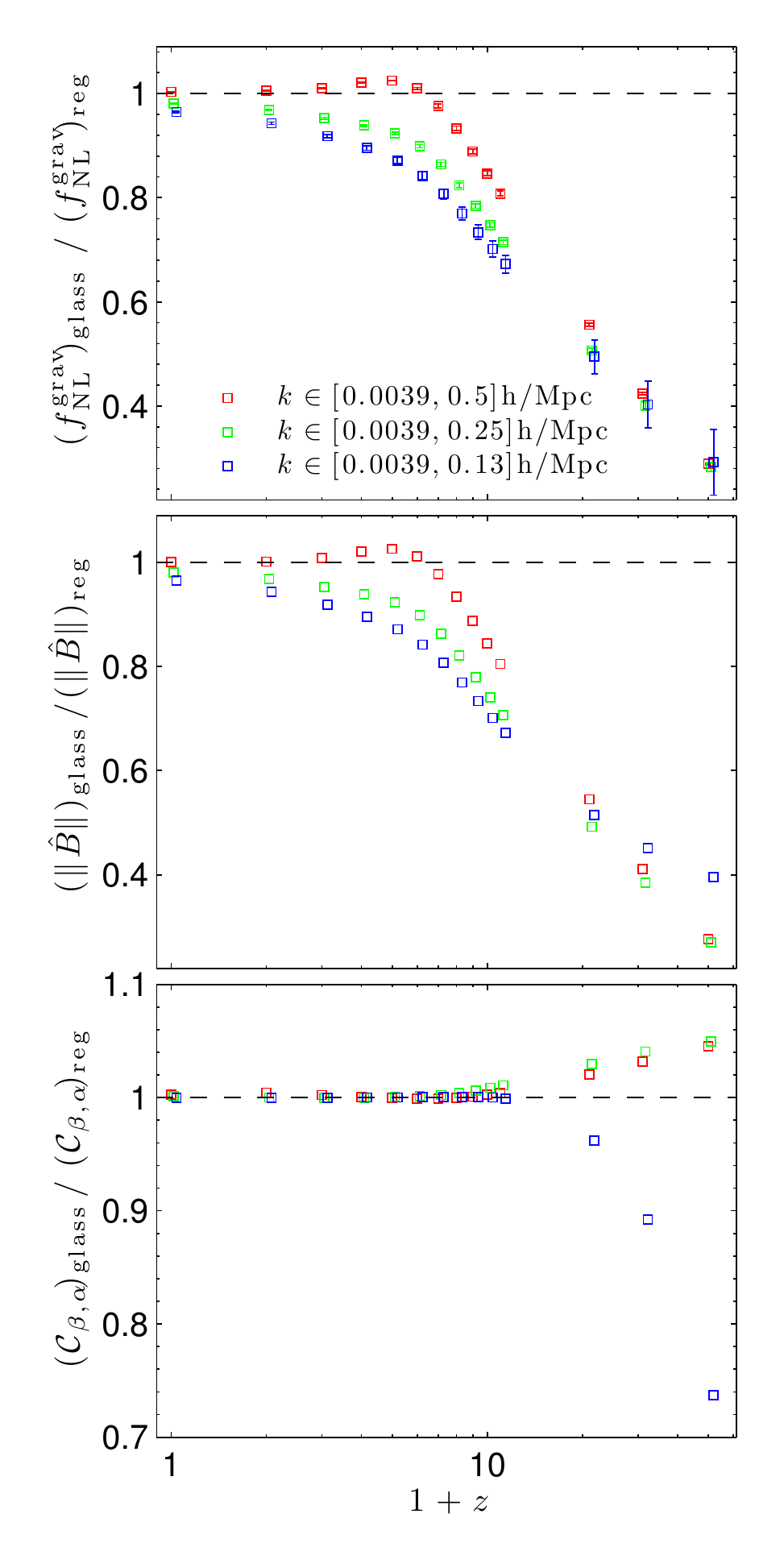}
\label{fig:bgrav_18Q_glassratio}}
\\
\subfloat[][G768]{
\includegraphics[height=0.45\textheight]{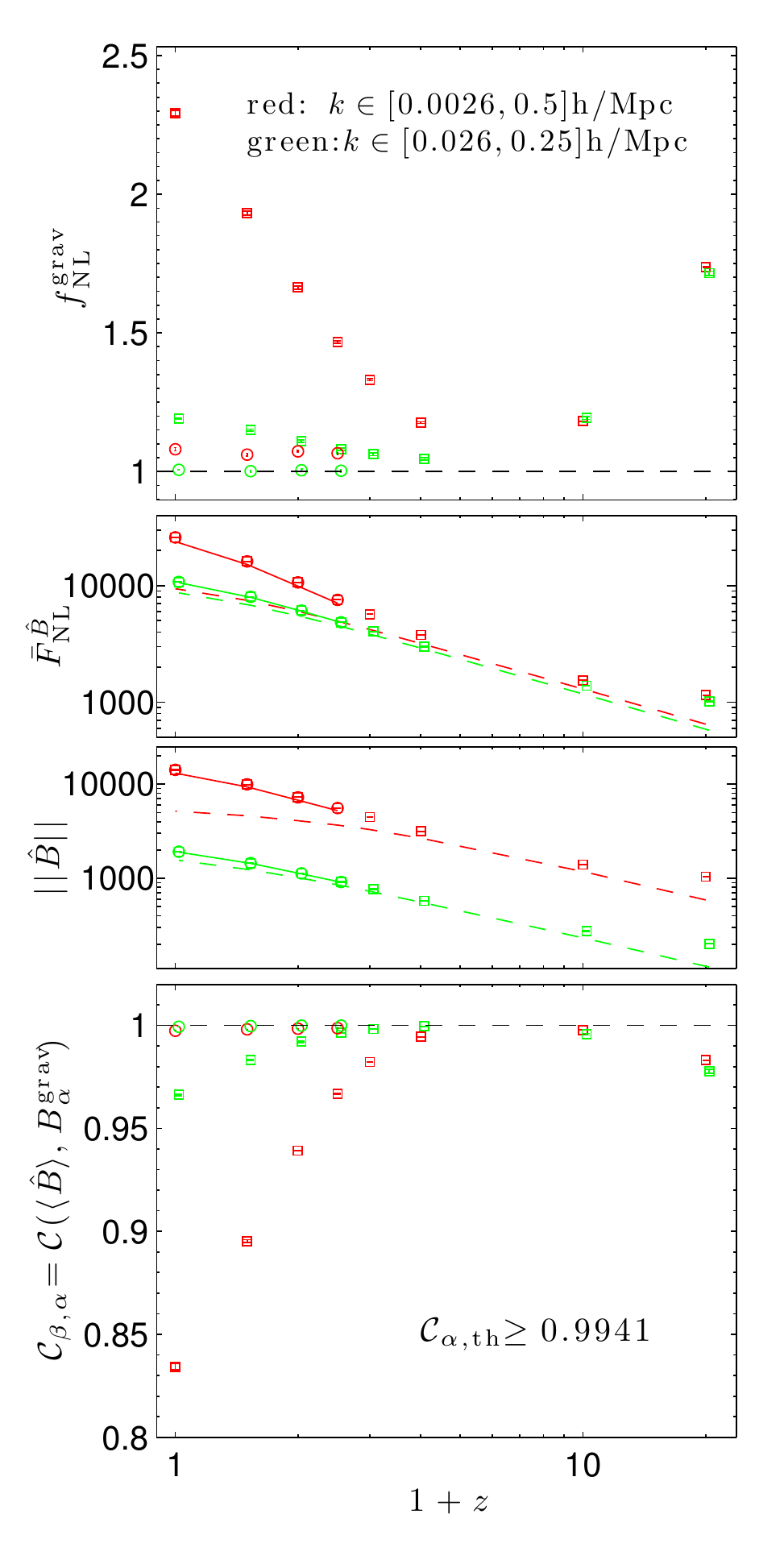}
\label{fig:bgrav_18S}}
\subfloat[][G1024]{
\includegraphics[height=0.45\textheight]{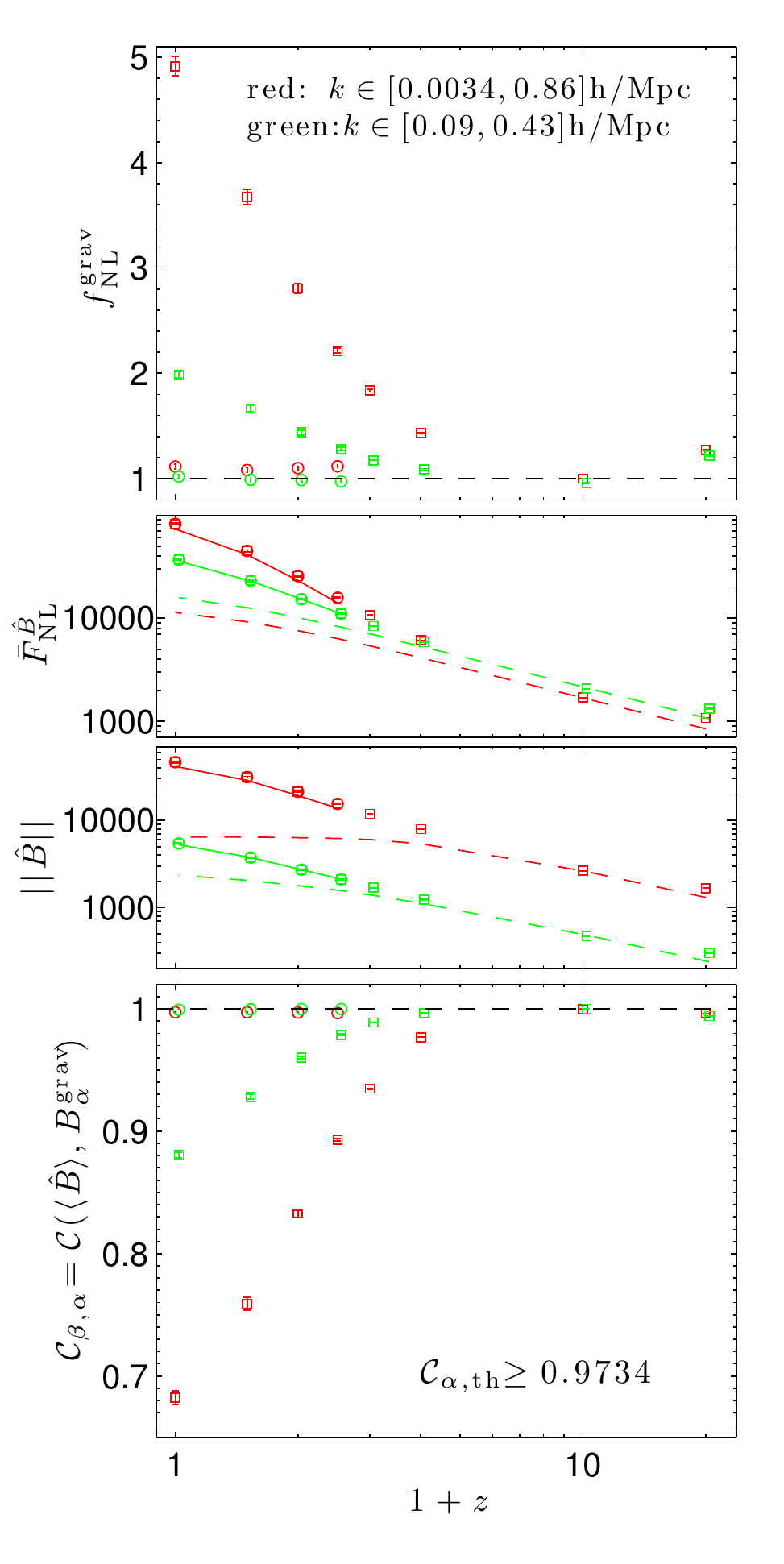}
\label{fig:bgrav_18V}}
$\;\;\;\;$
\subfloat[][Visualisation]{
\includegraphics[height=0.4\textheight]{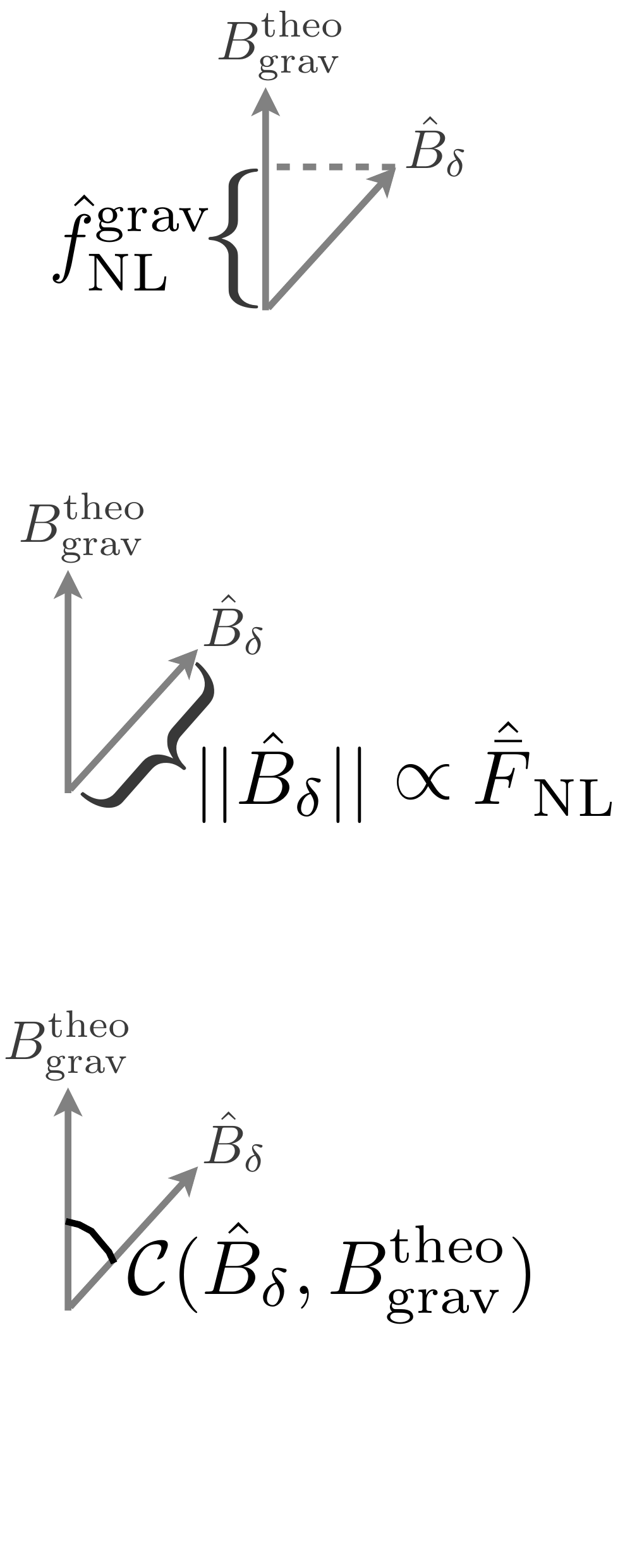}
\label{fig:visualise_scorrelz}} \mycaption{(a)-(e): Measured bispectra
in \emph{Gaussian} $N$-body simulations (see Table \ref{tab:nbody}) as a
function of redshift. \emph{Upper panels:} Amplitude
$f_\mathrm{NL}^\mathrm{grav}$ of the tree level gravitational
bispectrum \eqref{eq:Bgrav} (squares), normalised to unity if the tree
level prediction is correct, and amplitude of the fitting formula from
\cite{verde1111} (circles). \emph{Middle panels:} Total integrated
bispectrum \eqref{eq:FNL_reconstructed} of measured bispectrum
(symbols), tree level prediction (dashed lines) and fitting formula by
\cite{verde1111} (solid lines). \emph{Lower panels:} Shape correlation
\eqref{eq:shape-correl-alphabeta} of reconstructed bispectrum with
separable expansions of tree level theory (squares) and fitting
formula from \cite{verde1111} (circles). Colors indicate different $k$
ranges for the bispectrum estimation.
  The plotted quantities are
visualised in (f). }
\label{fig:bgrav_all}
\end{figure*}

To analyse the measured bispectrum more quantitatively and to
illustrate its time and scale dependence we show in \fig{bgrav_all}
the measured amplitude $f^\mathrm{grav}_\mathrm{NL}$ of the
gravitational bispectrum, its cumulative signal to noise $\|\hat B\|$
\eqref{eq:Fnl_norm} and the cumulative signal to noise normalised to
the local shape $ {\hat{\bar F}}_\mathrm{NL}$
\eqref{eq:FNL_localnorm}, as well as the shape correlation
$\mathcal{C}$ \eqref{eq:shape-correl} of the measured bispectra with
the theoretical expectation. The meaning of these quantities
is illustrated in \fig{visualise_scorrelz}. 
We plot them as functions of
redshift $z$ and use different colors for different $k$ ranges used
for the bispectrum measurements.  Note that
$f_\mathrm{NL}^\mathrm{grav}$ is the amplitude of the gravitational
bispectrum \eqref{eq:Bgrav}, i.e.~tree level perturbation theory
predicts $f_\mathrm{NL}^\mathrm{grav}=1$ in this convention.

\fig{bgrav_18BBfromglass} and \fig{bgrav_18Q} show that for
$0.0039h/\mathrm{Mpc}\le k\le 0.5h/\mathrm{Mpc}$ at $z=0$ the tree
level bispectrum underpredicts the total integrated bispectrum $\bar
F_\mathrm{NL}$ by a factor of $2.8$ and its shape correlation with the
measured bispectrum drops to $0.84$, implying that the projection of
the measured bispecrum on the tree level bispectrum is
$f_\mathrm{NL}^\mathrm{grav}=2.3$. At larger scales the tree level
bispectrum describes the measured bispectra much more accurately until late
times, as expected.
Similar plots for Gaussian $N$-body simulations with $768^3$ and $1024^3$
particles are shown in \fig{bgrav_18S} and \fig{bgrav_18V}, demonstrating
again the expected break down of tree level perturbation theory on
small scales at late times.

\subsection{Fitting formulae for Gaussian simulations}

\subsubsection{Separable polynomial expansion} 

The separable mode expansion allowed us to compress the measured
$N$-body dark matter bispectra to $n_\mathrm{max}=50$ numbers
$\beta^R_n$ at each redshift.  The first ten
 of these coefficients are
listed in Table \ref{tab:betas-nbody} and can be used as a polynomial
fitting formula of the matter bispectrum by evaluating
\eqref{eq:bisp-from-betas}.  The orthogonal contributions $B_n^R$
appearing in \eqref{eq:bisp-from-betas} were defined in \eqref{eq:B-n}
and contain the nonlinear power spectrum $P_\delta$ as well as
the orthogonal polynomials $R_n$, which can be obtained from $Q_n$,
defined in \eqref{eq:Qn_defn}, by the basis change\footnote{The transformation matrix
  $\lambda$ can be obtained from the authors upon request.}
$R_n=\sum_m\lambda_{nm}Q_m$.  The polynomials $Q_n$ were sorted with
the slice ordering defined in Eq.~(58) of \cite{shellard0912}.

\begin{table}[htb]
\begin{tabular}{|p{0.2cm}|p{1.2cm}|p{1.2cm}|p{1.2cm}|p{1.2cm}|p{1.2cm}|p{1.2cm}|}
\hline
$n$ & $\frac{(\beta^R_n)_\mathrm{G512}}{1000}$ & Loc10 & Eq100 & Orth100 & Flat10 &   
$\frac{(\beta^R_n)_{\mathrm{G}^{1024}_{1875}}}{10000}$ 
\\ \hline
0 & 283.8 & 846.8 & 3065 & -793.5 & 3544 & 167 \\ 
1 & 94.16 & 364.3 & 1390 & -240.4 & 1484 & 63.06 \\ 
2 & 0.6725 & 94.22 & 453 & 28.49 & 362.2 & 9.427 \\ 
3 & 17.04 & 18.58 & -183.5 & -238.7 & 128.1 & 0.2799 \\ 
4 & 18.89 & 21.7 & -47.71 & -173.6 & 163.5 & 5.774 \\ 
5 & -21.82 & 8.64 & 35.7 & 85.29 & -43.72 &  -7.911 \\ 
6 & 9.395 & -63.53 & -6.603 & 154.1 & -232.5 & 1.329 \\ 
7 & 1.624 & -17.96 & -2.818 & 53.24 & -77.86 & -0.4791 \\ 
8 & 15.04 & 21.4 & -57.18 & -186 & 171.9 & 5.825 \\ 
9 & -10.25 & -21.61 & 39.26 & 108.1 & -109.2 & -3.548 \\ 
\hline
\end{tabular}
\caption{First $10$ bispectrum expansion coefficients $\beta^R_n$
  from $N$-body simulations specified in Table \ref{tab:nbody} for $z=0$. For
  simulations with non-Gaussian initial conditions we show the
  difference to the Gaussian simulation,
  $(\beta^R_n)_\mathrm{X}=\beta^R_n-(\beta^R_n)_\mathrm{G512}$. The
  shape correlation between the full measured bispectrum
  \eqref{eq:bisp-from-betas} for $50$ modes and the bispectrum
  corresponding to the shown first $10$ modes is $99.7\%, 99.2\%,
  99.95\%, 95.8\%$, $99.7\%$ and $99.9\%$ for the columns from the left to
  the right.  We used $k_\mathrm{max}=0.86h/\mathrm{Mpc}$ for the
  $\mathrm{G}^{1024}_{1875}$ simulation in the
  column on the right and $k_\mathrm{max}=0.5h/\mathrm{Mpc}$ in all
  other columns.}\label{tab:betas-nbody}
\end{table}

\subsubsection{Time-shift model fit}

We find a remarkably accurate fit to the matter
bispectrum for Gaussian initial conditions by combining the modified tree level
gravity shape $B_{\delta,\mathrm{NL}}^\mathrm{grav}$, defined in
\eqref{eq:Bgrav_Pnonlinear}, with the `constant' model
$B_{\delta,\mathrm{const}}^\mathrm{grav}$, defined in
\eqref{eq:Bconst_grav}, which approximates the 1-halo shape, that is, 
\begin{align}
  \label{eq:msfit2_18DD}
  B_\delta^\mathrm{fit}(k_1,k_2,k_3)\equiv
  B_{\delta,\mathrm{NL}}^\mathrm{grav} 
+
B_{\delta,\mathrm{const}}^\mathrm{grav},
\end{align}
 with fitting parameters $c_1$ and $n_h$.

The combined shape \eqref{eq:msfit2_18DD} is dominated by the
perturbative gravity bispectrum at early times and by the constant or
(approximate) halo model prediction at late
times. We will demonstrate that this combination can achieve a good
fit of the matter bispectrum at all redshifts $z\leq 20$, while both
the perturbative and the halo model prediction individually break down
at intermediate redshifts, when nonlinearities are important but not
all dark matter particles can be treated as residing in halos.

\begin{table}[htb]
\begin{tabular}{|p{1.55cm}|p{0.95cm}|p{2.15cm}|p{0.85cm}|p{1.35cm}|p{1.2cm}|}
\hline
Simulation & $L[\frac{\rm{Mpc}}{\rm{h}}]$ & $c_{1,2}$ & $n_h^\mathrm{(prim)}$ & $\min\limits_{z\leq
  20}\mathcal{C}_{\beta,\alpha}$ &
${\mathcal{C}_{\beta,\alpha}}_{(z=0)}$
\\ \hline
G512g  & $1600$ & $4.1\times 10^6$ & $7$ & $99.8\%$ & $99.8\%$ \\
Loc10  & $1600$ & $2\times 10^3$ & $6$ & $99.7\%$ & $99.8\%$\\
Eq100 & $1600$ &  $8.6\times 10^2$ & $6$ & $97.9\%$ & $99.4\%$ \\
Flat10 & $1600$ & $1.2\times 10^4$ & $6$ & $98.8\%$ & $98.9\%$ \\
Orth100 & $1600$ & $-3.1\times 10^2$ & $5.5$ & $91.0\%$ & $91.0\%$ \\
\hline
$\mathrm{G}^{512}_{400}$ & $400$  & $ 1.0\times 10^7$ & $8$ & $99.8\%$ & $99.8\%$  \\
$\mathrm{Loc10}^{512}_{400}$ & $400$ & $2\times 10^3dD/da$ & $7$ & $98.2\%$ & $99.0\%$ \\
$\mathrm{Eq100}^{512}_{400}$ & $400$ & $8.6\times 10^2dD/da$ & $7$ & $94.4\%$ & $97.9\%$ \\
$\mathrm{Flat10}^{512}_{400}$ & $400$ & $1.2\times 10^4dD/da$ & $7$ & $97.7\%$ & $99.1\%$ \\
$\mathrm{Orth100}^{512}_{400}$ & $400$ & $-2.6\times 10^2$ & $6.5$ & $97.3\%$ & $98.9\%$ \\ 
\hline
\end{tabular}
\caption{
  Fitting parameters $c_1$ and $n_h$ for the fit
  \eqref{eq:msfit2_18DD} of the matter bispectrum for Gaussian 
  initial conditions (simulations G512g and $\mathrm{G}_{400}^{512}$) and $c_2$ and $n_h^\mathrm{prim}$ for the fit \eqref{eq:msfit3_NG} of the
  primordial bispectrum \eqref{eq:B_ng_meas}. The two columns on the right show the minimum shape correlation
  with the measured (excess) bispectrum in $N$-body
  simulations, which was measured at redshifts
  $z=49,30,20,10,9,8,\dots,0$, and the shape correlation at $z=0$.
  For the equilateral case the minimum shape correlation can be
  improved to $99.4\%$ if the term $4.6\times 10^{-5}
  f_\mathrm{NL}\bar D(z)^{0.5}\big[ 
  2P_{\delta}(k_1)P_{\delta}(k_2)F_2^{(s)}(\vk_1,\vk_2)
  + 2\mbox{ perms}\big]$ is added to \eqref{eq:msfit3_NG}.  
}\label{tab:msfit_NG_table}
\end{table}

\begin{figure*}[htb]
\centering
\subfloat[][Gaussian, $k_\mathrm{max}=0.5h/\mathrm{Mpc}$]{
\includegraphics[width=0.38\textwidth]{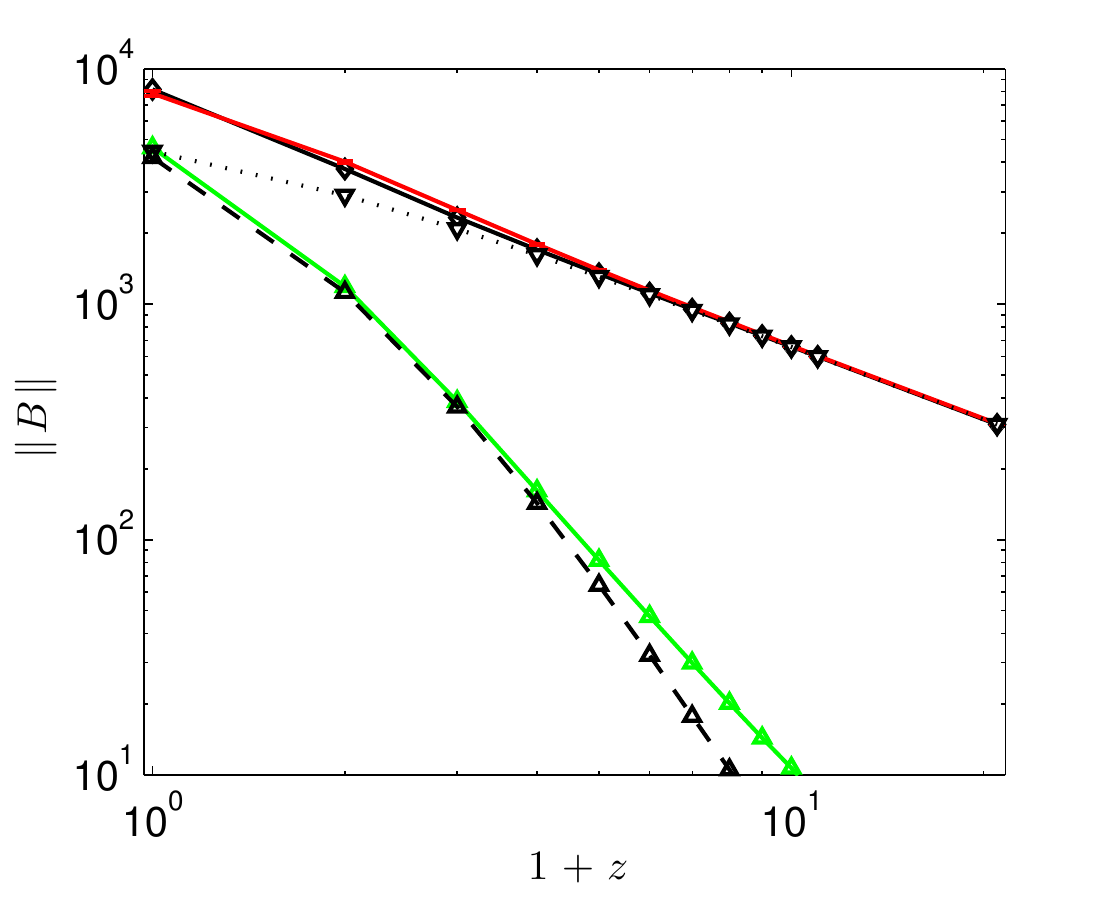}
\label{fig:msfit_18BBfromglassapril_tmp2}}
\subfloat[][Gaussian, $k_\mathrm{max}=2h/\mathrm{Mpc}$]{
\includegraphics[width=0.38\textwidth]{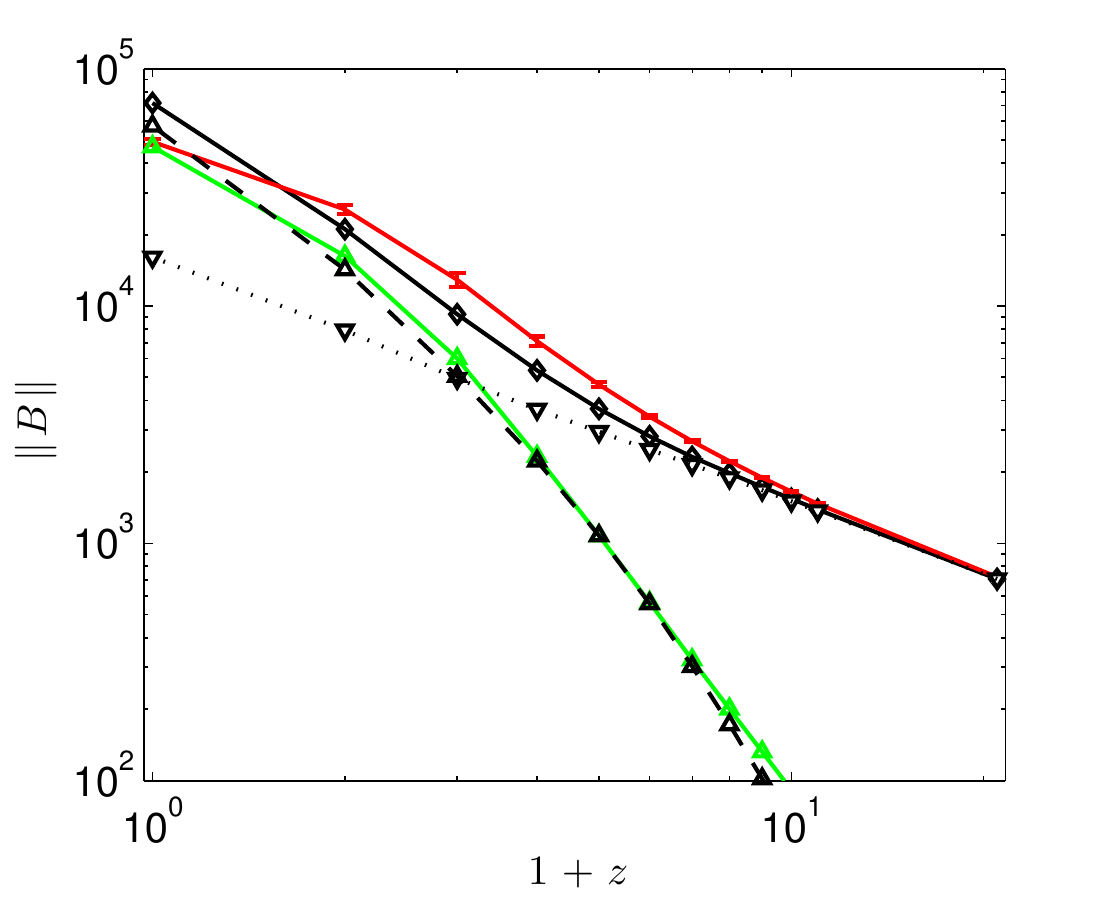}
\label{fig:msfit_18DD_tmp2}} \mycaption{Motivation for using the
growth function $\bar D$ in the simple fitting formula
\eqref{eq:msfit2_18DD}.  The arbitrary weight $w(z)$ in
$B_\delta^\mathrm{opt}=B_{\delta,\mathrm{NL}}^\mathrm{grav}+
w(z)(k_1+k_2+k_3)^\nu$ is determined analytically such that
$\mathcal{C}(\hat B_\delta,B_\delta^\mathrm{opt})$ is maximal (for
$\nu=-1.7$).  We plot $\|B_{\delta,\mathrm{NL}}^\mathrm{grav}\|$
(black dotted), $\|w(z)(k_1+k_2+k_3)^\nu\|$ (green) and
$B_{\delta,\mathrm{const}}^\mathrm{grav}$ (black dashed) as defined in
\eqref{eq:Bconst_grav} with fitting parameters given in Table
\ref{tab:msfit_NG_table}, illustrating that $w(z)=c_1\bar D^{n_h}(z)$
is a good approximation.  The continuous black and red curves show
$\|B_\delta^\mathrm{fit}\|$ from \eqref{eq:msfit2_18DD} and the
estimated bispectrum size $\|\hat
B_\delta\|$, respectively. The overall normalisation can be adjusted
with $N_\mathrm{fit}$ as explained in the main text.  }
\label{fig:msfit_grav_all_tmp2}
\end{figure*}

The fitting parameters $c_1$ and $n_h$ in \eqref{eq:Bconst_grav} are obtained as follows. At
each measured redshift the arbitrary weight $w(z)$ in
\begin{align}\label{eq:grav_fitting}
B_\delta^\mathrm{opt}=B_{\delta,\mathrm{NL}}^\mathrm{grav}+
w(z)(k_1+k_2+k_3)^\nu
\end{align}
 is analytically determined such that
the correlation with the measured bispectrum $\mathcal{C}(\hat B_\delta,B_\delta^\mathrm{opt})$ is maximal (see
green lines in \fig{msfit_grav_all_tmp2}).  Then
$c_1$ and $n_h$ are chosen such that $c_1\bar D^{n_h}(z)$
approximates the optimal weight $w(z)$ over all redshifts (see black
dashed lines in \fig{msfit_grav_all_tmp2}). Table \ref{tab:msfit_NG_table} lists the fitting parameters
obtained by this procedure.

\begin{figure*}[htb]
\centering
\hspace{-0.9cm}
\subfloat[][$k_\mathrm{max}=0.5h/\mathrm{Mpc}$]{
\includegraphics[width=0.42\textwidth]{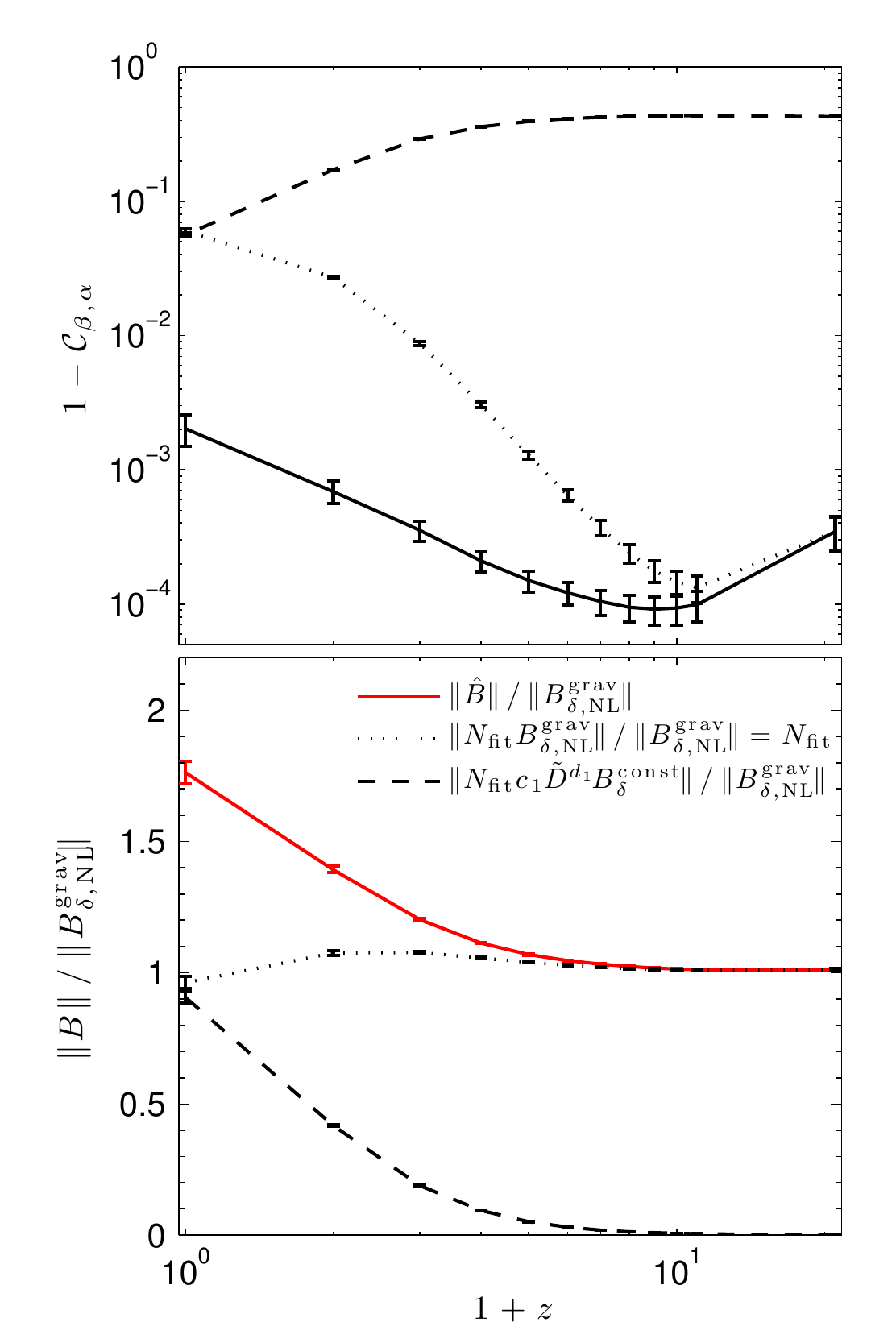}
\label{fig:msfit_18BBfromglassapril_46_51}}
\quad
\subfloat[][$k_\mathrm{max}=2h/\mathrm{Mpc}$]{
\includegraphics[width=0.42\textwidth]{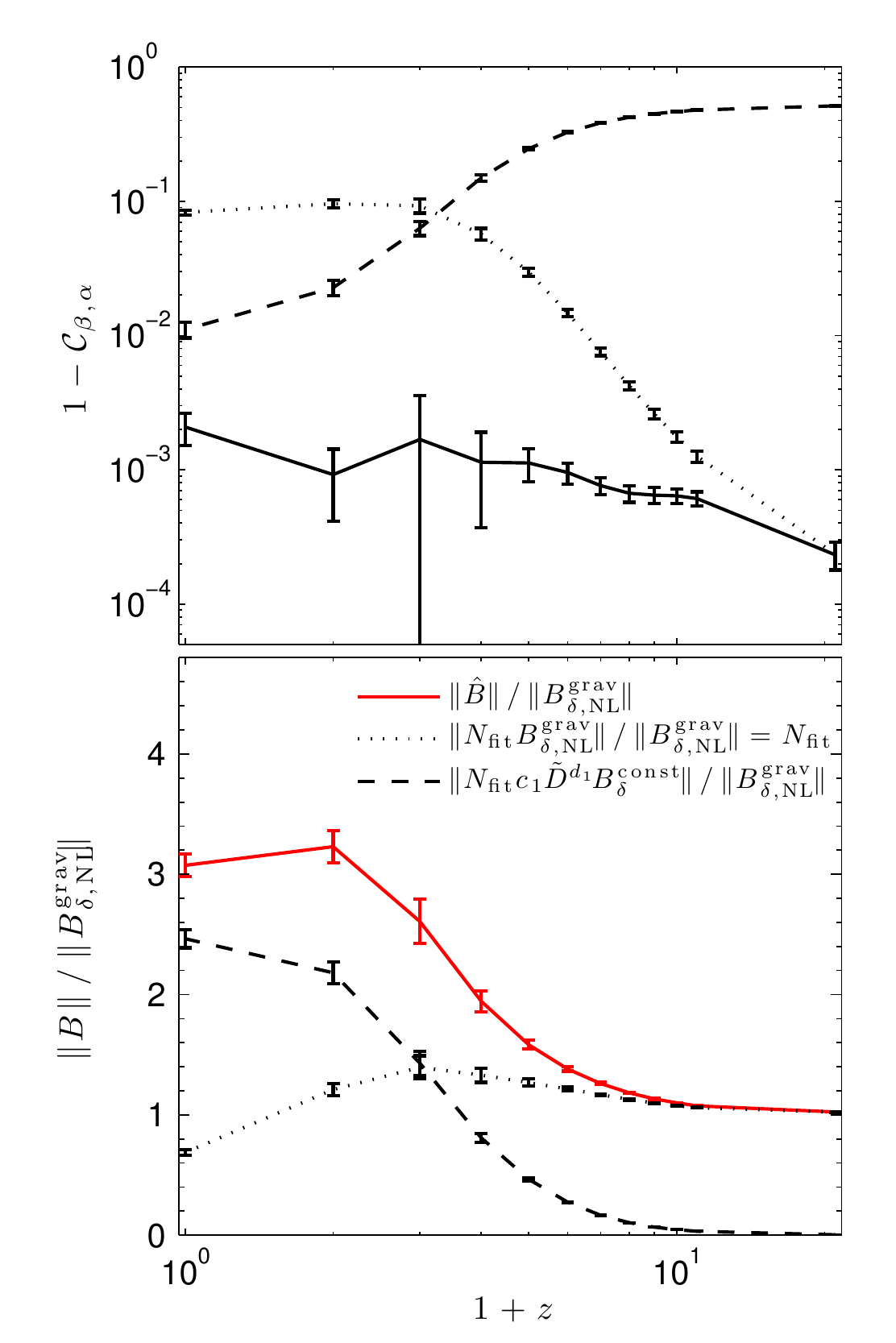}
\label{fig:msfit_18DD_46_51}}
\caption{
 \emph{Upper panels:} Shape correlation of estimated bispectrum in (a) G512g
 and (b) $\mathrm{G}_{400}^{512}$
  simulations with the full fitting formula \eqref{eq:msfit2_18DD} (solid
  black), the shape $B_{\delta,\mathrm{NL}}^\mathrm{grav}$ as defined in
  \eqref{eq:Bgrav_Pnonlinear} (dotted black) and the shape
  $B_{\delta,\mathrm{const}}^\mathrm{grav}$ as defined in \eqref{eq:Bconst_grav}
  (dashed black).
  All correlations were calculated with $120$ modes $Q_n$ for (a)
  $k\in [0.004,0.5]h/\mathrm{Mpc}$ and (b) $k\in
  [0.016,2]h/\mathrm{Mpc}$.  \emph{Lower panels:} Integrated size of
  the measured bispectrum (red) and the contributions
  $N_\mathrm{fit}B_{\delta,\mathrm{NL}}^\mathrm{grav}$ (dotted black)
  and 
  $N_\mathrm{fit} B_{\delta,\mathrm{const}}^\mathrm{grav}$
  (dashed black).
 All these bispectrum sizes are divided by the
  size of $B_{\delta,\mathrm{NL}}^\mathrm{grav}$ for
  convenience. The normalisation factor $N_\mathrm{fit}$ is defined in
  \eqref{eq:normfact_msfit2} and equals the dotted black curve. 
}
\label{fig:msfit_grav_all}
\end{figure*}

The shape correlation of the fitting formula \eqref{eq:msfit2_18DD}
with the measured bispectrum is $99.8\%$ or better at redshifts $z\leq
20$ for both $k_\mathrm{max}=0.5h/\mathrm{Mpc}$ and
$k_\mathrm{max}=2h/\mathrm{Mpc}$ as shown in the upper panels of
\fig{msfit_grav_all}.  The model \eqref{eq:msfit2_18DD}, therefore, contains 
all the meaningful bispectrum shape information.  All we require is the time dependence or
growth rate of the bispectrum amplitude.  As a first step, the fitting formula \eqref{eq:grav_fitting} 
can
be normalised to the measured bispectrum size by multiplying it with
the normalisation factor
\begin{align}
  \label{eq:normfact_msfit2}
  N_\mathrm{fit}\equiv \frac{\|\hat B\|}{\|B_\delta^\mathrm{fit}\|},
\end{align}
which is shown by the dotted line in the lower panels of
\fig{msfit_grav_all}. While it varies with redshift between $0.7$
and $1.4$ for $k_\mathrm{max}=2h/\mathrm{Mpc}$, it deviates by at most
$8\%$ from unity for $k_\mathrm{max}=0.5h/\mathrm{Mpc}$.  The lower
panels also show the measured integrated bispectrum size $\|\hat B\|$
and the two individual contributions to \eqref{eq:msfit2_18DD} when
the normalisation factor $N_\mathrm{fit}$ is included. These
quantities are divided by $\|B_{\delta,\mathrm{NL}}^\mathrm{grav}\|$
for convenience. At high redshifts the total bispectrum size is
essentially given by the contribution from
$B_{\delta,\mathrm{NL}}^\mathrm{grav}$, which equals the tree level
prediction for the gravitational bispectrum in this regime. The
contribution from $B_{\delta}^{\mathrm{const}}$ dominates at $z\leq 2$
for $k_\mathrm{max}=2h/\mathrm{Mpc}$ when filamentary and spherical
nonlinear structures 
are apparent. A similar transition can be seen at later times on
larger scales in \fig{msfit_18BBfromglassapril_46_51}, indicating
self-similar behaviour. 

It is worth noting that the high integrated correlation between the
simple fit \eqref{eq:msfit2_18DD} and measurements does not imply that
all triangle configurations agree perfectly and sub-percent level
differences between shape correlations can in principle contain
important information, e.g.~about the observationally relevant
squeezed limit which only makes a small contribution to the total tetrapyd
integral over the signal-to-noise weighted dark matter bispectrum.  
However, if we observed the dark matter bispectrum
directly, these shapes would be hard to distinguish because the shape
correlation contains the signal-to-noise weighting.  Modified shape
correlation weights and additional basis functions have been used for
better quantitative comparison of the squeezed limit of dark matter
bispectra, but this is left for a future publication.

\subsubsection{Alternative phenomenological fit}
An alternative fitting formula with $9$ fitting parameters and
calibrated on larger scales was given in \cite{verde1111} and
summarised in this paper in Section III.  In the range of validity
given by \cite{verde1111}, $0.03h/\mathrm{Mpc}\le k\le
0.4h/\mathrm{Mpc}$ at $0\le z\le 1.5$, we find good agreement with our
$N$-body measurements, see green circles in \fig{bgrav_18S} and
\fig{bgrav_18V}. Without having to run $N$-body simulations with
higher resolutions, we extended the bispectrum measurement to
$k_\mathrm{max}=0.86h/\mathrm{Mpc}$ with the fast separable estimator,
see red symbols in \fig{bgrav_18S} and \fig{bgrav_18V}.
In this extended regime the fitting formula of \cite{verde1111} still
has a shape correlation of $99.5\%$ or more with the measured
bispectrum at $z\le 1.5$, but underestimates the cumulative signal to
noise by up to $11\%$.
Our measured $\beta^R_n$ coefficients or our simple fit
\eqref{eq:msfit2_18DD} can be used as alternative fitting formulae for
the gravitational bispectrum valid to smaller scales,
$k_\mathrm{max}\leq 2h/\mathrm{Mpc}$, and for all redshifts $z\leq
20$.

\subsubsection{Halo model}

 The halo model prediction for the Gaussian
dark matter bispectrum yields a remarkably high shape correlation of
more than $99.7\%$ with the measured bispectrum at $z=0$ for
$k_\mathrm{max}=2h/\mathrm{Mpc}$. While the halo model bispectrum has
been tested on some one-dimensional slices in \cite{valageas1102} and
on larger scales in \cite{figueroa1205}, the result presented here
demonstrates that at $z=0$ the halo model shape is a good
representation of the shape measured in $N$-body simulations over the
full tetrapyd allowed by the triangle condition. At higher redshifts,
when less dark matter resides in halos, the halo model prediction
becomes worse and alternative phenomenological approaches like the ones
discussed above yield higher shape correlations with the measured bispectrum.
A more thorough examination of the halo model will be presented in \cite{Reganetal2012}.

\section{VII. Primordial Non-Gaussian Bispectrum Results}

\begin{figure*}[p]
\centering
\vspace{-0.4cm}
\subfloat[][Local]{
\includegraphics[height=0.5\textheight]{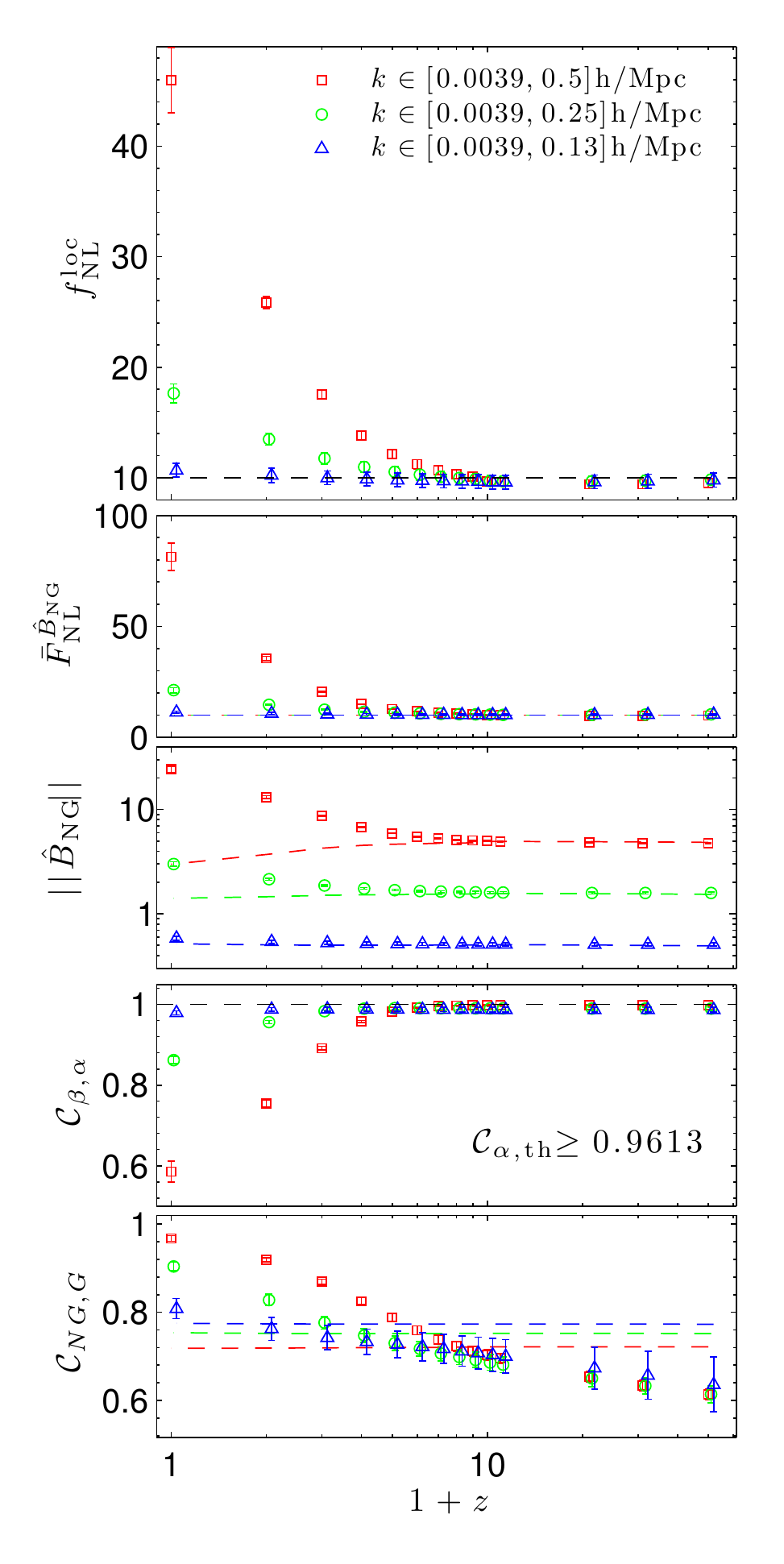}
\label{fig:scorrelz_18Qfnl10}}
\subfloat[][Equilateral]{
\includegraphics[height=0.5\textheight]{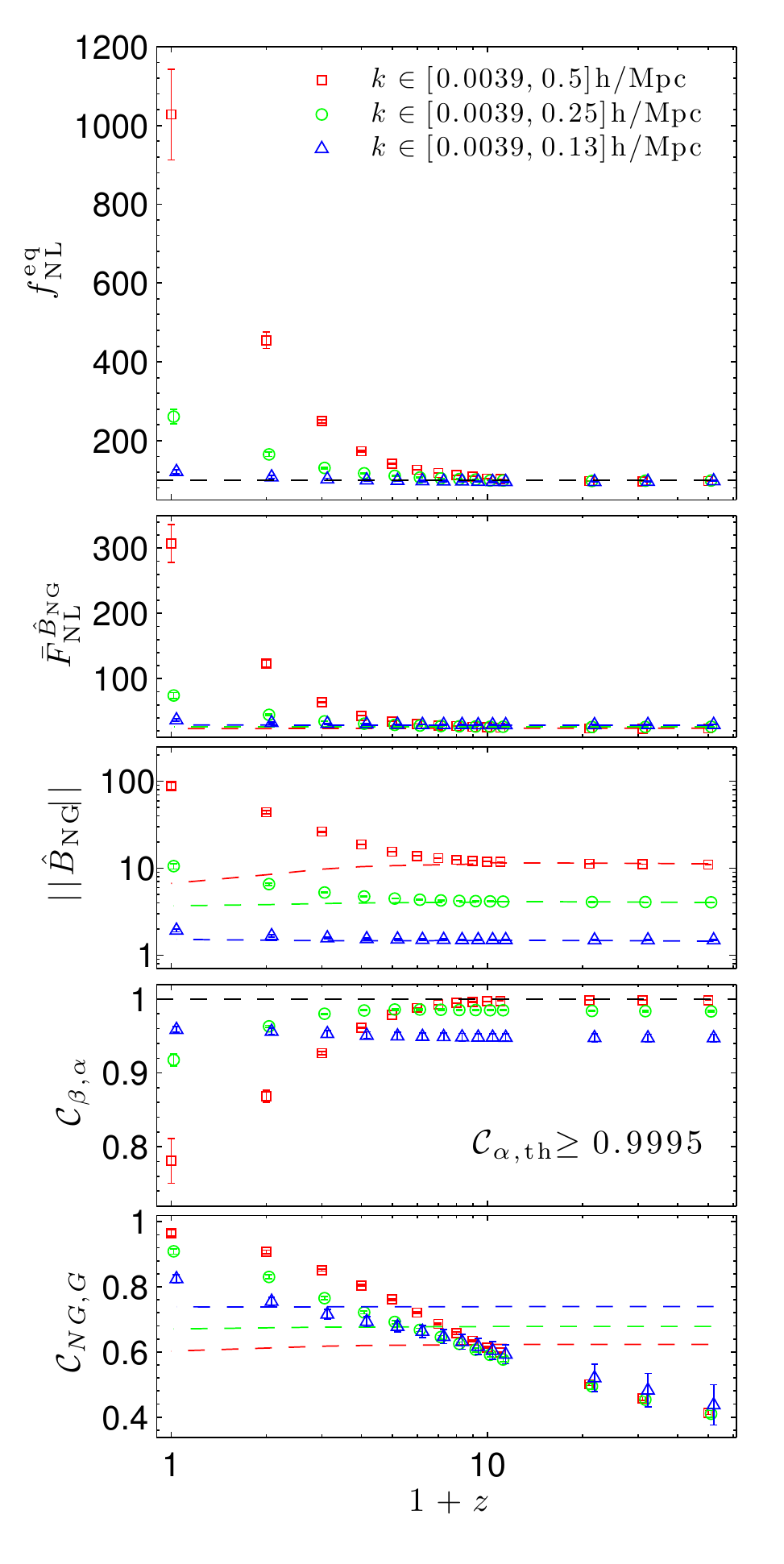}
\label{fig:scorrelz_18Qsp6fnl100}}
\subfloat[][Orthogonal]{
\includegraphics[height=0.5\textheight]{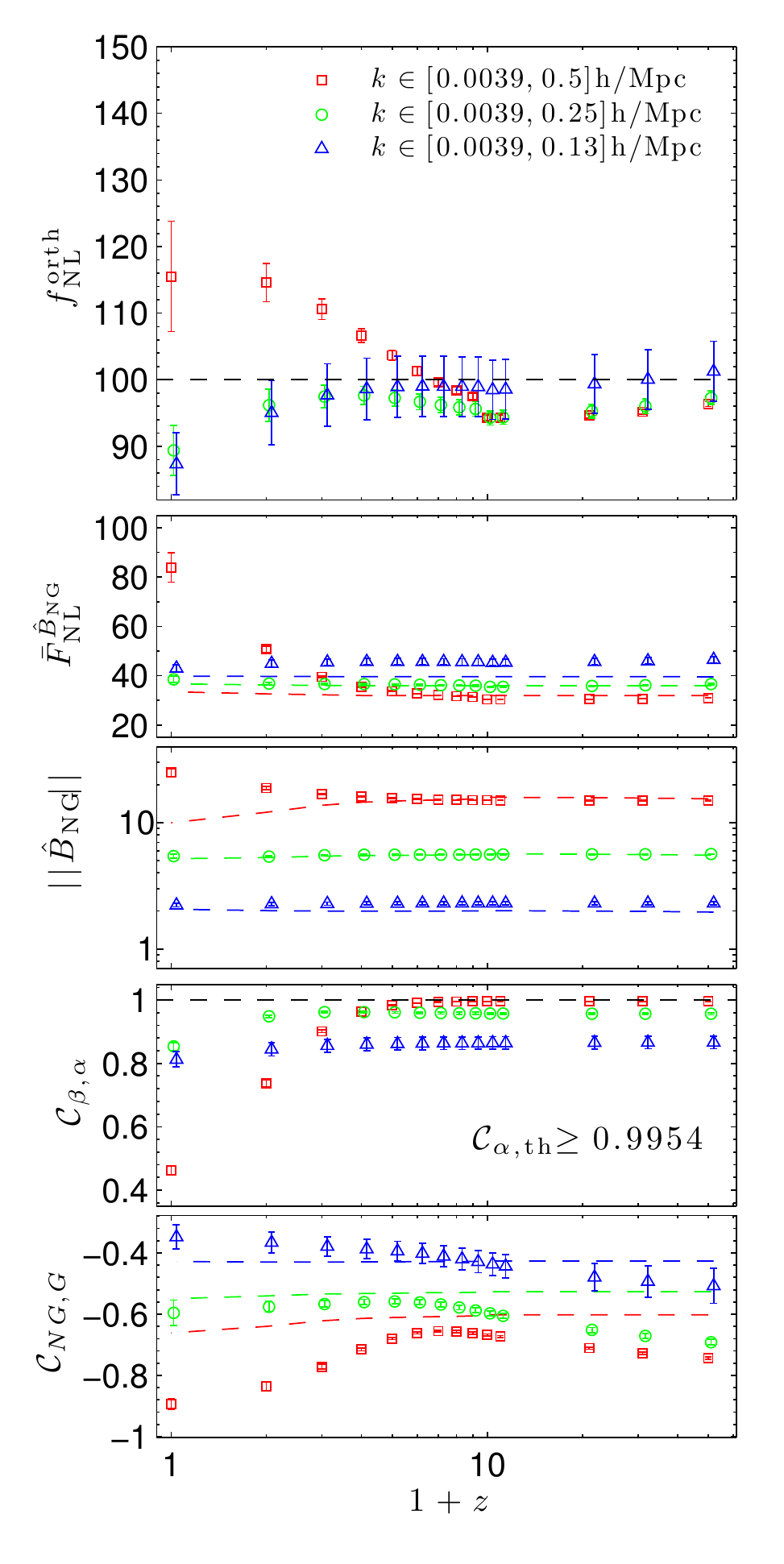}
\label{fig:scorrelz_18Qsp22fnl100}}
\mycaption{Dark matter bispectrum quantities as in  \fig{bgrav_all} but for the non-Gaussian simulations Loc10, Eq100 and
  Orth100 with local, equilateral and orthogonal initial
  conditions, respectively.  The non-Gaussian bispectrum is computed
  with \eqref{eq:B_ng_meas} and then compared to the tree level
  prediction \eqref{eq:bprim-tree} (dashed lines). The lowest panel
  shows the shape correlation of the measured non-Gaussian bispectrum
  with the measured bispectrum for Gaussian initial conditions. }
\label{fig:scorrelz_NG_all}
\end{figure*}

\subsection{Primordial bispectrum measurements}
We have set up and evolved non-Gaussian initial conditions for local models with 
$f_\mathrm{NL}^\mathrm{loc}=-10,10,20,50$ using \eqref{eq:local-phi-square},
for equilateral models with $f_\mathrm{NL}^\mathrm{eq}=\pm 100$, for orthogonal models with $f_\mathrm{NL}^\mathrm{orth}=\pm 100$ and for the flattened model with 
$f_\mathrm{NL}^\mathrm{flat}=10$ using separable expansions as in
\cite{shellard1008,shellard1108}.  Detailed parameters for the $N$-body
simulations are given in Table \ref{tab:nbody}.  

\subsubsection{Local shape}
In \fig{18Q_alphabeta}b we show comparisons of the measured
$\beta^R_n-(\beta^R_n)_\mathrm{Gauss}$ coefficients with the expansion
coefficients ${\alpha}^R_n$ of the linearly evolved primordial
bispectrum \eqref{eq:bprim-tree} for $f_\mathrm{NL}^\mathrm{loc}=10$,
finding good agreement at early times, but deviations at late times which are scale-dependent.
The most obvious feature 
as we consider increasingly nonlinear wavenumbers is the emergence of a large constant or equilateral 
signal.   
Indeed \fig{tet3dplots_grav_loc_eq_orth} shows this signal on small scales  with 
the late time bispectrum also exhibiting an enhanced signal in the squeezed limit.  

In \fig{scorrelz_18Qfnl10} we plot the projection
$f_\mathrm{NL}^\mathrm{loc}$ of the measured non-Gaussian bispectrum
$\hat B_\mathrm{NG}$ on the linearly evolved primordial bispectrum
(see \eqref{eq:bprim-tree} and \eqref{eq:Blocal}), their shape
correlation $\mathcal{C}_{\beta,\alpha}$, cumulative signal to
noise $\|\hat B_\mathrm{NG}\|$ and total integrated bispectrum
normalised to the local shape, $\bar F_\mathrm{NL}^{\hat B_\mathrm{NG}}$.  We also
show the shape correlation $\mathcal{C}_\mathrm{NG,G}$ between $\hat
B_\mathrm{NG}$ and the bispectrum measured in Gaussian simulations.
For $z=0$ and $k\in \[ 0.0039, 0.5\]h/\mathrm{Mpc}$ the linearly
evolved primordial bispectrum underpredicts the measured cumulative
signal to noise $\|\hat B_\mathrm{NG}\|$ by a factor of  $8$ and has a
shape correlation with the measured bispectrum of about $0.6$,
implying that the projection $f_\mathrm{NL}^\mathrm{loc}$ is about
$4.8$ times the input value of $f_\mathrm{NL}^\mathrm{loc}=10$.
The green and blue symbols in \fig{scorrelz_18Qfnl10}  show
that tree level perturbation theory improves if we consider
larger scales, as expected.

\subsubsection{Other shapes}
\begin{figure}[ht]
\centering
\includegraphics[height=0.55\textheight]{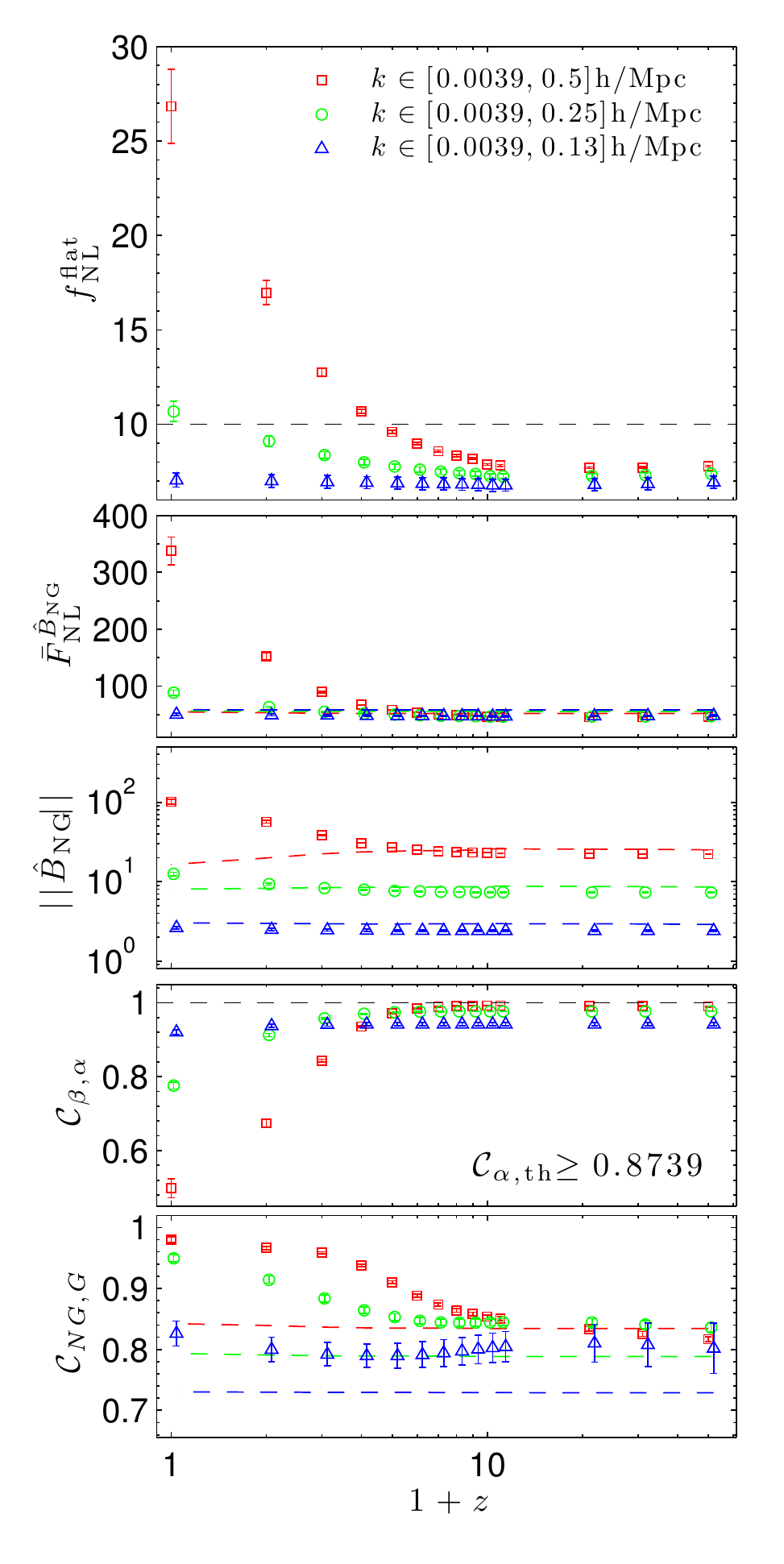}
\caption{Like \fig{scorrelz_NG_all} but for simulation Flat10 with
  flattened non-Gaussian initial conditions.}
\label{fig:scorrelz_18Qsp25fnl10}
\end{figure}

We also run simulations with equilateral, orthogonal and flattened
non-Gaussian initial conditions using the separable expansion method
\cite{shellard1008,shellard1108}.  Bispectrum measurements are shown
in Figures \ref{fig:tet3dplots_show_expansions_grav_loc_eq_orth}, \ref{fig:tet3dplots_grav_loc_eq_orth}, \ref{fig:scorrelz_NG_all} and
\ref{fig:scorrelz_18Qsp25fnl10}.  As in the local case we find agreement
with tree level perturbation theory at all times on large
scales. However on small scales and at late times,
\fig{tet3dplots_grav_loc_eq_orth} shows that the measured
bispectrum for equilateral initial conditions has additional
equilateral and flattened contributions compared to the linearly
evolved input bispectrum \eqref{eq:bprim-tree}.  For initial
conditions with the $f_\mathrm{NL}^\mathrm{orth}=100$ orthogonal shape, which is positive for
equilateral configurations and negative for flattened and squeezed
configurations, we find that the late time small scale bispectrum
turns slightly negative for equilateral configurations and is more
negative in flattened and squeezed configurations than the linearly
evolved input bispectrum.  The bispectrum from flattened initial
conditions shows additional contributions for squeezed, equilateral
and flattened configurations compared to the tree level prediction.
For orthogonal and flattened initial conditions the late time bispectra
have a large signal in the squeezed limit, leading to possible
confusion with local shape initial conditions (with negative
$f_\mathrm{NL}^\mathrm{loc}$ for positive input $f_\mathrm{NL}^{\rm{orth}}$
and positive $f_\mathrm{NL}^\mathrm{loc}$ for positive input
$f_\mathrm{NL}^{\rm{flat}}$).  However scale dependent halo bias is
sensitive to the scaling in the squeezed limit and can therefore help
to disentangle such shapes (see e.g.~\cite{dalal08,wagner-verde1102}).

It is worth noting that the shape correlation
$\mathcal{C}_{\beta,\alpha}$ between measured and input bispectra
drops as we reduce $k_\mathrm{max}$ already at the initial time of the
simulations, e.g.~$\mathcal{C}_{\beta,\alpha}\approx 0.95$ for
$k_\mathrm{max}=0.13h/\mathrm{Mpc}$ in the equilateral case. This
indicates that the initial conditions generated with a separable mode
expansion - which are generated at $k_\mathrm{max}=0.5h/\mathrm{Mpc}$ - are not a perfect fit to the shape when restricted to large scales. This is to be expected since
the expansion is optimised to fit the total cumulative signal to
noise up to $k_\mathrm{max}=0.5h/\mathrm{Mpc}$. There are however
several possibilities of further improving the
initial conditions on large scales. One could add basis functions which
are localised on large scales
or one could introduce optimised separable weighting functions in the
mode expansions.  This would be particularly important for examining
halo bias which is very sensitive to the squeezed limit and therefore
to large scale modes.  However for studying the global behaviour of
the dark matter bispectrum we find it acceptable that the shape of the
initial conditions is not perfect on very large scales and leave
further improvements of the initial conditions for future work.

\subsubsection{Loop corrections for primordial non-Gaussianity}
As is apparent from Figures \ref{fig:tet3dplots_grav_loc_eq_orth}, \ref{fig:scorrelz_NG_all} and \ref{fig:scorrelz_18Qsp25fnl10}, the tree level prediction for the
non-Gaussian matter bispectrum breaks down for small scales and low
redshift. The inclusion of loop corrections
described
in \cite{sefusatti09} 
 is expected to improve the fit at such scales
and redshifts.
 First results on the correlation between the observations and
theoretical prediction $\mathcal{C}_{\beta,\alpha}$ indicate that
$1$-loop corrections can vastly improve the shape correlation in the
nonlinear regime to $\mathcal{O}(0.8-0.9)$, while shape correlations of more than $0.9$
at $k=0.5h/\mathrm{Mpc}$ and $z=0$ may require higher order loop
corrections. However we defer a detailed analysis of loop corrections
and more phenomenological halo model approaches to a forthcoming
paper.

\subsubsection{Linearity in input $f_\mathrm{NL}$}
\begin{figure*}[p]
\centering
\subfloat[][Local]{
\includegraphics[height=0.5\textheight]{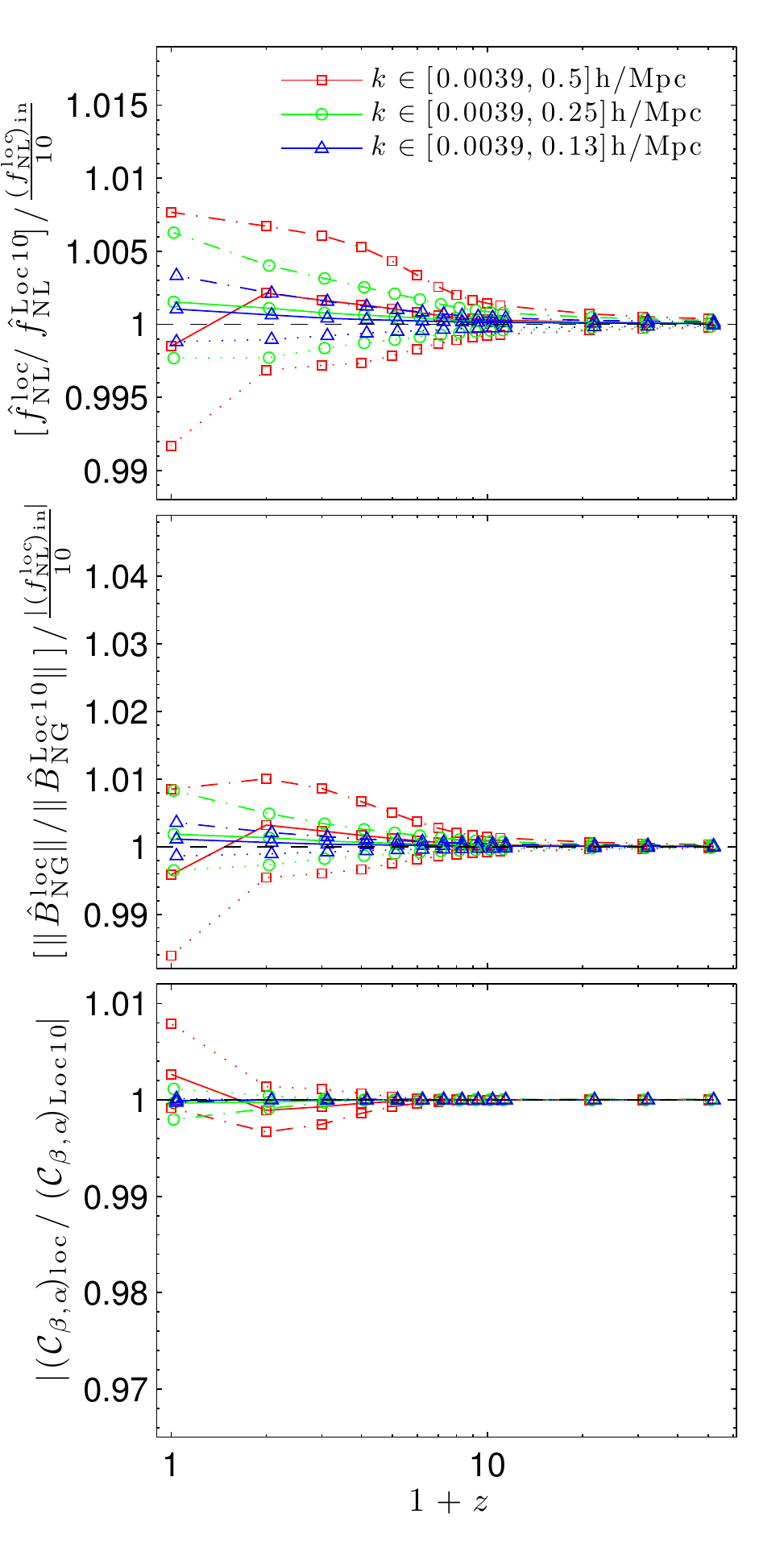}
\label{fig:linearity_loc}}
\subfloat[][Equilateral]{
\includegraphics[height=0.5\textheight]{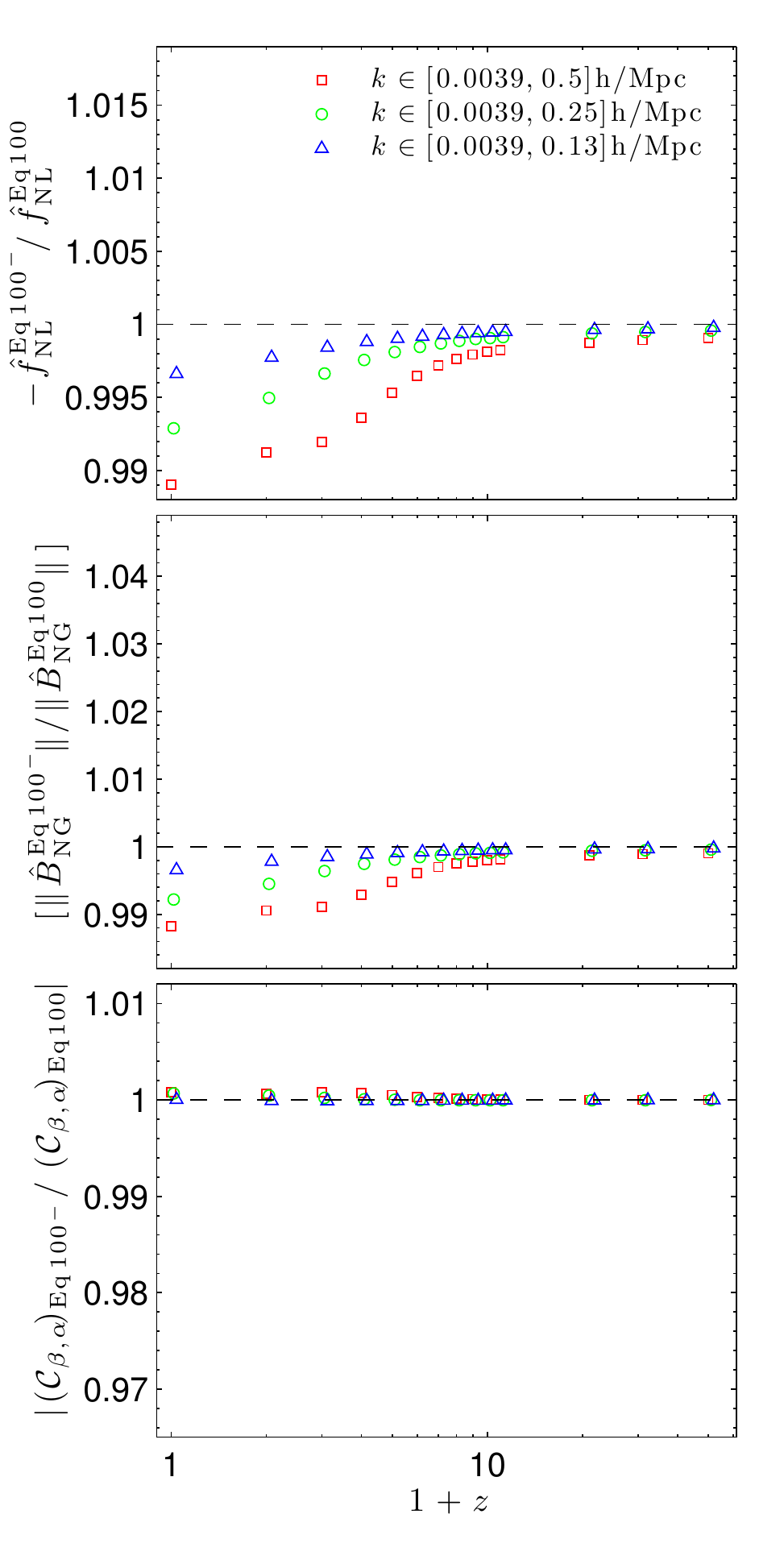}
\label{fig:linearity_eq}}
\subfloat[][Orthogonal]{
\includegraphics[height=0.5\textheight]{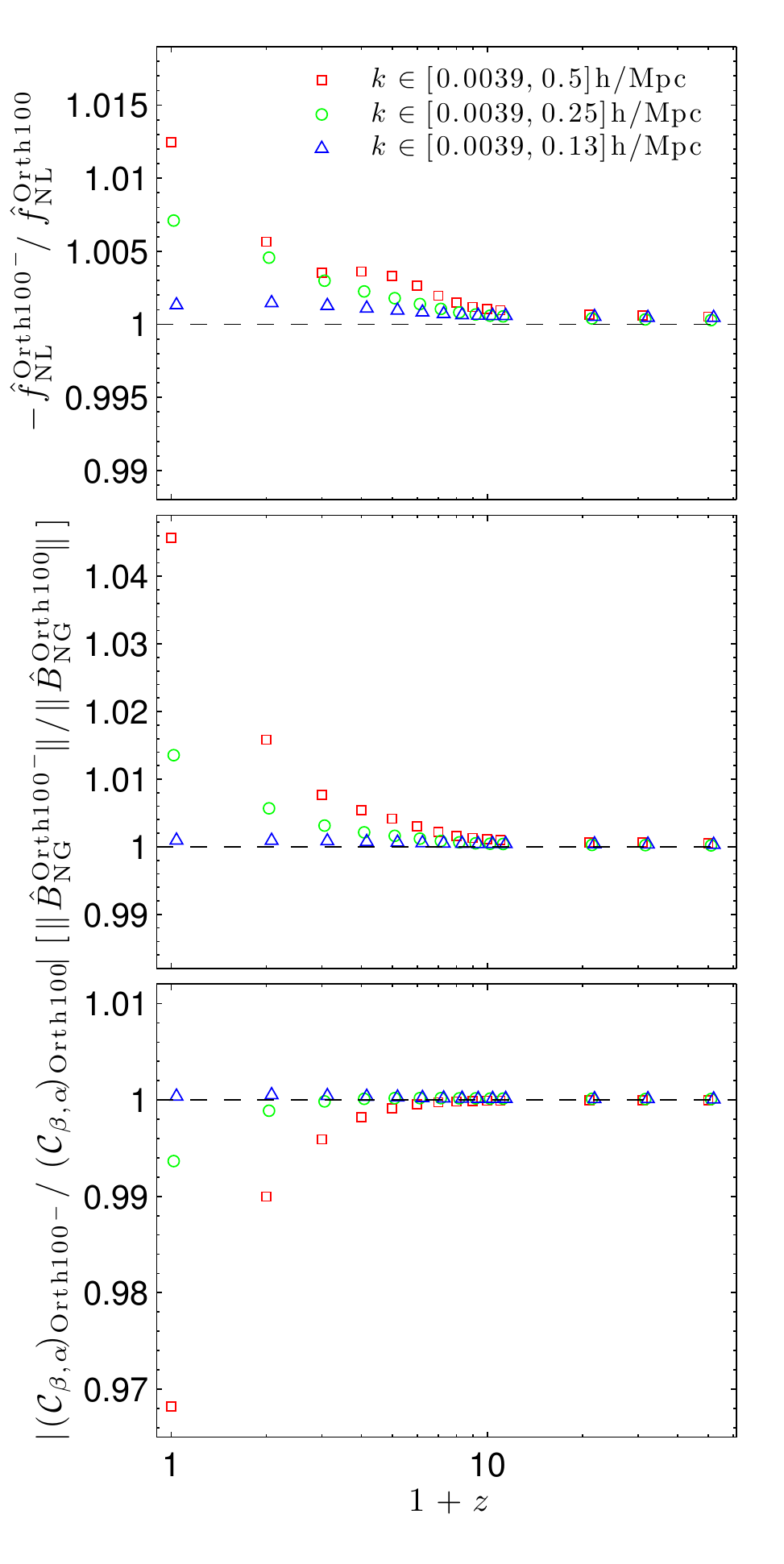}
\label{fig:linearity_orth}}
\mycaption{ Test if the bispectrum contributions \eqref{eq:B_ng_meas}
  from non-Gaussian initial conditions
 are linear in the input $f_\mathrm{NL}$.  (a) Ratios of
  Loc20 (solid line), Loc50 (dash-dot line) and $\mathrm{Loc10}^-$
  (dotted line) to Loc10 simulation. The label 'loc' stands for Loc20,
  Loc50 and $\mathrm{Loc10}^-$.  We scale the curves by
  $(f_\mathrm{NL}^\mathrm{loc})_\mathrm{in}/10$ so that in case of
  perfect linearity in $f_\mathrm{NL}$ all curves were unity. (b) Ratio of
  $\mathrm{Eq100}^-$ to Eq100 simulation. (c) Ratio of
  $\mathrm{Orth}^-$ to Orth100 simulation.}
\label{fig:linearity_fnl_all}
\end{figure*}

We test if $\mathcal{O}(f_\mathrm{NL}^2)$ corrections are important by
comparing simulations with $f_\mathrm{NL}^\mathrm{loc}=-10,20$ and
$50$ to the $f_\mathrm{NL}^\mathrm{loc}=10$ simulation in
\fig{linearity_loc} (with box size $L=1600\mathrm{Mpc}/h$). The curves
in all three panels would be exactly unity in case of perfect
linearity in the input $f_\mathrm{NL}$. Deviations from linearity are
at most $1\%$.

In case of equilateral and orthogonal initial conditions we compare
simulations with input $f_\mathrm{NL}=\pm 100$. The equilateral shape
gives similar results to the local shape, see \fig{linearity_eq}. For
the orthogonal shape the shape correlation between theory and
measurements deviates from linearity in the input $f_\mathrm{NL}$ by
about $3\%$ at $z=0$ for $k_\mathrm{max}=0.5h/\mathrm{Mpc}$ and the
total integrated bispectrum differs by less than $5\%$. These two
effects cancel each other approximately so that the measured
projection $f_\mathrm{NL}^\mathrm{orth}$ only deviates by less than
$1.5\%$ from linearity.

We conclude that  at the $\mathcal{O}(5\%)$ level the bispectra
measured in our large scale simulations can be scaled to other values
$f_\mathrm{NL}^\mathrm{new}$ fulfilling
$|f_\mathrm{NL}^\mathrm{loc}|\le 50$, $|f_\mathrm{NL}^{\rm{Eq}}|\le 100$
and  $|f_\mathrm{NL}^{\rm{Orth}}|\le 100$ using the linear scaling
\begin{align}
  \label{eq:fnl_linear_extrapolation}
  B_\mathrm{NG}(f_\mathrm{NL}^\mathrm{new}) = \frac{f_\mathrm{NL}^\mathrm{new}}{f_\mathrm{NL}}\hat B_\mathrm{NG}(f_\mathrm{NL}).
\end{align}

\subsection{Fitting formulae for non-Gaussian simulations}

\subsubsection{Separable polynomial fits}
Matter bispectra for non-Gaussian initial conditions of the local,
equilateral, orthogonal and flattened type are described by the
$\beta^R_n$ coefficients in Table \ref{tab:betas-nbody}.  These
polynomial fitting formulae can serve as a starting point
for future work that relies on non-Gaussian dark matter bispectra in
the nonlinear regime.  

\subsubsection{Time-shift model fits}

\begin{figure*}[ht]
\centering
\hspace{-0.9cm}
\subfloat[][]{
\includegraphics[width=0.49\textwidth]{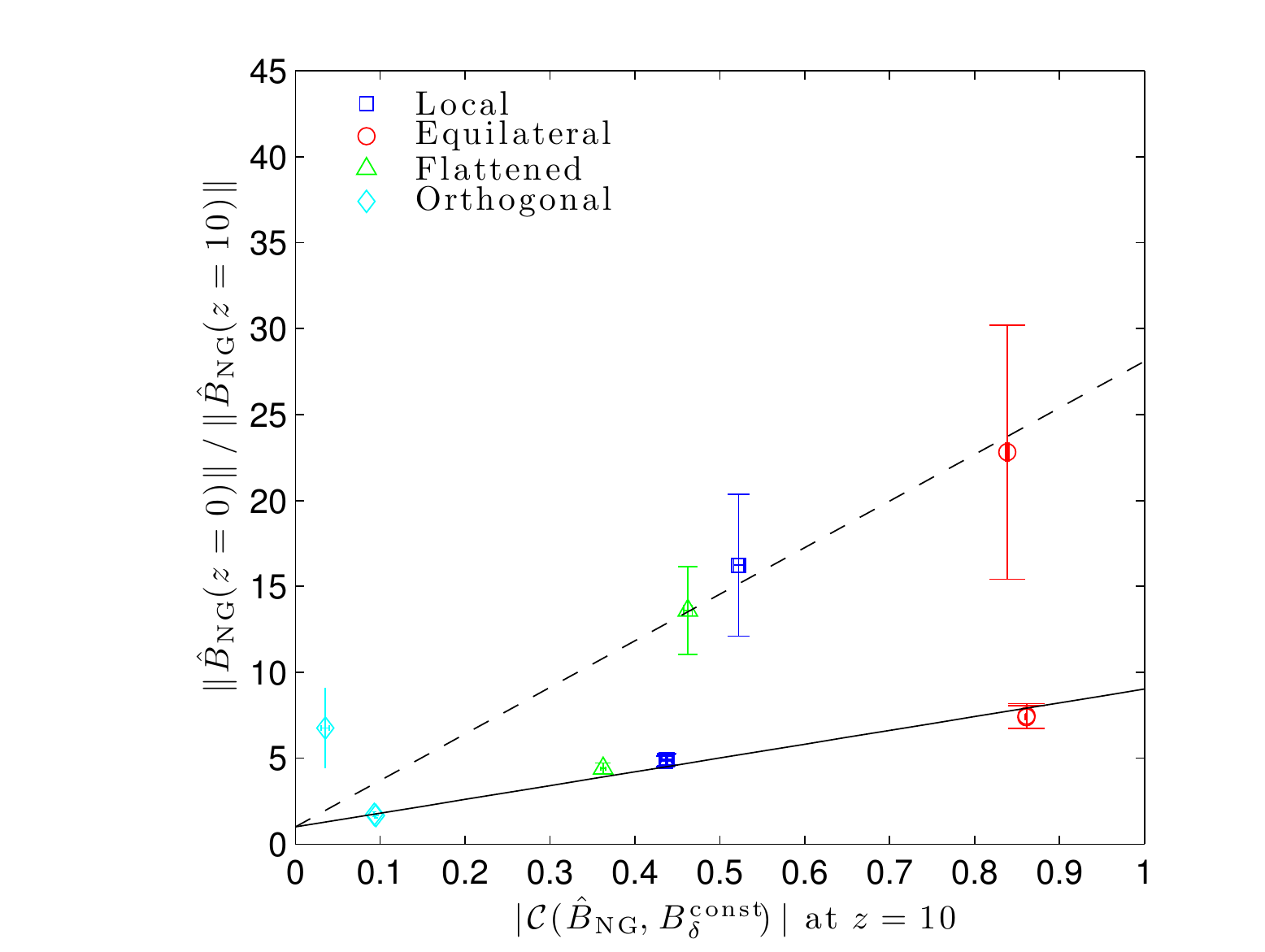}
\label{fig:NGgrowth_tmp1b}}
\subfloat[][]{
\includegraphics[width=0.49\textwidth]{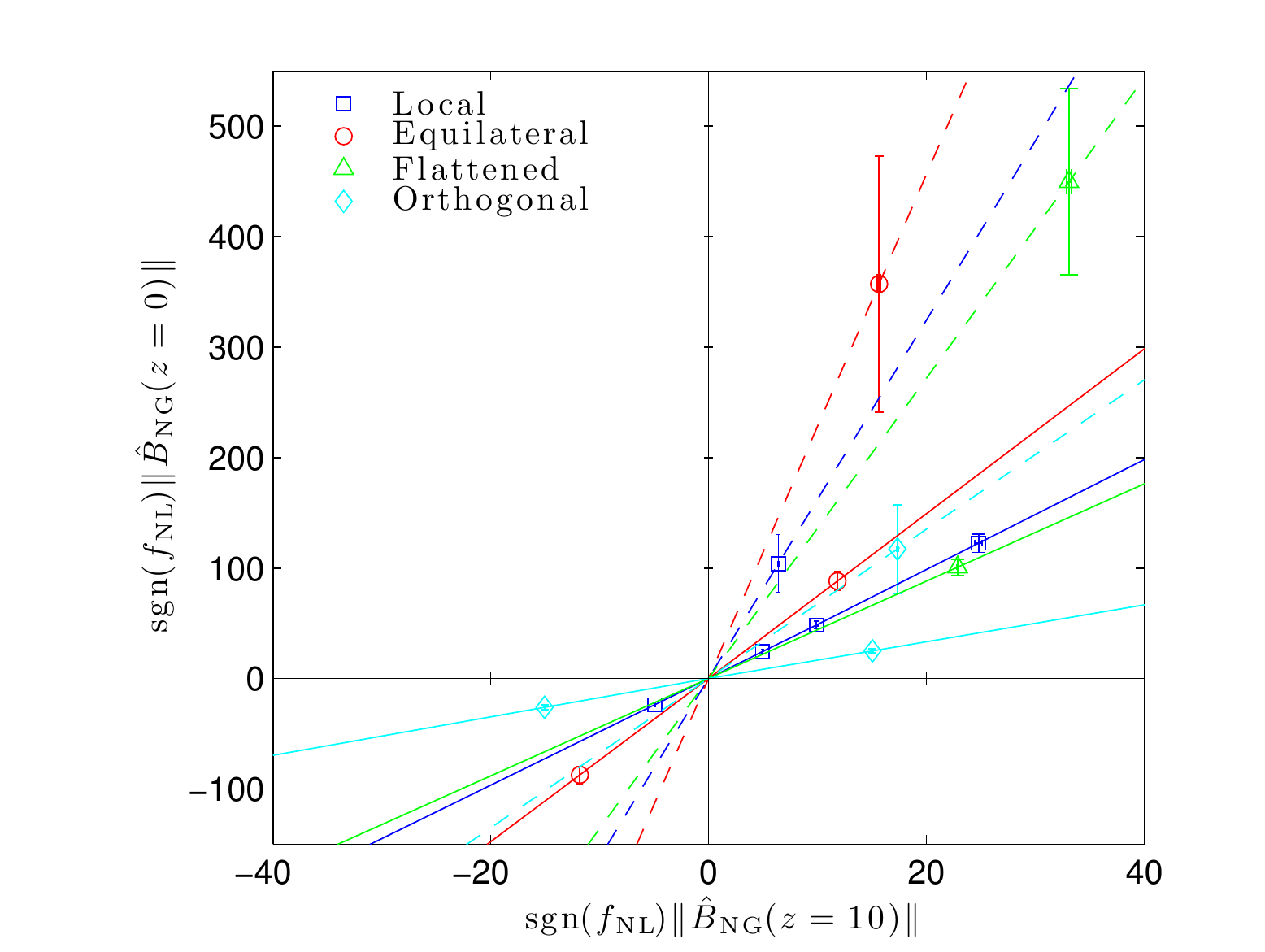}
\label{fig:NGgrowth_tmp1a_chaos}}
\caption{
(a)  Relative growth of bispectrum size between $z=10$ and $z=0$ as a
  function of the absolute value of the shape correlation between the measured non-Gaussian
  bispectrum $\hat B_\mathrm{NG}$ and the shape
  $B_\delta^\mathrm{const}$ defined in \eqref{eq:Bconst_grav} at
  redshift $z=10$, testing relation \eqref{eq:NGgrowth_motivation}.  The plot contains all simulations shown in (b) but
  data points which only differ in the input $f_\mathrm{NL}$ are
  almost indistinguishable. Regression lines through $(0,1)$ are shown for
  $k_\mathrm{max}=0.5h/\mathrm{Mpc}$ (continuous) and
  $k_\mathrm{max}=2h/\mathrm{Mpc}$ (dashed). 
(b) Bispectrum size at redshift $z=0$ as a function of the
  bispectrum size at high redshift, $z=10$, for different primordial
  shapes (see legend). Continuous lines correspond to
  $k_\mathrm{max}=0.5h/\mathrm{Mpc}$ while dashed lines correspond to
  $k_\mathrm{max}=2h/\mathrm{Mpc}$ simulations. Different points on
  one line show results for different input $f_\mathrm{NL}$. 
}
\label{fig:NGgrowth_all_tmp1}
\end{figure*}

Simple fitting formulae for the primordial
contribution to the matter bispectrum can be successfully obtained from the
halo time-shift model described in section III.  Before discussing this
 fit in detail, we perform a simple consistency check of the
basic idea of the model. The relatively fast growth of the constant
bispectrum implies that it constitutes the dominant contribution to
the non-Gaussian bispectrum $B_\mathrm{NG}$ at sufficiently small
scales and late time $z_\mathrm{late}$.  The amplitude of this constant
bispectrum is related to the projection of the non-Gaussian bispectrum
$B_\mathrm{NG}$ on the constant shape at the time $z_\mathrm{early}$
when halos start to form (and thereafter). Hence we expect that
\begin{align}
  \label{eq:NGgrowth_motivation}
  \|B_\mathrm{NG}(z_\mathrm{late})\| \propto\; 
&
  \mathcal{C}(B_\mathrm{NG}(z_\mathrm{early}),B_{\delta,\mathrm{const}}^\mathrm{grav}(z_\mathrm{early}))\nonumber\\ 
& \times \|B_\mathrm{NG}(z_\mathrm{early})\|.
\end{align}
This simple expectation is approximately seen in
\fig{NGgrowth_tmp1b} for the local, equilateral and flattened
shapes, confirming the basic idea of the time-shift model. The fact that the orthogonal shape deviates somewhat could be
related to the change of sign of the orthogonal shape for different
triangle configurations.  
Note that relation
\eqref{eq:NGgrowth_motivation} and \fig{NGgrowth_tmp1b} are
interesting results on their own because they show that the
relative growth of the non-Gaussian bispectrum can be predicted from
its correlation with the constant shape at early
times. \fig{NGgrowth_tmp1a_chaos} illustrates the absolute values of
the measured bispectrum sizes which were used to produce
\fig{NGgrowth_tmp1b}.

\begin{figure*}[htp]
\centering
\subfloat[][local, $k_\mathrm{max}=0.5h/\mathrm{Mpc}$]{
\includegraphics[width=0.31\textwidth]{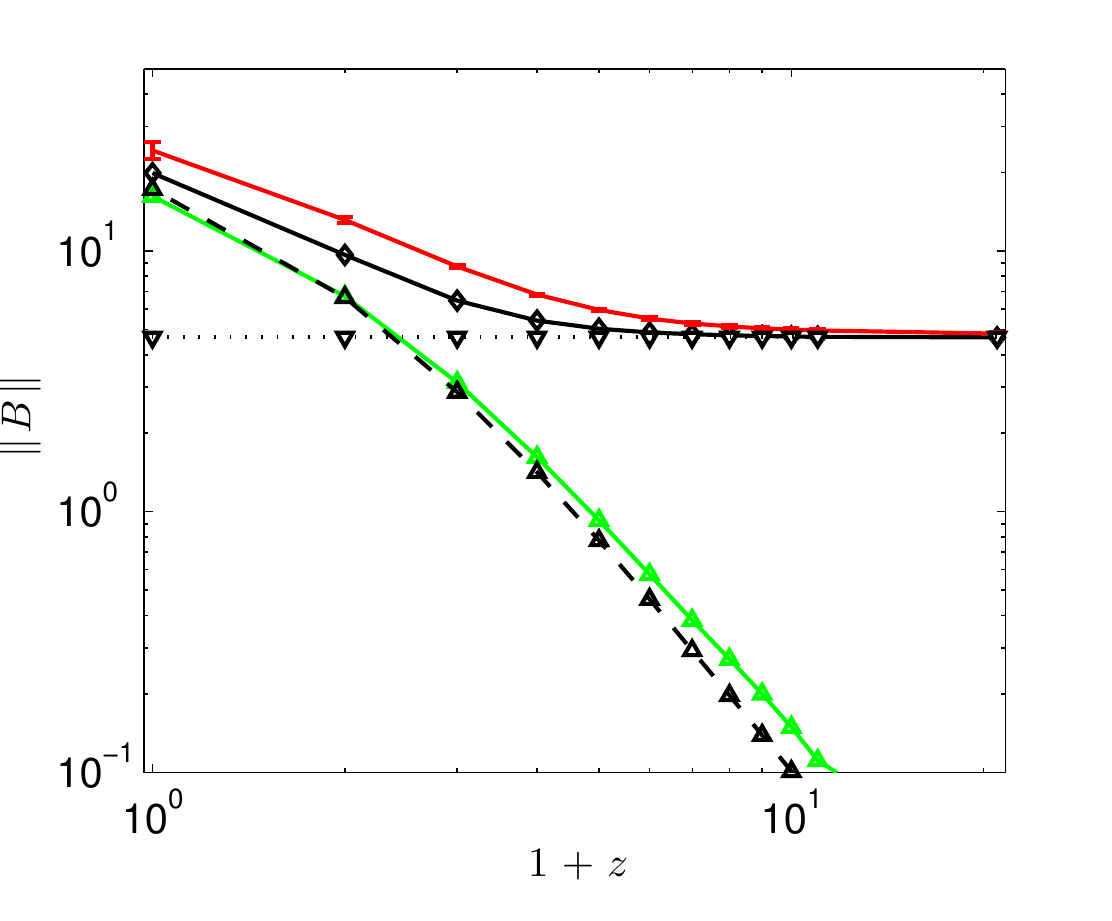}
\label{fig:msfitNG_tmp1a}}
\subfloat[][local, $k_\mathrm{max}=2h/\mathrm{Mpc}$]{
\includegraphics[width=0.31\textwidth]{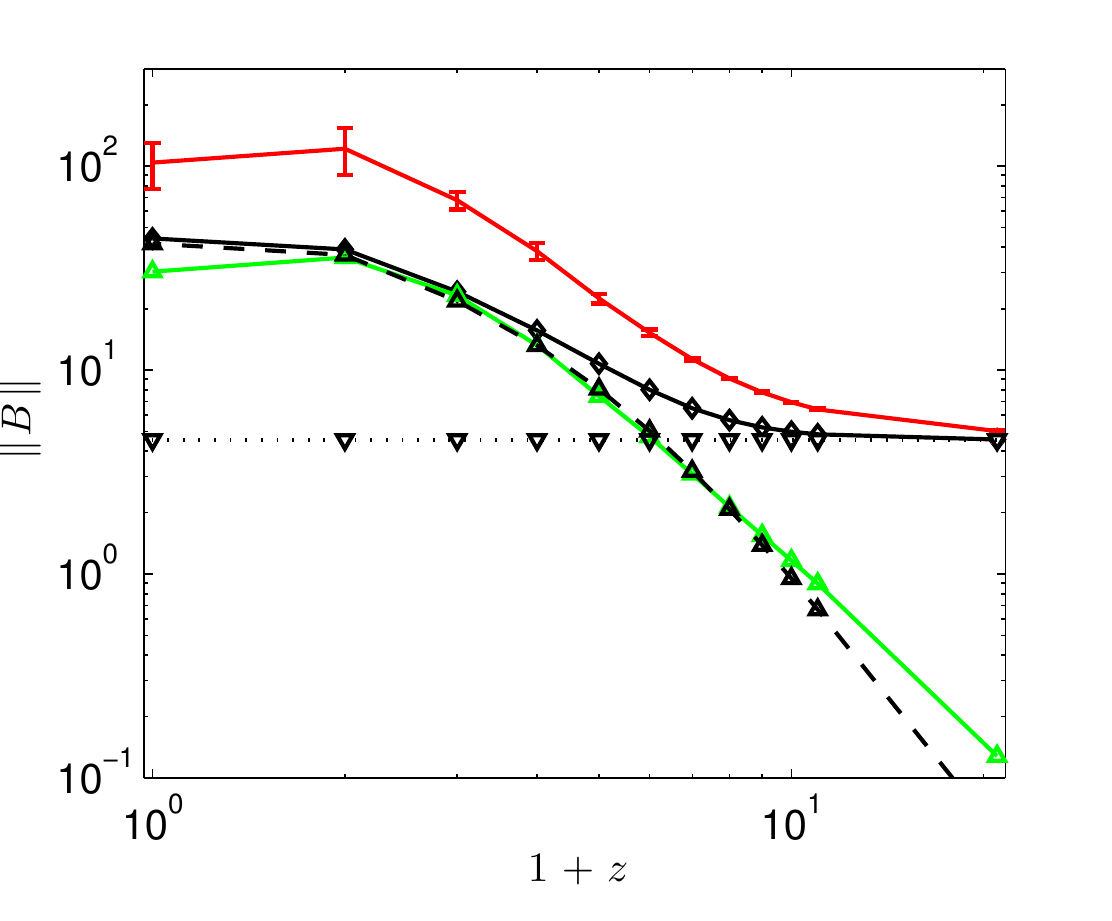}
\label{fig:msfitNG_kmax2_tmp1a}}
\\
\subfloat[][equilateral, $k_\mathrm{max}=0.5h/\mathrm{Mpc}$]{
\includegraphics[width=0.31\textwidth]{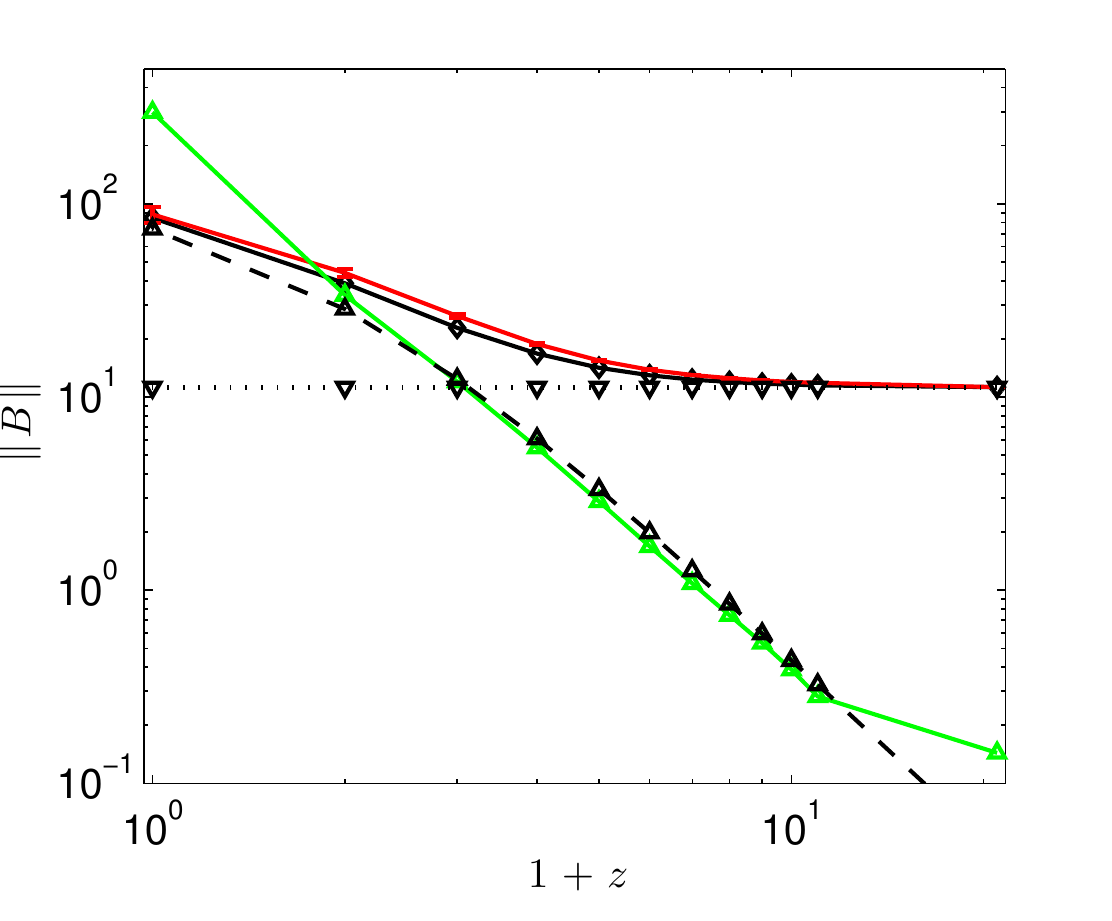}
\label{fig:msfitNG_tmp1b}}
\subfloat[][equilateral, $k_\mathrm{max}=2h/\mathrm{Mpc}$]{
\includegraphics[width=0.31\textwidth]{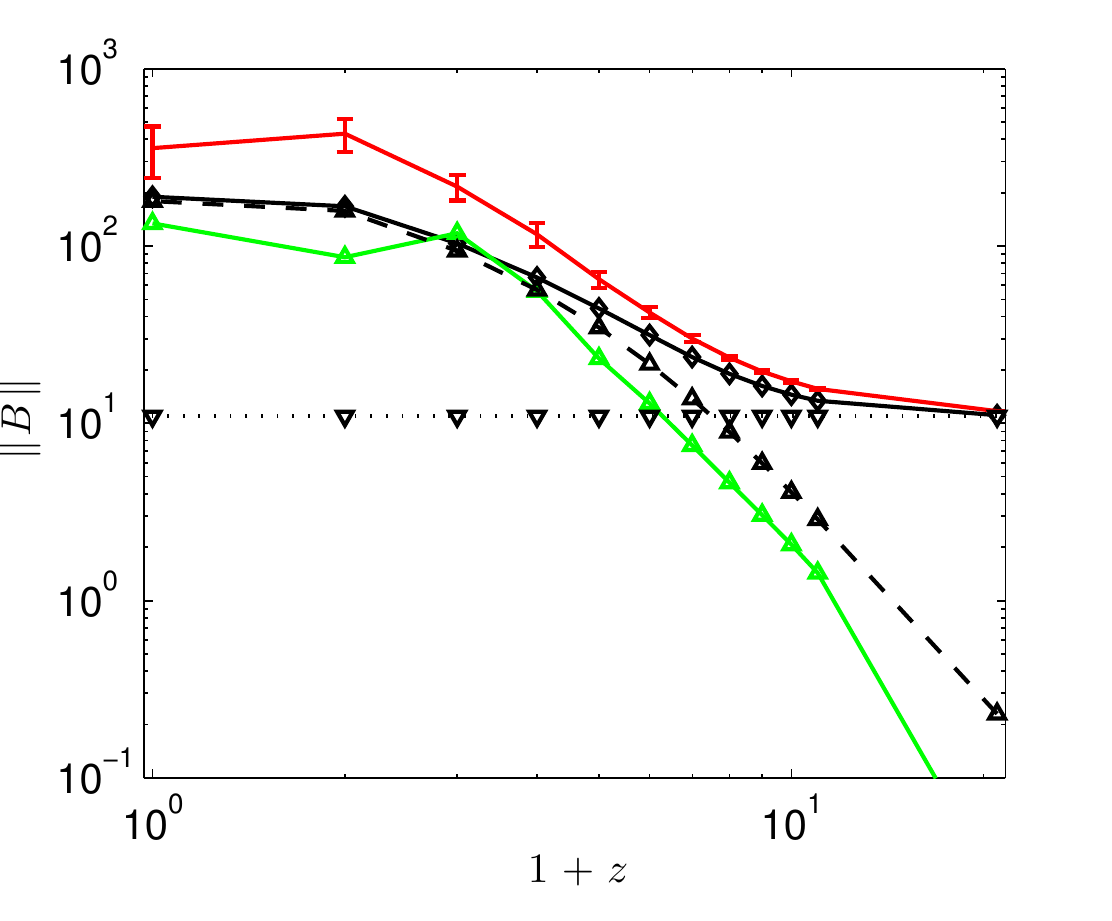}
\label{fig:msfitNG_kmax2_tmp1b}}
\\
\subfloat[][flattened, $k_\mathrm{max}=0.5h/\mathrm{Mpc}$]{
\includegraphics[width=0.31\textwidth]{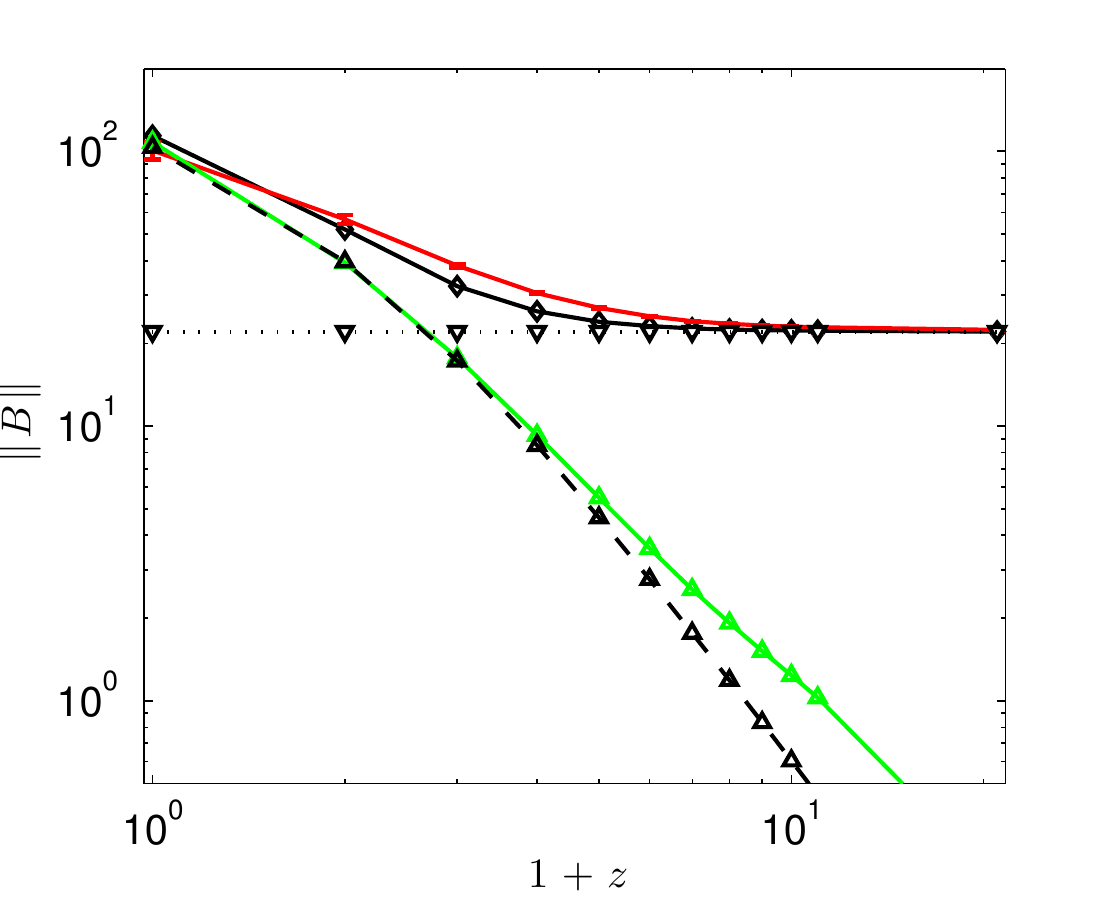}
\label{fig:msfitNG_tmp1c}}
\subfloat[][flattened, $k_\mathrm{max}=2h/\mathrm{Mpc}$]{
\includegraphics[width=0.31\textwidth]{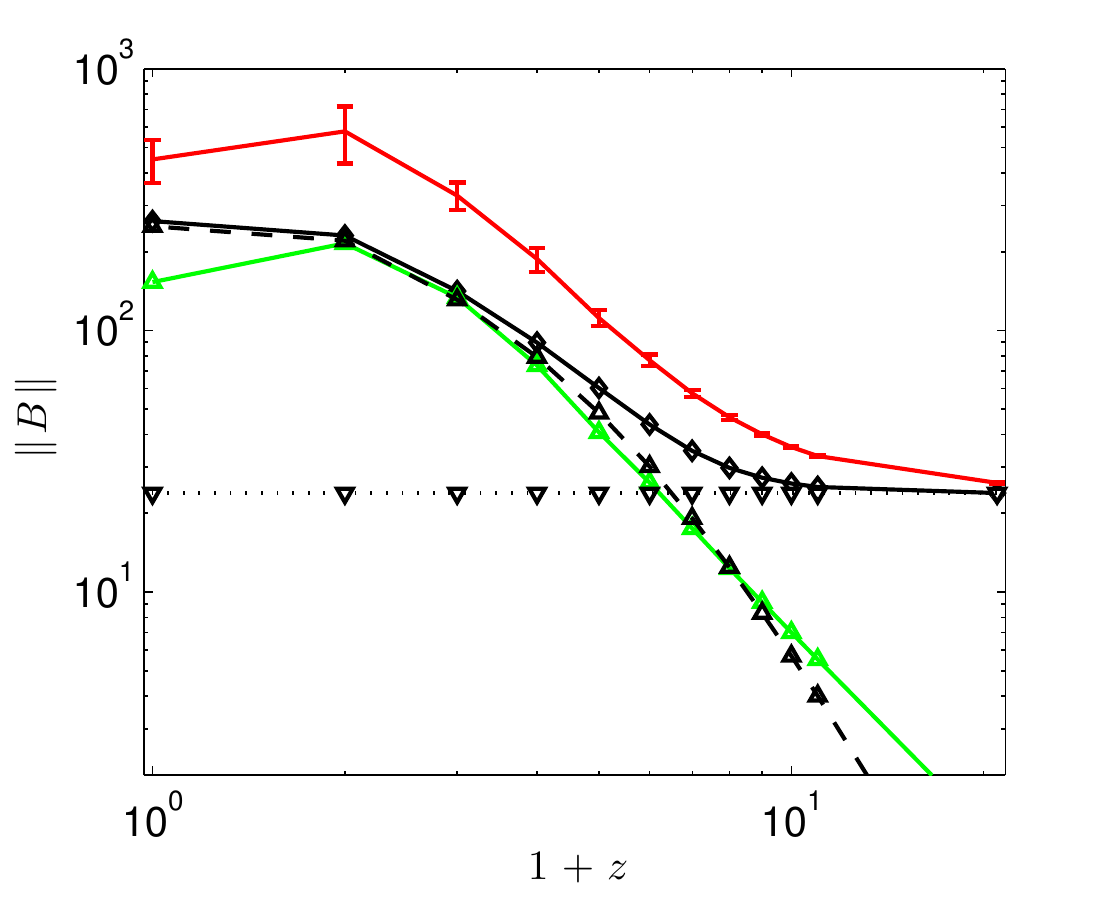}
\label{fig:msfitNG_kmax2_tmp1c}}
\\
\subfloat[][orthogonal, $k_\mathrm{max}=0.5h/\mathrm{Mpc}$]{
\includegraphics[width=0.31\textwidth]{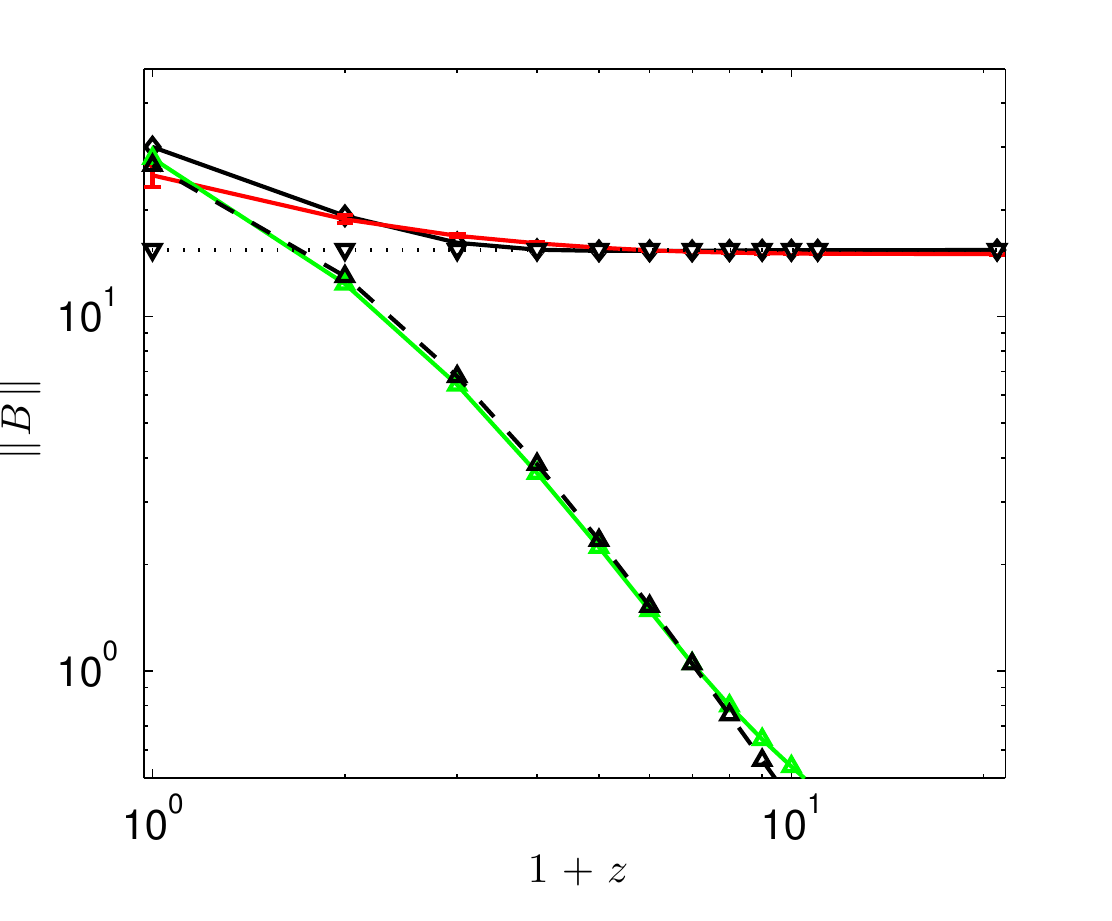}
\label{fig:msfitNG_tmp1d}}
\subfloat[][orthogonal, $k_\mathrm{max}=2h/\mathrm{Mpc}$]{
\includegraphics[width=0.31\textwidth]{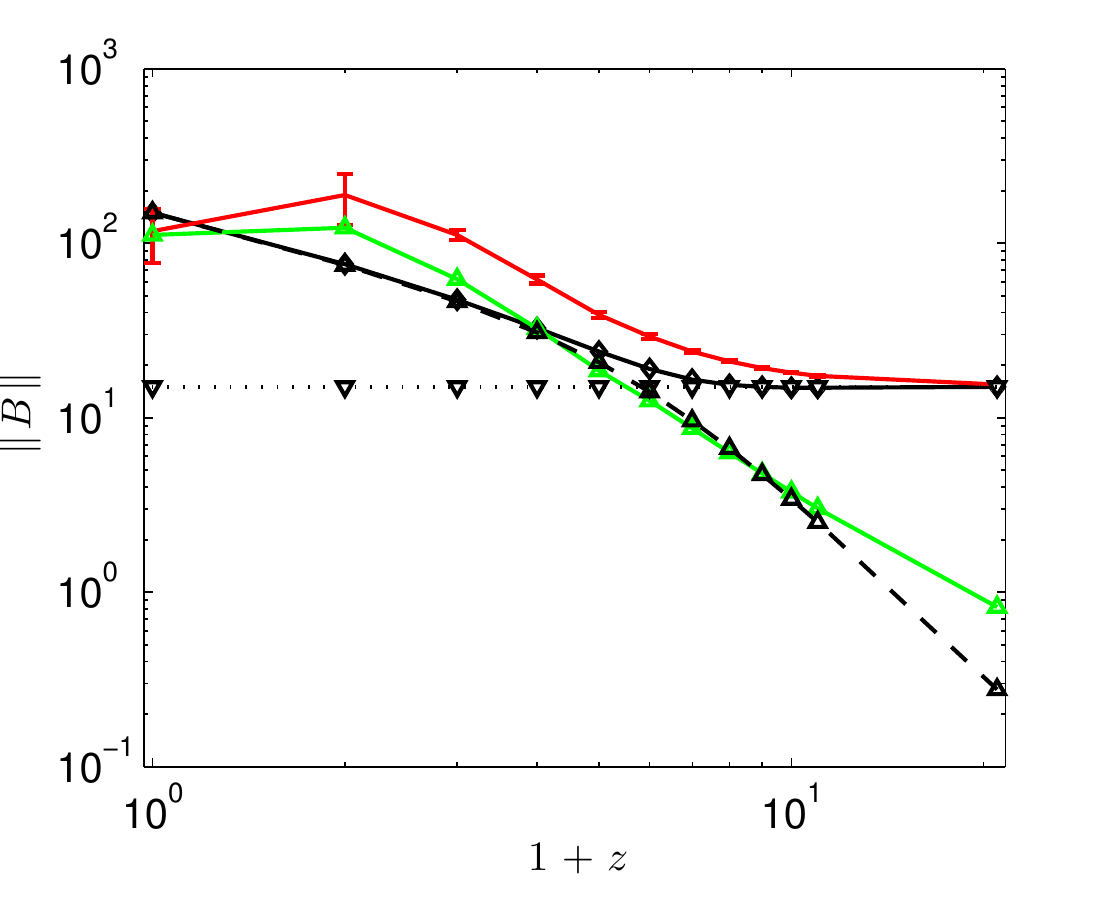}
\label{fig:msfitNG_kmax2_tmp1d}}
\caption{Contributions to
  simple non-Gaussian fits \eqref{eq:msfit3_NG}.  The arbitrary weight
  $w(z)$ in
  $B_\mathrm{NG}^\mathrm{opt}=B_{\delta,\mathrm{NL}}^\mathrm{prim}+
  w(z)(k_1+k_2+k_3)^\nu$ is determined analytically such that
  $\mathcal{C}(\hat B_\mathrm{NG},B_\mathrm{NG}^\mathrm{opt})$ is maximal (for
  $\nu=-1.7$).  We plot $\|B_{\delta,\mathrm{NL}}^\mathrm{prim}\|$
  (black dotted), $\|w(z)(k_1+k_2+k_3)^\nu\|$ (green) and
  $B_{\delta,\mathrm{const}}^\mathrm{prim}$ (black dashed) as defined
  in \eqref{eq:Bconst_prim} with fitting parameters given in Table
  \ref{tab:msfit_NG_table}.  The continuous black and red
  curves show $\|B_\mathrm{NG}^\mathrm{fit}\|$ from \eqref{eq:msfit3_NG}
  and the estimated primordial bispectrum size $\|\hat B_\mathrm{NG}\|$,
  respectively. }
\label{fig:non-gaussian-msfit-all_kmax05}
\end{figure*}

In detail, the simple fitting formulae for the non-Gaussian bispectra
are obtained by combining the partially loop-corrected tree level
expression \eqref{eq:bprim-nonlinear-evolved} with the constant shape
\eqref{eq:Bconst_prim} as 
\begin{align}
  B_\mathrm{NG}^\mathrm{fit}(k_1,k_2,k_3)\equiv f_\mathrm{NL}\big[
B_{\delta,\mathrm{NL}}^\mathrm{prim} 
 +
B_{\delta,\mathrm{const}}^\mathrm{prim}
\big].
  \label{eq:msfit3_NG}
\end{align}
The fitting parameters $c_2$ and $n_h^\mathrm{prim}$ in
\eqref{eq:Bconst_prim} are listed in Table
\ref{tab:msfit_NG_table}. Similar to the Gaussian case they
were obtained by analytically determining the optimal weight $w(z)$ for the
`constant' $(k_1+k_2+k_3)^\nu$ contribution and approximating this with
$c_2\bar D(z)^{n_h^\mathrm{prim}}$ (see green and black dashed lines in
\fig{non-gaussian-msfit-all_kmax05}). As expected from Section III we
find $n_h^\mathrm{prim}=n_h-1$ for local, equilateral and flattened
initial conditions.

Table \ref{tab:msfit_NG_table} also shows the shape correlation with
the measured excess bispectrum \eqref{eq:B_ng_meas}. These shape
correlations are remarkably good given the simplicity of
\eqref{eq:msfit3_NG}, especially for local, equilateral and flattened
initial conditions. The impact of the orthogonal shape on the matter
bispectrum seems to be somewhat harder to model, which is reflected in
the fact that our simple model performs worse for this shape.  
This is expected because the orthogonal shape contains almost no
constant component, the dominant evolution of which is the basis for 
the simple time-shift model.  
While the integrated bispectrum size $\bar F_\mathrm{NL}$ of the fitting
formulae is consistent with the measured bispectrum size at high
redshift and at $z=0$, it underestimates the measured size at the
$10-20\%$ level at intermediate redshifts for
$k_\mathrm{max}=0.5h/\mathrm{Mpc}$.  This could be improved by adding
more shapes or by using a redshift-dependent normalisation factor in
\eqref{eq:msfit3_NG} similar to the $N_\mathrm{fit}$ factor in the
Gaussian case. On smaller scales, $k_\mathrm{max}=2h/\mathrm{Mpc}$,
the non-Gaussian fitting formulae provide a less accurate overall fit, 
but we also list them in Table \ref{tab:msfit_NG_table} for
completeness and because the correlations at $z=0$ are quite high. 

It should be noted that simple fits like the ones presented here may be
somewhat more important in the mildly nonlinear regime than in the
strongly nonlinear regime because the halo model (extended to non-Gaussian
initial conditions) should be able to describe the strongly nonlinear regime.
We leave more accurate extensions of the simple fits for non-Gaussian
initial conditions and more quantitative comparisons with loop
corrections and halo model predictions for future work \cite{Reganetal2012}.

\subsubsection{Redshift at which 1-halo contribution becomes
  important}
We determine the redshift $z_*$ at which the approximate 1-halo bispectrum
$B_\delta^\mathrm{const}$ in the simple fits of the bispectrum for
non-Gaussian initial conditions becomes important by
matching 
\begin{align}
  \label{eq:zhaloformation_matching}
  \|\mathcal{C}(B_{\delta,\mathrm{NL}}^\mathrm{prim}(z_*),
  B_{\delta,\mathrm{const}}^\mathrm{prim}(z_*)) B_{\delta,\mathrm{NL}}^\mathrm{prim}(z_*)
  \|
= \| B_{\delta,\mathrm{const}}^\mathrm{prim}(z_*) \|.
\end{align}
For $k_\mathrm{max}=0.5h/\mathrm{Mpc}$ we find $z_*=3.5, 3.3, 4.3$ for
local, equilateral and flattened initial
conditions, respectively. For $k_\mathrm{max}=2h/\mathrm{Mpc}$ we find
instead $z_*=8.5, 8$ and  $9.6$ for the same initial conditions.
The orthogonal shape has almost no constant part. We therefore get
 somewhat different values of $z_*=5.5$ and $11$ for
$k_\mathrm{max}=0.5h/\mathrm{Mpc}$ and
$k_\mathrm{max}=2h/\mathrm{Mpc}$, respectively.  This demonstrates
that the time at which nonlinear structures contribute significantly
to the perturbative prediction for the primordial bispectrum is
earlier on smaller scales.

\section{VIII. Summary and Conclusions}

We have presented an implementation of a bispectrum estimator for
$N$-body simulations using a separable modal expansion of the bispectrum as
described in \cite{shellard1008}.  While a brute force estimation of the full
bispectrum is computationally expensive, requiring $\mathcal{O}(N^6)$
operations for $N$ particles per dimension, we find that the gravitational and the
most prominent primordial bispectra can be approximated by only
$n_\mathrm{max}=\mathcal{O}(50)$ separable basis functions (for the range
of scales relevant for $N$-body simulations).  The
bispectrum projection on the corresponding subspace of all possible
bispectra is estimated with $\mathcal{O}(n_\mathrm{max}N^3)$
operations, which is faster than brute force estimation by a
factor of $\mathcal{O}(N^3/n_\mathrm{max})\sim \mathcal{O}(10^7)$ for
typical simulations.  Thus the computational cost for accurate 3D bispectrum
estimation is almost negligible compared to the cost for running
the $N$-body simulations (e.g.~we can estimate the full bispectrum of a
$1024^3$ grid up to $k_\mathrm{max}=\tfrac{N}{4}\tfrac{2\pi}{L}$ in
one hour on only $6$ cores). This allows us to estimate the bispectrum as a 
standard simulation diagnostic
whenever the power spectrum is measured. 
  Expressing the bispectrum using its
$n_\mathrm{max}$ separable components yields a radical compression of
data which simplifies further analysis like comparisons between theory
and simulations. 
The separable bispectrum estimator therefore provides a
very useful additional statistic characterising the formation of
structures in $N$-body simulations, with high sensitivity to different
shapes of primordial non-Gaussianity corresponding to different models
of inflation.

We have performed many $N$-body simulations with Gaussian initial conditions as well as non-Gaussian
initial conditions of the local, equilateral, orthogonal and
(non-separable) flattened shape, exploiting the separable bispectrum
expansion for efficient generation of initial conditions as described
for primordial fields in 
\cite{shellard1008,shellard1108}.
On large scales the measured gravitational and primordial bispectra
agree with leading order perturbation theory and with measurements for
Gaussian initial conditions by \cite{verde1111} in the mildly
nonlinear regime, demonstrating the unbiasedness  of the
estimator and the initial conditions.  In the nonlinear regime, the
gravitational bispectrum becomes dominated by a large `constant' signal 
receiving elongated and equilateral
contributions not captured by tree level perturbation theory. However, it 
remains suppressed in the squeezed limit, where primordial bispectrum
signals can peak (see \fig{tet3dplots_grav_loc_eq_orth}).

Our measured $N$-body bispectra for Gaussian and non-Gaussian simulations
can be expressed by $50$ components
$\beta^R_n$, which we provide in Table \ref{tab:betas-nbody} for the key 
models.  They can be used as fitting formulae for the
gravitational and primordial bispectrum in the nonlinear regime. 

Less
accurate but simpler fitting formulae are obtained by modeling the
bispectrum as a combination of partially loop-corrected perturbative
bispectra and a simple `constant' bispectrum, which is constant on
slices $\sum k_i= \mathrm{const.}$ and
which is
obtained as an approximation to the $1$-halo bispectrum.  
While the former contribution dominates on large scales and
early times, the latter constant contribution dominates in the
nonlinear regime. 
Interpreting the effect of primordial non-Gaussianity on the
constant bispectrum contribution as a time-shift with respect to
Gaussian simulations allows us to model the time dependence of the
constant bispectrum contribution for non-Gaussian initial conditions.
For Gaussian initial conditions, the simple fits achieve a shape
correlation of at least $99.8\%$ with the measured gravitational
bispectrum for $z\leq 20$ and
$k_\mathrm{max}=\{0.5,2\}h/\mathrm{Mpc}$. For local, equilateral and
flattened non-Gaussian initial conditions the primordial bispectrum is
fitted with a shape correlation of at least $97.9\%$ at $z=0$ and at
least $94.4\%$ for $z\leq 20$ (with correlation typically $\gtrsim
98\%$ for most shapes and redshifts, see Table \ref{tab:msfit_NG_table}).
The impact of the orthogonal shape seems to be somewhat harder to
model, because it does not have a constant component initially, 
but our simple fit still achieves correlations of at least $91\%$ for $z\leq 20$.

Throughout this work we have visualised the measured bispectra in
three-dimensional tetrapyd plots \cite{shellard0912}, which show the
bispectrum shape and amplitude at different length scales and
generalise commonly used plots of one- or two-dimensional slices
through the tetrapyd volume. For a more quantitative analysis,
particularly to test analytic predictions and fitting formulae,
 we
have made extensive use of full three-dimensional shape correlations,
the cumulative signal-to-noise of the bispectra and their projection
$f_\mathrm{NL}$ on theoretical shapes. These quantities have been
evaluated extremely efficiently using the bispectrum components
obtained from the separable estimator.

We find that regular grid initial conditions produce an initial
spurious bispectrum due to the anisotropy of the regular grid, which
can be avoided by using glass initial conditions. However the
difference between regular grid and glass initial conditions decreases
with time as gravitational perturbations grow such that both initial
conditions yield similar results at late times. Effects of order
$f_\mathrm{NL}^2$ were shown to affect the bispectrum measurements by
less than $5\%$ in our large scale simulations.

Clearly, further work is required to study the effects of general
primordial non-Gaussianity on observable quantities like the
bispectrum of galaxies, particularly in the nonlinear
regime. However the present work represents, we believe, an important
step forward in the understanding of structure formation in the
presence of primordial non-Gaussianity and the search for primordial
non-Gaussianity in large scale structures.

\section*{Acknowledgements}
We are especially grateful to James Fergusson for his work on the
development of the separable bispectrum estimation formalism, for
sharing his experience with the application to the CMB and for many
informative and illuminating discussions regarding the application to
large scale structures. We thank Anthony Challinor for comments on the
manuscript and Daniel Baumann,
Francis Bernardeau, Anthony Challinor, Hiro Funakoshi, Helge Gruetjen,
Martin Haehnelt, Antony Lewis and Emiliano Sefusatti for very helpful
discussions.  We are also very grateful to Andrey Kaliazin for his
invaluable computational help.  We thank Volker Springel for providing
us the Gadget-3 $N$-body code (see \cite{gadget1,gadget2} for older
versions). We also thank Edwin Sirko for his initial conditions code
\cite{sirko}, parts of which we use for setting up initial conditions,
Sebastian Pueblas and Roman Scoccimarro for their public 2LPT code for
Gaussian initial conditions
\cite{scocci-transients0606} and Antony Lewis and
Anthony Challinor for CAMB \cite{camb}. We acknowledge the use of the
FFTW3 library and SPLASH \cite{splash} for visualising particle distributions.  Simulations were performed on the COSMOS supercomputer
(an Altix 4700) which is funded by STFC, HEFCE and SGI. MMS was
supported by the Science and Technology Facilities Council, DAMTP
Cambridge and St John's College Cambridge. DMR was supported by the
Science and Technology Facilities Council grant ST/I000976/1.  EPSS
was supported by the STFC grant ST/F002998/1 and the
Centre for Theoretical Cosmology.

\bibliography{LSS_ms}

\end{document}